\documentclass[prd,12pt,preprintnumbers,nofootinbib,showpacs,twoside]{report}

\usepackage[latin1]{inputenc}			
\usepackage[english]{babel}
\usepackage{latexsym}
\usepackage{graphicx}
\usepackage{epsfig,amsmath,amssymb}

\begin{document}

\begin{titlepage}
\pagestyle{empty}
\pagenumbering{Roman}
\begin{large}

\begin{figure}[htbp]
\begin{center}
\includegraphics[width=2cm]{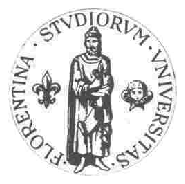}
\end{center}
\end{figure}

\begin{center}
Universit\`a degli studi di Firenze\\
Dipartimento di Fisica\\
{ Tesi di Dottorato in Fisica }\\
\vspace{1.0cm}
\begin{LARGE}
\textbf{A study of meson condensation and \\\vspace{.16cm} of  the QCD critical line\\\vspace{.4cm}}
\vspace{2cm}\bf{Lorenzo Ravagli}
\end{LARGE}
\end{center}
\end{large}
\begin{large}
\vspace{1.8cm}
\begin{flushleft}
\textit{Supervisore:}\ \ \ \ \ \ \ \  \ \ \ \ \ \ \ \ \ \ \ \ \ \ \ \ \ \ \ \ \ \ \ \ \ \ \ \\   
\vspace{.4cm}
{\bf Dr. Giulio Pettini} 
\vskip0.9cm

\vspace{0.6cm}
\end{flushleft}
\vspace{0.6cm}
\centerline{Anno Accademico 2005-2006}
\end{large}
\end{titlepage}
\begin{large}
\thispagestyle{empty}
\begin{center}
\textbf{~} 
\end{center}
\vspace{2cm}
\end{large}

\baselineskip 19pt
\textheight 22cm
\begin{large}
\begin{center}
\textbf{Ringraziamenti} 
\end{center}
\end{large}
Ringrazio anzitutto la mia famiglia per aver supportato la mia avventura nel campo della ricerca scientifica.
Voglio ringraziare il mio Supervisore, Dr. Giulio Pettini, per aver seguito con passione e cura questo lavoro di tesi.
Devo essere grato in modo particolare verso il Prof. Andrea Barducci per l'aiuto
offertomi nei numerosi momenti di bisogno.
Ringrazio il Prof. Roberto Casalbuoni per aver fornito lo spunto iniziale a questo progetto e per avere partecipato con interesse agli sviluppi dello stesso.
Ringrazio inoltre il Dr. Francesco Becattini per l'interessamento mostrato in questi anni verso la mia attivit\`a scientifica.
Saluto inoltre tutti i compagni/e con cui ho condiviso anni ed anni nel dipartimento di Fisica di Firenze.

\pagenumbering{Roman}
\setcounter{page}{2}
\tableofcontents
\newpage
\pagenumbering{arabic}
\setcounter{page}{1}

\addcontentsline{toc}{chapter}{Preface}
\chapter*{Preface}
\label{Preface}

Quantum Chromodynamics (QCD) is the theory thought as fundamental \footnote{In the framework of standard model. QCD can be otherwise considered as an effective theory deriving from more fundamental interactions (SuperSymmetry, String Theory), but in the following work we will not deal with such theories.} for describing strong interactions. The invariance under $U(1)$ gauge transformations of Quantum Electro-Dynamics (QED) is extended to a non-abelian gauge one $SU(3)$; this fact implies that gluons, the intermediate bosons of strong interactions, are charged, and thus self-interacting. This feature makes QCD deeply different from the abelian theory.
The study under the renormalization group of the strong coupling, hence of the $\beta$ function, must take into account  all of the gluons self-interacting diagrams. These terms give an opposite contribution to the screening phenomenon present in QED, and the consequence is a color charge growing with the distance. On the other hand, in the short distance regime the strong coupling goes to zero, and the theory displays the mechanism of asymptotic freedom (this aspect of the theory caused the Nobel Prize 2004 to the american physicists Gross, Wilczek and Politzer who had anticipated this feature in the early seventies \cite{Gross:1973id,Politzer:1973fx,Gross:1973ju}).
Therefore, in QCD we can distinguish two opposite limits: the one of high energies, where it is possible to apply a perturbative expansion, and the one of low energies. In the latter regime, the perturbative expansion becomes meaningless and in order to compute physical quantities we must use alternative techniques or effective models.
The physical scale separating the two aforementioned regimes, $\Lambda_{QCD}$, is of the order of a few hundred $\mbox{MeV}$ (from recent determinations $\Lambda_{QCD}\sim 200~\mbox{MeV}$). Scattering processes characterized by energies  $>1~\mbox{GeV}$ have been checked to be in agreement with perturbative QCD predictions. 
In this work, we will face mainly the non-perturbative
aspects of the theory. 

 The main QCD open problem, not yet resolved in a satisfactory way, is confinement: quarks are not physical states which can be observed as quasi-particles, but only as bound states which correspond to the hadrons; these objects are forced to be color neutral. In practice, the potential energy for a couple $q\bar{q}$ grows indefinitely with the separation distance, so that it is impossible to separate them unless another quark-antiquark couple is created from the vacuum.
Therefore, the only way to break a color string is to divide it into two shorter ones. Confinement deals with the long distances regime of the theory, and is therefore intrinsecally a non-perturbative phenomenon.

Another non-perturbative feature of the theory is the dynamical breaking of chiral symmetry. If we consider QCD with $N_f$ massless fermions, the Lagrangian possesses a global symmetry $SU(N_f)_L \otimes SU(N_f)_R$.
Chiral symmetry is not observed in nature, since quantum interactions lead to the formation of a chiral condensate $\langle\bar{\Psi}\Psi\rangle$ which spontaneously breaks the symmetry $SU(N_f)_L \otimes SU(N_f)_R$ to its subgroup $SU(N_f)_V$. In this case we speak of dynamical breaking since it is the dynamics of the interactions which is responsible for the breaking.

Chiral symmetry breaking implies many theoretical and phenomenological consequences; moroeover, it is closely connected to the confinement problem, because both phenomena are related to special configurations of the gauge field. These issues will be discussed in the Introduction.
Here we stress that confinement and chiral symmetry breaking are both non-perturbative effects, and we cannot deal with them by means of a perturbative diagrammatic expansion. 
We can analytically explain confinement only in suitable extensions of the fundamental theory.
For this reason, a study of such behaviours starting from QCD itself is a hard goal, which has not been satisfactory resolved yet. On the other hand, a class of effective models are able to mimic the basic features of the non-perturbative regime, even if not starting from first principles. When dealing with effective models, which are obtained as simplified versions of the fundamental theory, the study with different approaches of a physical problem is useful to make a result stable and possibly model independent. In the following chapters some of these effective models will be presented; in particular, we will focus our attention to the ladder-QCD and Nambu-Jona-Lasinio (NJL) models, which have been used to carry out this specific work.

A powerful tool to deal with the non-perturbative problem of strong interactions is to consider QCD on a space-temporal lattice; starting from the QCD Lagrangian, and by means of numerical Monte-Carlo simulations, it should be possible to derive all the physical quantities related to QCD without adding new free parameters. The problems related to this kind of approach are mostly of numerical nature\footnote{The difficulty of considering finite quark chemical potentials in lattice simulations will be dicussed later on in this work.}; first, we have to check that finite lattice spacing does not infer the determination of physical observables; second, the needed power of calculus is huge, and present supercomputers are still unable to push the analyses to the final step. In any case, waiting for an improvement of resources with new generation processors, we can think of lattice QCD as the closest ``effective model'' to the fundamental theory.

Up to now, we have considered properties of the QCD vacuum. On the other hand, the introduction of finite temperature and charge densities 
can cause the restoration of symmetries which are broken in the vacuum, namely at zero temperature and density. Actually, the restoration of chiral symmetry in the regime of high temperatures and/or densities is a stable feature amongst several models.
Moreover, from the early seventies, it has been argued that the restoration of chiral symmetry may coincide with the transition to the deconfined phase of the theory, the QGP (Quark-Gluon Plasma) regime, a plasma of weakly interacting quarks and gluons. In this new regime, the constraint of color neutrality relaxes, and it should possible to think of quark and gluons as quasi-particles moving almost freely through the medium. 

The possibility of considering QCD beyond the vacuum state is fundamental for establishing a comparison with realistic experimental settings.
QCD in the regime of high temperatures regards the physics of Relativistic Heavy-Ion collisions, where we can assume the formation of clusters characterized by a high amount of kinetic energy. The investigation of this context is stimulated by the perspective of getting an experimental proof of deconfinement.
On the other hand, the low temperatures and high densities regime is relevant for studying the physics of compact stars. In that regime,  chromomagnetic forces induce an attractive quark-quark interaction which can lead to an instability of the Fermi surface and to the formation of Cooper pairs. 
If we replace electrons with quarks, and phonon exchange with gluon exchange, we find a complete analogy with superconductivity in ordinary materials.
For this reason, we refer to this phenomenon as Color Superconductivity (CSC). 

The physical scenario arising in these regimes is very different, and will be thoroughly discussed in the Introduction.
As for water we find a liquid, a solid, or a gaseous phase, depending on the values of the thermodynamic parameters, in QCD we can have a hadron phase, a QGP, or a CSC regime according to the temperature and densities involved: the study of the different regimes of QCD by varying the thermodynamic parameters defines the phase diagram of the theory. As a further analogy with water, we will find a critical point and a critical line of first order transitions. The critical point should belong to the universality class of the 3-d Ising model.
These are spectacular proofs of universality in Physics: we find a similar critical behaviour for systems which differ by about ten orders of magnitude in temperature, and which are characterized by radically different interactions.

In year 2000, a paper of Son and Stephanov \cite{Son:2000xc} has opened a new field of research in literature.
In the framework of a low energy chiral model, they have analyzed
the phenomenon of pion condensation in the regime of low temperatures and high isospin densities. Several analyses expect kaon condensation as well in other low temperatures regimes.
There are many reasons why this physics is intriguing.
First, from the theoretical point of view,
meson condensation can regard the physics of compact stars, and is closely related to color superconductivity.
Second, the effect of a low but finite isospin chemical potential is important in heavy ion experiments to give a signature of the expected transition between the hadronic and the QGP phases.  
Third, following the idea of Alford, Kapustin and Wilczek \cite{Alford:1998sd}, QCD at finite isospin densities can be simulated on the lattice by means of Monte-Carlo simulations, whereas standard techniques are unable to describe the regime of high baryon densities. 

These issues will be discussed more carefully in the Introduction and in Chapter II, but we anticipate here the motivation of the present work: 
the study of chiral symmetry restoration and meson condensation by using a microscopic model. 
 A model with quarks as fundamental degrees of freedom has a wider range of applicability than low energy models.
In addition, we can display chiral symmetry breaking and restoration as a dynamical effect.
In order to carry out this project, we have employed the models ladder-QCD and NJL.
Our goal has been to investigate a large slice of the space of thermodynamic parameters. 
In the following, a rapid outline of the work is presented.

In the Introduction, we present a general survey on the QCD phase diagram. We will discuss more carefully the influence of chiral symmetry breaking and restoration on the structure of the phase diagram, in particular in relation with the confinement problem.
We will discuss the phases of high and low temperatures, and the related phenomenology. We stress the presence of a critical point in the regime of intermediate temperatures and densities, since it is a promising feature to be experimentally highlighted.

In Chapter II, we will review the recent literature which has stimulated this work. In particular, we will be interested in QCD at finite isospin densities, where, for high enough values of isospin charges, the condensation of charged pions is expected. The formalism for describing Bose Einstein condensation in Quantum Field Theory is presented. 
In order to clarify the physical problems we have dealt with, we will discuss some theoretical issues concerning QCD at finite chemical potentials. 

From Chapter III on, the more specific research part of this work starts. 
Here we present a study of two-flavors QCD at non vanishing isospin density which has been carried out in the framework of the ladder-QCD model. The steps for deriving the two-loops effective action, encompassing a non-trivial non-perturbative dynamics, are elucidated.
The possibility of pion condensation at high isospin densities is investigated, as well as the phenomenological consequences of a non vanishing isospin chemical potential for heavy ions experiments.

In Chapter IV, we will perform the same analysis of Chapter II by using the NJL model. Due to technical reasons, this model is simpler to work with than ladder-QCD, and from now on will be used as the main tool for our studies.
A detailed investigation of the two-flavors QCD phase diagram, by varying arbitrarily the two light quark chemical potentials, has been done.

In Chapter V, we will extend the analysis to the strange flavor. We have studied the possibility of pion and kaon condensation in the space of thermodynamic parameters, by means of a three-flavors NJL model, by making a comparison with previous analyses performed within low-energy models. An original proposal for kaon condensation in the regime of high baryon densities has been suggested.
  
In Chapter VI we will focus our attention to the critical line for chiral symmetry restoration. The study will be performed in a three-flavors NJL model with an instanton-induced interaction. Three main issues are discussed: the dependence of the critical curve on baryon and isospin chemical potential; the sensitivity of the order of the chiral transition on the value of the strange quark mass; the dependence of the curve $T(\mu)$ on the number of massless flavors $N_f$.
Some other recent developments are briefly discussed too.
These analyses are motivated mainly by the growing interest that the lattice community has shown for this subject.
In this way, the consistency between effective models results and lattice QCD simulations can be checked.

In the Conclusions, an overview of the work done is afforded.

\chapter{Introduction: A survey of the QCD phase diagram}

In this introductory Chapter, we give an overall view on the properties of QCD at finite temperature and densities. To start with, we continue the treatment of the zero temperature features which will be most relevant for further applications.

\section{Vacuum properties of QCD}

We analyze the phenomenon of chiral symmetry breaking. If we start from the chiral limit, namely where the three light quarks are exactly massless, the QCD Lagrangian admits an invariance under chiral transformations 
\begin{equation}
U(3)_L \otimes U(3)_R\equiv SU(3)_L \otimes SU(3)_R \otimes U(1)_L \otimes U(1)_R
\end{equation}

Actually, in the chiral limit, left/right chirality components $\Psi_{L,R}=(1\pm\gamma_5)\Psi/2$ ($\Psi=(u,d,s)$ is the spinor in the Dirac and flavor space) decouple, and the global $U(3)$ flavor symmetry doubles into $U(3)_L\otimes U(3)_R$. 
A mass term in the QCD Lagrangian would lead to the mixing of left/right components
\begin{equation}\label{massterm}
m\bar{\Psi}\Psi=m(\bar{\Psi}_L\Psi_R+\bar{\Psi}_R\Psi_L)
\end{equation}
As far as the group $SU(3)_L \otimes SU(3)_R$ is concerned, the symmetry reads in the following invariance for $\Psi_{L,R}$
\begin{equation}
\Psi_{L,R}\rightarrow exp~(-i\lambda_a\alpha_a^{L,R})\Psi_{L,R}
\end{equation}
where $\lambda_a,~ a=1\dots 8$ are the Gell-Mann matrices (representatives of SU(3) generators) and $\alpha_a^{L,R}$ the corresponding arbitrary parameters.
For every multiplet of particles, this symmetry would imply the existence of a mirror multiplet with opposite parity; this is not a phenomenon observed in nature.
Actually, quantum interactions lead to the formation of a chiral condensate $\langle\bar{\Psi}\Psi\rangle$ which spontaneously breaks the symmetry $SU(3)_L \otimes SU(3)_R$ to its diagonal subgroup $SU(3)_V$. The phenomenology which follows from the spontaneous breaking of a global continuous symmetry is characterized by the presence of a set of Goldstone bosons, corresponding to the broken generators of the symmetry group \footnote{The Goldstone theorem applies only when a global symmetry is broken. In the case of breaking of a local symmetry, the degrees of freedom of the would-be massless particle give mass to the intermediate vector bosons. Since in particle Physics only approximate global symmetries are actually broken, genuine massless bosons have not been observed. Massless modes can occur instead in condensed matter Physics. We will encounter examples of this kind in the following of this work.}: in this case the three pions ($m_{\pi}\sim 140~\mbox{MeV}$), the four kaons ($m_{K}\sim 500~\mbox{MeV}$) and the $\eta$ particle ($m_{\eta}\sim 550~\mbox{MeV}$), the meson octet, can be interpreted as the pseudo-Goldstone bosons corresponding to the breaking of $SU(3)_A$ \footnote{This is a commonly used, but not quite exact notation. In fact $SU(3)_A$ is not properly a group (the commutator of two axyal charges is a vectorial one), but the counting of Goldstone bosons is in agreement with this naive analysis. However, in the following of this work we will use sometimes this notation.}. Their mass is little with respect to that of the other hadrons, and derives solely from the current mass term present in the QCD Lagrangian. Since the explicit breaking term is small if compared to the dynamical generated mass, we can still treat chiral symmetry as an approximate symmetry.

The formation of a chiral condensate would break in a similar way the residual symmetry $U(1)_L\otimes U(1)_R$ down to $U(1)_V$.
This would imply the spontaneous breaking of the extra $U(1)_A$ symmetry.
Actually $U(1)_A$ is not a symmetry observed in nature, because there not exists for every hadron a partner with the same mass and opposite parity. Moreover a ninth particle with a mass comparable with that of the octet and interpretable as pseudo-Goldstone for the breaking of $U(1)_A$ is not observed.
$U(1)_A$ is in fact an anomaly (it is called {\it axial} anomaly), that is a symmetry present at the classical level but violated by quantum interactions. This problem was resolved by Adler-Bell-Jackiw, who noticed the singular nature of the operator $\bar{\Psi}(y)\Psi(z)$ when considering $(y-z)\rightarrow 0$. 
The corresponding axial current $J_5^{\mu}$, which would be conserved at the classical level, satisfies 
\begin{equation}\label{axialcharge}\partial_{\mu}J_5^{\mu}=\frac{g^2}{16\pi^2}n_f~F^C_{\alpha\beta}\tilde{F}_{\alpha\beta}^C=2N_F~Q(x)\end{equation}
 where $N_f$ is the number of fermion flavors, $g$ the coupling constant and $F^C_{\alpha\beta}$ the gluon field strength tensor ($\tilde{F}_{\mu\nu}^A=\epsilon_{\mu\nu\rho\sigma}F_{\rho\sigma}^{A}$ is the dual gluon field strength tensor);\\ $Q(x)=\frac{g^2}{32\pi^2}~F^C_{\alpha\beta}\tilde{F}_{\alpha\beta}^C$ is the topological charge density.
The four-divergence of the axial current is not vanishing but depends on the gluon configurations; this fact implies that this classical symmetry is {\it explicitly} broken, and that therefore no Goldstone boson appears.
This is the reason why the $\eta^{\prime}$ ($m_{\eta^{\prime}}\sim 1~\mbox{GeV}$) is much heavier than the mesons of the octet.

A class of gauge field solutions which give rise to a violation of axial symmetry are instantons.
Instantons can be interpreted as classical solutions of the euclidean action, and are related to non trivial topological configurations (with non vanishing winding number) of the gauge field: 
actually, they connect asymptotic configurations characterized by different topological charges. 
Similar solutions must tunnel under a potential barrier (a double well potential is a toy description for instantons in quantum mechanics), and are characterized by a finite action; therefore, they do give a finite contribution to physical quantities. 
By reminding eq.(\ref{axialcharge}), it is clear that instantons 
break explicitly $U(1)_A$. 
These solutions cannot be obtained by smooth perturbation of the vacuum, due to their non trivial topological structure, and are therefore intrinsically non-perturbative.
 For the case of a $U(N_f)_L \otimes U(N_f)_R$ invariant theory, 't Hooft realized \cite{'tHooft:1976fv} that the coupling of quarks to such localized topological solutions of the pure gluonic fields produces an effective interaction between quarks which is still $SU(N_f)_L \otimes SU(N_f)_R$ but explicitly breaks the $U(1)_A$ symmetry. 
This interaction gives rise to a mixing between different flavors; 
as a  consequence the wave functions of physical isosinglet bosons $\eta,\eta'$ are mixed with respect to the $U(1)_A$ invariant case (in that case we would have $\eta\sim\bar{\Psi}\lambda_8\Psi,~\eta'\sim\bar{\Psi}\lambda_0\Psi$).
Simultaneously, the $\eta'$ mass is pushed up by this term, in agreement with phenomenological observations.
A detailed analyses of instanton effects in QCD, and a presentation of the so-called instanton liquid model has been done in \cite{Schafer:1996wv}.
However, Witten \cite{Witten:1979vv} and Veneziano \cite{Veneziano:1979ec} argued that the solution of the $U(1)_A$ problem should not necessarily come from instanton solutions but rather from some other type of gluonic configurations in QCD (e.g. the ones related to confinement); this resulted in the famous Witten-Veneziano formula which relates the flavor-singlet (the $\eta'$ particle) mass to the topological susceptibility $\chi$
\begin{equation}\label{etaprimmass}
m^2_{\eta}+m^2_{\eta'}-2m^2_K=\frac{\chi}{f^2_{\pi}}
\end{equation}
The topological susceptibility is defined as $\chi=\frac{\partial \epsilon^2}{\partial \theta^2}|_{\theta=0}$; 
$\epsilon$ is the vacuum energy density as a function of the $\theta$ parameter. Actually in QCD a topological term 
\begin{equation}\label{topologicalterm}\mathcal{L}_{\theta}=i~\theta~Q(x)\end{equation}
 allowed by the QCD gauge invariance, can be added to the standard QCD Lagrangian. This parameter defines the choice of vacuum among an infinity of possible distinct and generally inequivalent vacua. Each $\theta$ represents a possible true vacuum and there are in general an infinity of distinct theories arising from any given Lagrangian.
Due to the finite value of $\chi\sim(170~\mbox{MeV})^4$ in the vacuum, the $\eta'$ mass does not vanish even in the chiral limit (where $m_K=m_{\eta}=0$). 

The spontaneous breaking of chiral symmetry produces an extra contribution to the effective quark mass, due to interactions. 
Actually, the quark condensate in QCD is given by the trace of the full quark propagator $S_F$:
\begin{equation}\label{quarkpropag}
\langle\bar{\Psi}\Psi\rangle=-i~ \mbox{lim}_{y\rightarrow x+}~\mbox{Tr}~S_F(x,y)
\end{equation}
Since $\bar{\Psi}\Psi$ is a gauge invariant quantity, one can take any gauge to evaluate $S_F(x,y)$ which will have a general form in the momentum space as
\begin{equation}\label{quarkpropagimpuls}
S_F(p)=\frac{A(p^2)}{p\cdot\gamma-B(p^2)}
\end{equation}
Due to the property $\mbox{Tr}~\gamma_{\mu}=0$, only a non zero B, i.e. a mass term, can induce a non vanishing quark condensate. 
According to this constituent particles description, the main effect of interactions is the redefinition of an effective constituent mass:
quarks behave as quasi-particles\footnote{Obviously, confinement prevents to directly observe quarks outside of hadrons.} with an effective mass, given by the sum of the current mass (the one present in the Lagrangian) and of the correction coming from interactions. 
This fact allows us to consider the quarks inside the nucleons as objects characterized by a mass $\sim~350~\mbox{MeV}$, instead of by the current mass, which is of the order of $5\div10~\mbox{MeV}$ (at the $1~\mbox{GeV}$ scale). 
As we can see, for the light quarks the dynamical breaking effect is dominant with respect to the explicit chiral violating term.
The phenomenon of mass generation cannot be seen at any perturbative order; actually, within perturbation theory in the chiral limit, there is no mixing between left handed and right handed quarks, and any diagram would imply a vanishing mass. 
Therefore, to highlight this feature
it is necessary to sum the contribution of an infinite class of Feynman diagrams.\\
For this reason, a study of such features starting from the fundamental theory requires some attention. One possibility to face this problem is to consider a non perturbative loop expansion of the effective action, following a method proposed by Cornwall-Jackiw-Tomboulis \cite{Cornwall:vz}, and to apply it to the QCD case. This approach led to an effective theory, sometimes referred to in literature as ladder-QCD, which displays the dynamical breaking of chiral symmetry, and whose phase diagram is expected to be consistent with that of QCD. However, strong assumptions are necessary to write down an effective action, and in this spirit ladder-QCD must be considered as an effective theory, althought derived directly from first principles. The most used microscopic model for studying chiral symmetry breaking is the Nambu-Jona-Lasinio model (NJL) \cite{Nambu:1961tp,Nambu:1961fr}. 
The model was proposed in 1960 in order to describe the generation of a mass gap for fermionic excitations, in analogy with superconductivity, and to motivate the smallness of pion mass with respect to baryon mass.
In superconductivity, the gap is caused due to the fact that the attractive phonon-mediated interaction between electrons produces correlated pairs of electrons with opposite momenta and spin near the Fermi surface, and it takes a finite amount of energy to break this correlation. Excitations in a superconductor can be conveniently described by means of a coherent mixture of electrons and holes. 
 After the proposal of QCD as the fundamental theory
NJL has been used at the quark level. 
In analogy with the Fermi model for weak interactions, the exchange of a gluon is replaced with an effective four fermion vertex; here it is the strength of interaction, and not the mass of the intermediate boson, which allows a similar 
identification.
The attractive interaction leads to the formation of a fermionic condensate $\langle\bar{\Psi}\Psi\rangle$; we can think of Bogoliubov fermionic quasi-particles as collective excitations provided with a constituent mass. 
The model is reliable in the regime where QCD interactions are ``strong'', at a scale $\sim\Lambda_{QCD}$, see e.g. \cite{Hatsuda:1994pi}; therefore, NJL can be considered as a good model for studying chiral symmetry breaking. 
On the other hand, since gluons are not present in the model, we cannot employ it for describing the perturbative high-energy regime.
The interaction respects the global symmetries of the original Lagrangian, and we can model the $U(1)_A$ breaking with a suitable determinantal term in flavor space ('t Hooft term). 

In this work, we will deal with the two aforementioned models. 
However, there is a larger class of models that can be employed for studying the non perturbative features of QCD. 
In the Preface we have mentioned lattice QCD as the closest model to the fundamental theory. Otherwise, in the spirit of the renormalization group, we can write down an effective theory integrating out the heavy degrees of freedom.
We can study a theory where only the low energy degrees of freedom are taken into account: due to confinement, these states must necessarily be color-neutral hadrons, and due to the breaking of chiral symmetry, these are the pseudo-Goldstone bosons corresponding to the breaking of the global symmetry.  The interactions of these low energy states are given by symmetries of QCD Lagrangian. These low-energy models, called chiral models, are basically non linear sigma model (the non linear constraint fixes the $\sigma\sim\bar{\Psi}\Psi$ field); they are effective in the sense that they display chiral symmetry breaking already at the tree level. Perturbative calculation can be done with respect to the external momentum.
Chiral models receive a large application in nuclear physics and in hadron spectroscopy; the lightest baryons can be included in the theory too.
A random matrix model can be used as a QCD effective theory as well: the idea is to replace the matrix elements of the fermionic Dirac operator by Gaussian distributed random variables, while respecting all the global symmetries of QCD. We will often refer to these approaches in the following of the present work.

\section{QCD at high temperature and Deconfinement}

We have anticipated in the Preface the existence of a phase transition associated with chiral symmetry. According to the Landau classification of phase transitions, an order parameter may separate the low temperature-broken symmetry phase from the high temperature-restored symmetry phase.
The fermion condensate $\langle\bar{\Psi}\Psi\rangle$ is the order parameter for the chiral transition. However, as far as chiral symmetry is concerned, a change in the symmetry properties across the transition is associated only in the chiral limit, where chiral transformations are exact invariances of the QCD Lagrangian. The value of the critical temperature is about $150~\mbox{MeV}$.

Conversely, in the pure gauge limit, namely when quarks are infinitely heavy, 
there exists a rigorous phase transition associated with deconfinement. In this case QCD possesses a $Z(3)$ center symmetry ($Z(N)$ in general for a $SU(N)$ gauge theory), which is spontaneously broken by the QCD vacuum; all of the dynamics regarding confinement is in this case ruled by gluons \footnote{This is the reason why a model without gluons like NJL cannot account for confinement.}.
It is still an open question what kind of gluon configuration are responsible for confinement: in particular, the connection between confinement, chiral symmetry and $U(1)_A$ breaking must be clearly understood \cite{Glozman:2002cp}.
 In the pure gauge limit QCD is strictly confining, in the sense that the static potential between a quark and an antiquark grows with their distance.
For finite quark masses the discrete $Z(3)$ symmetry is explicitly broken, and
in the infinite distance regime the static potential becomes flat: this means the creation of another couple $q\bar{q}$ for breaking the color string.
The order parameter related to confinement is the Polyakov loop $L$; $L$ is charged under the center symmetry, therefore $L\neq 0$ breaks $Z(3)$.
The Polyakov loop belongs to the class of Wilson loops, which are gauge-invariant observables, and which constitute the fundamental tools for decribing a gauge theory on the lattice.
At finite temperature $L(T)\sim~exp({-F_{q\bar{q}}/T})$ where $F_{q\bar{q}}$ is the free energy of a static $q\bar{q}$ pair for $r\rightarrow\infty$. Therefore (in the $m\rightarrow\infty$ limit) $L(T)=0$ for $T<T_d$ and $L(T)\neq0$ for $T>T_d$, where $T_d$ is the deconfinement temperature. 
Up to now, the only proof of confinement comes from numerical studies of the Polyakov loop on the lattice.

In the physical case, with finite values of the quark masses, neither the chiral nor the $Z(3)$ center symmetries are exact. It is however possible to define an approximate phase transition by studying the behaviour of the susceptibilities associated with the respective order parameter; for realistic values of quark masses the transition should be a cross-over.

Moreover, several lattice studies have found that the peaks in the chiral and Polyakov loop susceptibilities are reached at the same temperature.
This means that deconfinement and chiral symmetry restoration are strictly related to each other. Therefore, the increase of temperature and/of quark density beyond a line of critical values could cause two effects. Deconfinement implies that hadrons dissolve into their elementary constituents.
Chiral symmetry restoration implies that quark are characterized by their current mass, rather than by the constituent one.
In this work we will be dealing with models concerning the study of chiral symmetry, having in mind its connection with deconfinement.

The experimental investigation of a QGP is a challenging enterprise. Ultra Relativistic Heavy Ion Collisions are a tool for getting a proof of the deconfined phase: the energies involved allow the formation of very hot objects, with temperatures higher than the deconfinement one, in the state following the collision. 
The cooling of these {\it clusters}, and the consequent formation of hadronic matter in the final state, must however bring the memory of the deconfined phase the system has passed through.
It is very important to identify an observable which can be identified as a signature of the QGP. A possible proof for the formation of a QGP is strangeness enhancement: in a deconfined and chirally restored plasma the ratio strange/light quarks is sensibly higher than in the vacuum. The latter fact reflects in the production of strange hadrons after hadronization.
Heavy-flavored hadrons production and charmonium ($J/\psi$) suppression in the final channel are probes for the QGP formation as well. 

\begin{figure}[htbp]
\begin{center}
\includegraphics[width=12cm]{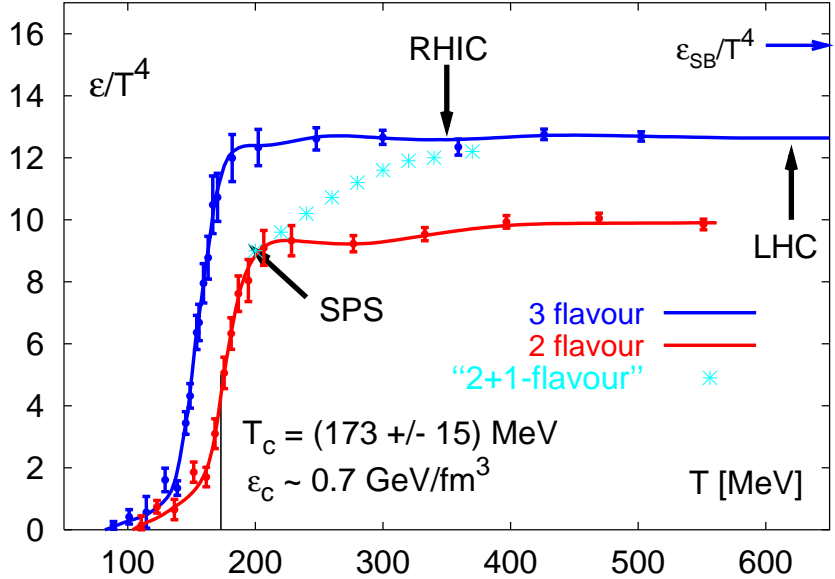}
\end{center}
\caption{$\epsilon/T^4$ computed from the lattice as a function of temperature. The discrepancy between the Stefan-Boltzmann limit is still about 20\% at $T=600~\mbox{MeV}$. The picture is taken from \cite{Karsch:2003jg}.}
\label{fig:eqstatoKarsch}
\end{figure}

In comparison with earlier expectations, in recent years the study of QCD at high temperature has obtained remarkable progresses. Lattice studies have provided a prediction for the dependence of pressure/energy density  on temperature, see fig. \ref{fig:eqstatoKarsch}. For an ideal gas ${\epsilon}/{T^4}={3p}/{T^4}\equiv{\epsilon_{SB}}/{T^4}$ (where $\epsilon$ is the energy density, $p$ the pressure and $\epsilon_{SB}$ the Stefan-Boltzmann value).
For a weakly interacting plasma the discrepancy between the Stefan-Boltzmann limit must be negligible, and the system is almost in the ideal gas regime.
Actually, perturbative methods can be applied for studying QCD at high temperatures and densities; $\alpha_s$ goes to zero for asymptotic values and the system behaves like an ideal gas \cite{Gross:1980br}. In an intermediate regime where $T,\mu\sim\Lambda_{QCD}$ the interactions can still be important.
According to recent lattice analyses, see fig. \ref{fig:eqstatoKarsch}, until temperature is lower than a few critical temperature $T_c$ (where $T_c$ is the critical temperature for chiral symmetry restoration), the effect of interaction appears to be not negligible at all, and the system seems to behave as a perfect liquid, with zero viscosity, instead as a perfect gas. This is the picture of a strongly interacting QGP (sQGP) \cite{Shuryak:2004kh}. In this regime, the presence of bound states is still allowed, as they are expected to dissolve only for $T>4\div5~T_c$. The restoration of chiral symmetry could coincide with the passage to this phase and not to the ideal QGP.

When referring to the experimental settings, it is important to stress that a thermodynamic equilibrium formalism must be adopted with a certain care. The collision of ultrarelativistic heavy ions causes the formation of intermediate objects, characterized by a huge amount of kinetic energies. The typical scale of time involved in these processes are of the order of a few $fm/c$.
Since we are dealing with strong interactions only (the other forces can be turned off at this step of the reaction), we can admit that these times are large enough to favour a local equilibration; hence, one can introduce temperature as a useful parameter, still reminding that these clusters expand very rapidly.
A hydrodynamic description of the expanding system can be given, and significant evidence of this phenomenon can be extracted from RHIC data \cite{Heinz:2004ar}. In particular, elliptic flows have been detected as stable signatures of this fluid behaviour \cite{Bhalerao:2005mm}. The energy densities extracted from RHIC data seem to confirm the idea that a sQGP has formed (it behaves like an almost perfect fluid).
Thermalization effects (also in the framework of a non equilibrium Quantum Field Theory \cite{Berges:2000ur}), in the spirit of the Boltzmann equation, can be considered to take into account that the system is not at the complete thermodynamic equilibrium.

As the fluid expands, its temperature decreases: when the deconfinement temperature is reached, hadronization takes place, and hadrons are formed. 
Since the typical scale related to the hadronic regime ($\sim$ GeV) is of the same order as $\Lambda_{QCD}$, the hadronization mechanism is intrinsecally non-perturbative; again, some effective model must be used.
Hadrons are the only objects detectable after the collision, therefore hadronic models are largely used to reproduce final molteplicities. In recent years, by following previous ideas of Fermi and Hagedorn, a simple Statistical Hadronization Model \cite{Becattini:1997rv,Becattini:2000jw} has been able to to reproduce the most part of experimental data by using a few free parameters. In this model, hadrons are considered as a non interacting gas, and thermodynamic parameters are fitted in order to reproduce experimental data. 
As far as total multiplicities are concerned, hadron models are able to reproduce the most part of the data; hydrodynamic models are more appropriate to justify the geometrical distributions of final hadrons.

\section{The chiral transition and the critical point}
 Coming back to the chiral transition, the first analyses of the phase diagram consisted in the study of the order parameter $\langle\bar{\Psi}\Psi\rangle$ as a function of temperature and quark density, in the framework of a thermodynamic equilibrium formalism. 
To this end, a gran-canonical approach can be applied to a model for strong interactions.
A group of pioneer works which contributed to the determination of the general structure of the phase diagram is found in refs. \cite{Asakawa:1989bq,Barducci:1989wi,Barducci:1989eu}:
those authors identified for the first time the presence of a tricritical point in the phase diagram, separating first order from continuous phase transitions along the critical line. The models used were NJL \cite{Asakawa:1989bq} and ladder-QCD \cite{Barducci:1989wi,Barducci:1989eu}.
The presence of a tricritical point is a promising feature to highlight experimentally the passage to the deconfined phase. For a general treatment of the theory of tricritical points see \cite{Domb:xyz}. In the following, some general aspects of the chiral transition in QCD are reviewed.

\begin{figure}[htbp]
\begin{center}
\includegraphics[width=10cm]{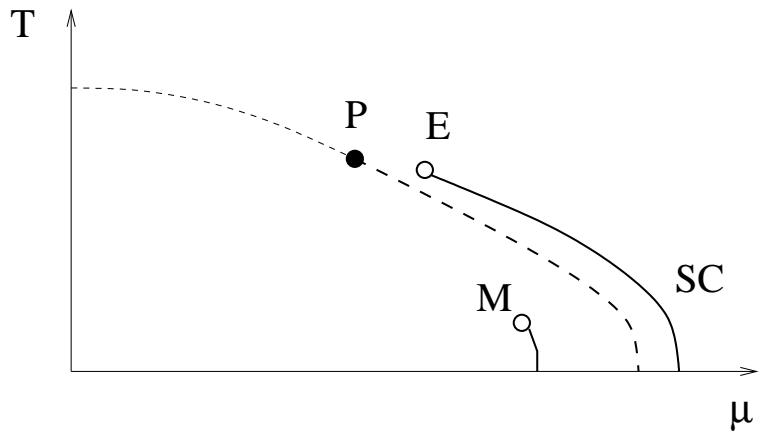}
\end{center}
\caption{A schematic phase diagram of QCD, from \cite{Stephanov:1998dy}. The dashed lines represent the boundary of the phase with spontaneously broken chiral symmetry in QCD with two massless flavors. The point P is tricritical. The solid line with critical end-point E is the line of first order transitions in QCD with 2 quarks of small mass. The point M is the end-point of the nuclear liquid-gas transition. SC labels the regime of color superconductivity.}
\label{fig:Diagenold}
\end{figure}

In QCD with two massless quarks, chiral symmetry, spontaneously broken at $T=0$, is restored at finite temperature. The phase transition is supposed to be of second order \cite{Pisarski:1983ms} and to belong to the universality class of $O(4)$ spin models in 3 dimensions. If the transition is second order in the chiral case, when we consider two light quarks with a non-zero mass there is only a smooth crossover as a function of T. This picture is consistent with lattice simulations and several effective models, however the problem is still matter of debate in the scientific community.

On the other hand, at T=0 several models suggest (NJL \cite{Asakawa:1989bq}, Ladder-QCD \cite{Barducci:1989wi}, Random Matrices \cite{Halasz:1998qr}) that the restoration of chiral symmetry at finite $\mu$ is realized through a first order transition. Assuming that this is the case in real QCD, one can argue that there must be a tricritical point in the $(T,\mu)$ plane, where the transition changes smoothly from first to second order. Using the notation of \cite{Stephanov:1998dy} in fig. \ref{fig:Diagenold}, we will refer to it as $P$.
By analogy with an ordinary critical points, where {\it two} distinct coexisting phases become identical, it is possible to identify a tricritical point where {\it three} coexisting phases become simultaneously identical. A tricritical point marks an end-point of three-phase coexistence. In order to highlight this in QCD, it is necessary to consider another dimension in the space of parameters - that of the quark mass $m$. In such a three-dimensional space, there are two surfaces (symmetric under $m\rightarrow-m$) of first-order phase transitions originating from the first-order line at $m=0$. On these surfaces, or wings, with $m\neq0$, two phases coexist: a low density phase and a high density phase. There is no symmetry distinguishing these two phases, chiral symmetry being explicitly broken at $m\neq0$. Therefore, the surfaces can have a hedge which is a line of critical points: along this line, the discontinuity in the order parameter between the high and low density phases vanishes. These wing lines start from the tricritical point. The first-order phase transition line can be identified as a line where three phases coexist: one high $T$ and $\mu$ phase and two low $\mu$ and $T$ phases (with opposite signs of $m$ and consequently of $\langle\bar{\Psi}\Psi\rangle$). This line is called a triple line.

\begin{figure}[htb]
\setlength{\unitlength}{2.4in}
\centerline{\psfig{file=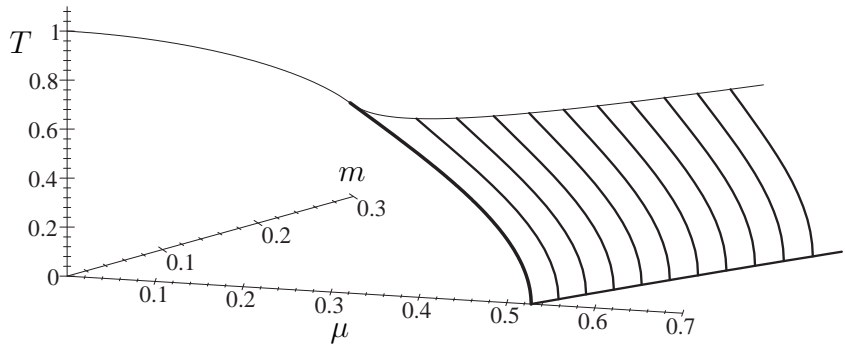,width=2\unitlength}}
\vspace{-\unitlength}
\begin{picture}(2,1)
\put(.26,.66){ $T$}
\put(.96,.04){ $\mu$}
\put(1.00,.38){$m$}
\end{picture}
\caption{Phase diagram of QCD in the space $(T,\mu,m)$ for two light flavors, from \cite{Halasz:1998qr}. In the figure only the wing with positive $m$ is shown (the other is symmetric with respect to the plane $m=0$). Marked lines label first-order transitions.}
\label{fig:tricrit}
\end{figure}

The appearance of a tricritical point can be understood by considering a Landau-Ginzburg expansion for the order parameter of chiral symmetry breaking, $\phi=(\sigma$,$\boldsymbol{\pi}$$)\sim\langle\bar{\Psi}\Psi\rangle$ ($\sigma$ is the scalar isosinglet, $\boldsymbol{\pi}$ the pseudoscalar isovector):

\begin{equation}\label{eqn:GinzLandau}
\Omega(\phi)=a\phi^2+b(\phi^2)^2+c(\phi^2)^3
\end{equation}
\\
being $\Omega(\phi)$ the appropriate effective potential. The coefficients $a,b$ and $c>0$ are functions of $\mu$ and $T$. 
The second order phase transition line described by $a=0$ at $b>0$ becomes first order when $b$ changes sign; therefore, the condition $a=b=0$ locates the tricritical point.
The critical properties of this point can be derived from universality arguments \cite{Halasz:1998qr,Berges:1998rc} and the critical exponents are those of the three-dimensional mean field theory (apart from logarithmic correction arising from the renormalization group analysis \cite{Domb:xyz}). 
Actually, at finite temperature the long-wavelength modes live in a 3-d space, the temporal direction being compactified.
In \cite{Halasz:1998qr,Stephanov:1998dy,Berges:1998rc} a general survey of the properties of the critical point in QCD is given; the great merit of those papers is to have extracted a universal picture from a description which is in general model-dependent.

In real QCD with nonzero quark masses the second order phase transition becomes a crossover and the tricritical point becomes a standard critical end-point of a first order phase transition line. This point is labeled in fig. \ref{fig:Diagenold} as $E$.
In the expansion of eq. (\ref{eqn:GinzLandau}) a current mass enters as a term $\sim-m\sigma$.
It can be argued that the point E is in the universality class of the Ising model in 3 dimensions, because the $\sigma$ is the only genuine massless field at this point (the pions remain massive because of the explicit breaking of chiral symmetry). 
It is a general property of liquid-gas type transitions (like that of water), not being associated with a change of the symmetry properties across the transition, to terminate with  endpoints which belong to the Ising-3d universality class.
Universality arguments also predict that the end-point E in QCD with small quark masses is shifted with respect to the tricritical point P towards larger $\mu$ as shown in Fig. \ref{fig:Diagenold}. 

The strange quark plays an important role on the location of the points P and E. At $\mu=0$, if the strange quark mass $m_s$ is less than a critical value $m_{s3}$, the finite temperature transition changes from second to first order. This leads to a tricritical point, $P_s$, in the $(T,m_s)$ plane. In terms of eq. (\ref{eqn:GinzLandau}) the effect of decreasing $m_s$ could favor the coefficient $b$ to become negative \cite{Stephanov:1998dy}. 
The instanton induced interaction for three massless quark would cause in the expansion a $\phi^3$ term, forcing the transition to be first-order even at $\mu=0$ \cite{Gavin:1993yk}.

\begin{figure}[htbp]
\begin{center}
\includegraphics[width=8cm]{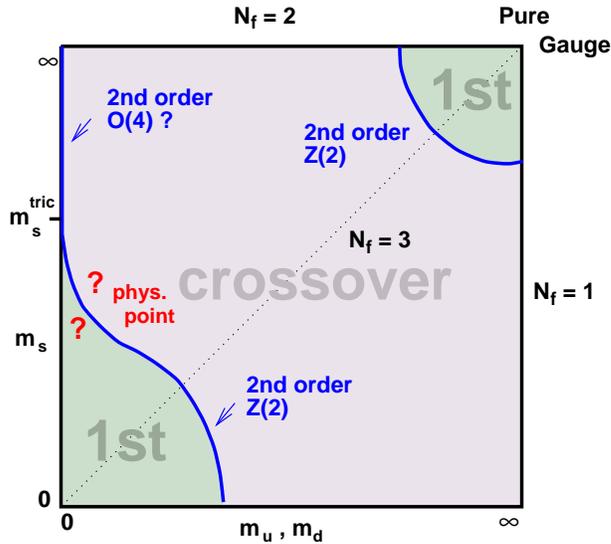}
\end{center}
\caption{QCD phase diagram in the $m_{u,d},m_s$ plane, from \cite{Laermann:2003cv}. The finite temperature, zero chemical potential transition is here considered. For realistic values of quark masses, the $\mu=0$ transition should be a crossover.}
\label{fig:m_us_plane}
\end{figure}

Obviously, the physical picture in the $(T,\mu)$ plane is that of Fig. \ref{fig:Diagenold} only for $m_s>m_{s3}$. As $m_s$ is reduced starting from infinity, the tricritical point P of Fig. \ref{fig:Diagenold} moves to lower $\mu$ until, at $m_s=m_{s3}$, it reaches the $T-$axis and can be identified with the tricritical point in the $(T,m_s)$ plane \cite{Gavin:1993yk,Rajagopal:1992qz}. The two tricritical points, $P$ and $P_s$, are continuously connected. 
For $m_s<m_{s3}$ the transition should remain first order at all nonzero $\mu$; in this case there would not be any critical end point along the line for chiral symmetry restoration. 
In the real world, with physical values of quark masses, it is generally assumed that $m_s>m_{s3}$, in agreement with several lattice results and effective model calculations. It is important to stress that the qualitative effect of the strange quark is to reduce the value of $\mu_P$, and thus of $\mu_E$, with respect to the pure two flavor case, being $\mu_P=0$ at $m_s=m_{s3}$. This shift may be significant, since lattice studies show that the physical value of $m_s$ is only slightly higher than $m_{s3}$. Therefore, a reliable study of the phase diagram must take the strange quark into account. In fig. \ref{fig:m_us_plane} we plot a schematic picture of the phase diagram at $\mu=0$ in the $(m_{u,d},m_s)$ plane.

The critical end point seems to be a rather stable feature for the QCD phase diagram; however its exact location in the $(T,\mu)$ plane is much more uncertain and model dependent. For instance, ladder-QCD provides an estimate which lies at higher temperatures and lower chemical potentials than that of NJL. Recent lattice simulations \cite{Fodor:2001au,Fodor:2001pe} seem to favor a high temperature location.

RHIC and LHC (whose realization is forecast for 2007) should be able to detect the critical point. Since E is a genuine critical point, one expects phenomena related to the enhancement of long wave-length fluctuations \cite{Stephanov:1998dy}. Event-by-event fluctuations in heavy ion collisions can be studied as probes of the critical point \cite{Stephanov:1999zu}.
At the critical point the $\sigma$ mass is rigorously zero even in the massive case, so there must be a region around E where the decay $\sigma\rightarrow\pi^+~\pi^-$ is forbidden. This fact should lead to a non-thermal enhancement of low $p_T$ pion spectrum in the final channel.
Moreover, pion properties in the final channel are strongly modified, because in this region the inter-hadron interactions should be dramatically changed \cite{Shuryak:2005vk}.

\section{The regime of low temperatures and high densities}

After the papers \cite{Alford:1997zt} and \cite{Rapp:1997zu}, the interest for the study of the QCD phase diagram has grown up considerably. It is by now a commonly accepted picture that in the high density and low temperature regime 
new phases, different from that of high temperatures can develop.
Actually, Cooper pairs of quarks could form, namely pairs of fermion bound together by the chromomagnetic force (see for instance \cite{Rajagopal:2000wf}). 

The basic idea is the following: in the high density regime, where the Fermi momentum is higher than $\Lambda_{QCD}$, we can interpret quarks as weakly interacting quasi particles. Furthermore, as we know from the theory of superconductivity in ordinary materials, the Fermi surface is unstable in presence of an arbitrarily weak attractive interaction.
Actually, in QCD it turns out that the effective interaction between two quarks mediated by the exchange of a gluon is attractive in the colour antisymmetric $\bar{3}$ channel whereas it is repulsive in the symmetric $6$ channel ($3\otimes3=6\oplus\bar{3}$). However, by studying the renormalization group dependence of these effective couplings, one recognizes that at the Fermi surface the attraction in the $\bar{3}$ channel is dominant \cite{Schafer:1999fe}; this may induce the formation of $qq$ pairs. Of course, this is the dominant mechanism in the regime where the coupling gets weak, and multi-gluon exchange are suppressed. Therefore, the mechanism of coupling between quarks on the Fermi surface, with opposite spatial momentum, is expected. The pair behaves like a boson: this is the reason why we speak of {\it color superconductivity}, and of {\it di-quark} condensation: actually, superconductivity can be interpreted as a condensation of fermionic pairs. The most studied case is that of a spin zero condensate, since it allows a better use of the Fermi surface. Recently, the case of a spin one condensate has been considered \cite{Schmitt:2004hg}.
In order to study the ground state, we have to consider the following matrix element:

\begin{equation}\label{cscordparam}
\langle\Psi_{i}^{a}C\gamma_5\Psi_{j}^{b}\rangle=\phi_{ij}^{ab}
\end{equation}

Where $\phi_{ij}^{ab}$ is a $N_f\times N_f$ matrix in flavor space (with indices $i,j$) and a $N_c\times N_c$ matrix in color space (with indices $a,b$), $C$ is the charge conjugation matrix; Dirac indices are not explicitly shown. Symmetry considerations are useful to study the order parameter:

\begin{itemize}
\item{The condensate must be antisymmetric in color indices $(a,b)$ in order to have attraction.}
\item{The condensate must be antisymmetric in Dirac indices in order to have spin zero.}
\item{Pauli principles requires antisymmetry in flavor indices, fixed the color and Dirac structure.}
\end{itemize}

Therefore, the structure of the ground state depends on the number of colors $N_c$, which is equal to three in the real case (even though effective theories with $N_c$ different than three are often considered), and of degenerate light flavors $N_f$.

A di-quark condensate (apart from the case of two colors) is not invariant under color transformations: this implies the breakdown of color gauge invariance. As usual, this statement has to be interpreted with care. Local gauge invariance cannot really be broken, according to the Elitzur theorem \cite{Elitzur:1975im}. Nevertheless, the spontaneous breaking of gauge invariance is a useful concept. We can fix the gauge, introduce a gauge non-invariant order parameter and study its effect on gauge invariant correlation functions. The most important gauge invariant consequence of superconductivity is the appearance of a mass gap; gluons aquire a mass through the Meissner-Anderson-Higgs mechanism, and there is a gap $\Delta$ in fermions spectrum for exciting a single particle from the condensed ground state. 
Color superconductivity may also lead to the spontaneous breaking of global symmetries; broken global symmetries imply the appearance of Goldstone bosons, and determine the low energy effective description of the system.

The qualitative structure of these superconducting phases strongly depends on the densities that are at stake. For high enough densities, so that we can neglect the mass difference between the strange and the two light quarks (and the value itself of the masses), the fermionic condensates which form should be those involving $u,d,s$, as though if they were massless: this phase is said CFL ({\it Color-Flavor Locking}), corresponding to $N_f=3$. For lower densities, the 2SC ({\it Two Flavor Superconductor}) phase appears to be more favored, and only the two light flavors are involved, hence $N_f=2$. Other more exotic possibilities can be studied but are less likely, see e.g. \cite{Schafer:2003vz}.

The two situations are quite different from each other.
In the case of CFL we have a condensate (in the simplest case where the coupling in the $\bar{3}$ color sector is considered only):

\begin{eqnarray}\label{condCFL}
\langle\Psi_{iL}^{a}\Psi_{jL}^b\rangle=-\langle\Psi_{iR}^{a}\Psi_{jR}^b\rangle=\Delta\sum_{C=1}^3\epsilon^{abC}\epsilon_{ijC}
\end{eqnarray}
(indices $L/R$ label left/right chirality components of the spinor)
\begin{eqnarray}\label{pattsymmbreakCFL}
SU(3)_c\otimes SU(3)_L\otimes SU(3)_R\otimes U(1)_B\otimes U(1)_A\rightarrow SU(3)_{C+L+R}\otimes Z_2\otimes Z_2
\end{eqnarray}

We can actually believe that $U(1)_A$ is restored at high densities, therefore we consider an interaction which respects that symmetry. The $Z_2$ symmetries arise since the condensate is invariant under a change of sign of the left- and/or right-handed fields. We observe that CFL implies that chiral symmetry is broken. This mechanism is rather unusual: the order parameter involves no coupling between left and right handed fermions. Nevertheless, both left and right handed flavors are locked to color, and because of the vectorial coupling of the gluon left handed components are effectively locked to right handed components. The pattern of symmetry breaking of CFL phase (apart from the spontaneous breaking of $U(1)_A$) looks quite similar to that of zero density; therefore, we might expect that the low density phase and the high density phase could be continuously connected without a phase transition. Electric charge is broken but there exists a linear combination with a broken color generator which is still unbroken. Baryon number is broken. Since the color group is completely broken, all gauge particles get a mass, and all fermions are gapped. As far as Goldstone bosons are concerned, they are 8+2 corresponding to the broken global symmetries.

For 2SC we have a quite different situation; the condensate is
\begin{eqnarray}\label{cond2SC}
\langle\Psi_{iL}^{a}\Psi_{jL}^b\rangle=\Delta\epsilon^{ab3}\epsilon_{ij}
\end{eqnarray}
Due to the antisymmetry in color the condensate must necessarily choose a direction in color space. The condensate breaks the color group but it does not break any flavor symmetry:
\begin{eqnarray}\label{pattsymmbreak2SC}
SU(3)_c\otimes SU(2)_L\otimes SU(2)_R\otimes U(1)_B\rightarrow SU(2)_{C}\otimes SU(2)_L \otimes SU(2)_R \otimes U(1)\otimes Z_2\nonumber\\
\end{eqnarray}
therefore chiral symmetry remains unbroken in the 2SC phase. Baryon number is broken, but there exists a linear combination of baryon number and of a broken local generator which is still unbroken, therefore no Goldstone bosons appear. Due to the breaking of the local symmetry $SU(3)_c$ down to $SU(2)_c$, five of the eight gluons acquire mass. Up and down quarks with two out of the three colors become gapped.

In order to construct an effective theory it is necessary to establish a hierarchy of the masses that are into play. In any case, the lowest mass degrees of freedom are Goldstone bosons, ungapped fermions and holes and massless gauge fields.
The corresponding low-energy effective theory has been formulated in \cite{Casalbuoni:1999wu} for CFL and in \cite{Casalbuoni:2000cn,Casalbuoni:2000jn} for 2SC phases; in \cite{Nardulli:2002ma} a general treatment of this topic is proposed.
If we increase energy, we must consider the degrees of freedom whose typical energy is of the order of the gap $\Delta$.
For $\mu\gg \Lambda_{QCD}$ the theory is perturbative, and quarks can be treated as almost free particles. In that regime, we can consider as light the particles staying in a thin shell around the Fermi sphere, and as heavy all of the others (in particular antiquarks for $\mu>0$). A high density effective theory can be derived integrating out all of the heavy degrees of freedom \cite{Casalbuoni:2000na};
the larger it is the chemical potential the better it is the approximation.

In the case where the two fermion participating to the pair do not have equal momentum in magnitude, ${\mathbf p_1}+{\mathbf p_2}=2~{\mathbf q}\neq 0$, the condensate carries a non vanishing total momentum, breaking rotational and translational invariance; this
can be the case when we consider a system where neutrality under electric charge is imposed, and the two light quark chemical potentials (namely Fermi energies) are consequently different. 
The most simple form of the condensate compatible with a non zero ${\mathbf q}$ is just a plane wave, $\langle\Psi(x)\Psi(x)\rangle\sim\Delta~e^{2i{\mathbf q} \cdot {\mathbf x}}$, although more general cristalline structures could be considered. This non homogeneous structure of the order parameter corresponds to the so called LOFF phase, see e.g. \cite{Alford:2000ze}. If the difference of radii of Fermi spheres is high enough, the formation of the Cooper pair becomes unfavored and the condensate vanishes.

Recent studies have proposed the possibility of kaon condensation in superconducting regimes, before the pure CFL phase takes over; in that case, the non vanishing value of strange quark mass plays the role of an effective chemical potential for strangeness \cite{Schafer:2003vz} and favors kaon condensation.

\begin{figure}[htbp]
\begin{center}
\includegraphics[width=12cm]{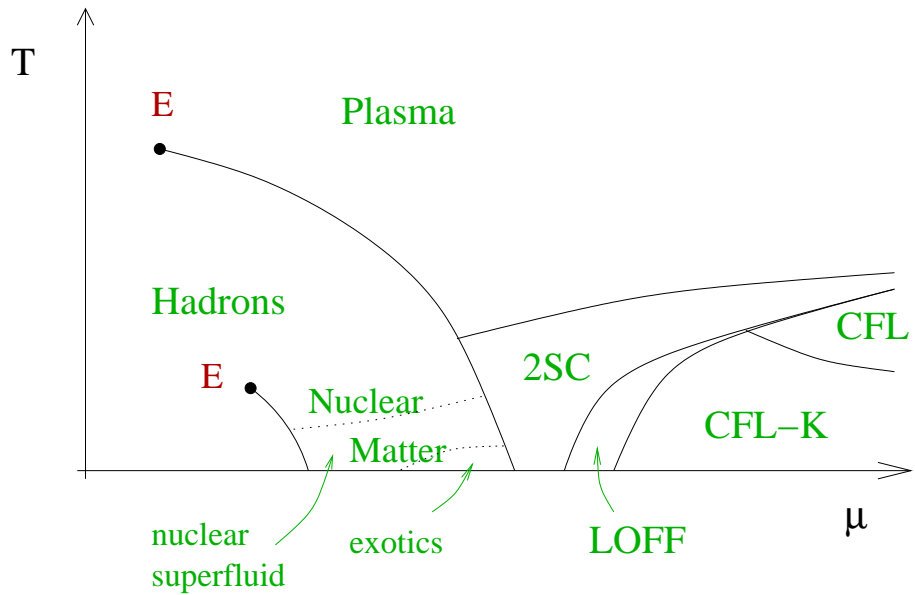}
\end{center}
\caption{Pictorial representation of QCD phase diagram, from \cite{Schafer:2003vz}}
\label{fig:Diagen}
\end{figure}

The study of the QCD phase diagram turns out to be therefore remarkably complicated by those new phases, with respect to the pioneering studies of the eighties. In Fig.\ref{fig:Diagen} a qualitative picture of the situation under examination is shown.

Heavy ion experiments are not intended to reproduce a scenario compatible with superconducting effects: the most natural context this matter of studies is devoted to is the physics of compact stars.
In astronomy, a compact star (or compact object) is a star that is a white dwarf, a neutron star, a quark star, or a black hole. Compact star is a name often used when the exact nature of the star is not known, but when there are suggestions that it is very massive and with a small radius, thus leaving the above-mentioned possibilities. For example, a typical radius of a neutron star can be estimated to be about 10 Km (1/100.000 the solar radius), with a mass which can be comparable with that of the Sun.

When a star has exhausted all its energy (stellar death) the gas pressure of the hot interior cannot support anymore the weight of the star (the gravitational pull) and the star collapses to a denser state: a compact star.
In particular, when the star mass is above 1.44 solar masses (the so-called Chandrasekhar limit) enough energy is available from the gravitational collapse to force the combination of electrons and protons to form neutrons. As the stars contracts further, all the lowest neutron energy levels are filled and the neutrons are forced into higher and higher energy levels. This creates an effective pressure, due to the Fermi statistics,
which prevents further gravitational collapse, forming a neutron star. 
However, for masses greater than two or three solar masses, even neutron degeneracy can't prevent further collapse and it continues toward the black hole state. Quark stars are thought to represent an intermediate regime between the two latter, with a core of deconfined strange matter.
We can therefore believe that these astrophysical objects can be characterized by high enough densities to favor the appearance of superconducting phases. 
In any case, the expected temperatures in the core of such a system ($\sim$ keV) are much lower than the critical temperatures for melting the superconducting regimes ($\sim$ tens of MeV), so that the latter would not be destroyed by thermal effects \cite{Rajagopal:2000wf}. 

The condensed matter regime of QCD has been indicated as responsible of interesting and up to now unresolved astrophysical problems, as {\it gamma ray burst} \cite{Ouyed:2001cg}.
The signals emitted from the astrophysical objects called {\it pulsars}, characterized by a well definite period of oscillation, can be explained 
as generated from a point source of radiation on a spinning neutron star.
The occasional ``glitches'' observed in the period of oscillation can be attributed to the presence of vortices in the superfluid regime within the star.
Other cosmological questions, like baryon asymmetry and CP-violation in the universe,
can be faced within this formalism.
QCD in the high densities regime represents a bridge between historically disconnected branches of physics like quantum field theory, condensed matter, astrophysics and gravity. It is a relatively new field of application that will receive further attention as soon as stable signatures of compact stars will be obtained.

\chapter{QCD at finite isospin density and meson condensation}

Recently (2000) Son and Stephanov \cite{Son:2000xc} have considered the possibility of pion condensation in the region of high isospin densities of the QCD phase diagram. After that paper, the possibility of meson condensation, initially proposed by Migdal \cite{Migdal:1978az}, Kaplan and Nelson \cite{Kaplan:1986yq} and Politzer and Wise \cite{Politzer:1991ev} starting from the 70's in the context of nuclear physics, has been reformulated in the framework of QCD effective theories, 
 in presence and not of superconducting effects. 
A flavor-asymmetrical composition of quark densities favours pion and/or kaon condensates.
In the core of a neutron star, due to the charge neutrality $Q=0$ constraint, the up and down chemical potential are not equal, since they carry different electrical charge.
The linear combinations $\mu_q=(\mu_u+\mu_d)/2$, associated with the conservation of the baryon number, and  $\mu_I=(\mu_u-\mu_d)/2$, fixed by imposing $Q=0$, are introduced. An isospin chemical potential $\mu_I$ measures the different chemical composition of u,d quarks; in a neutron star $\mu_I$ is negative since neutrons dominate over protons. At $T=\mu_q=0$ for $|\mu_I|>m_{\pi}/2$, according to \cite{Son:2000xc}, there should be the condensed pions regime.
The effect of a $\mu_I\neq 0$, besides astrophysics, could regard also the heavy ion collision experiments; in that context (being unlikely the phenomenon of pion condensation) a $\mu_I\neq0$ could induce the splitting of the light quark critical lines. Moreover, QCD at $\mu_I\neq 0$ has a positive definite fermion determinant and can be safely simulated on the lattice \cite{Alford:1998sd}.\\
In the first section of this Chapter, we consider Bose-Einstein condensation within the formalism of Quantum Field Theory. In the second section, we focus our attention to QCD at finite isospin density. In the third section, we consider some theoretical issues concerning QCD at finite chemical potential, which are of interest for the present study.

\section{Bose-Einstein condensation in Quantum Field Theory}
\label{BEC-QFT}

In this section we will review the phenomenon of Bose-Einstein condensation (BEC) in the framework of Quantum Field Theory. Let us consider a relativistic system of charged, interacting bosons, described by the following Lagrangian density

\begin{equation}\label{eqcharged}
\mathcal{L}_0=(\partial^{\mu}\Phi^*)(\partial_{\mu}\Phi)-m^2\Phi^*\Phi-\lambda(\Phi^*\Phi)^2
\end{equation}
where a positive $\lambda$ (necessary in order to have a well defined theory) describes a repulsive interaction. The system possesses a $U(1)$ invariance related to the charge conservation. We can introduce a chemical potential $\mu$ coupled to the conserved current, and thus we obtain an effective Lagrangian density in Minkowski space

\begin{equation}\label{efflagra}
\mathcal{L}=(\partial_0+i\mu)\Phi^*(\partial_0-i\mu)\Phi-(\vec{\nabla}\Phi^*)(\vec{\nabla}\Phi)-m^2\Phi^*\Phi-\lambda(\Phi^*\Phi)^2
\end{equation}
The chemical potential can be introduced by noticing that it enters into the Lagrangian as the zero-component of a gauge field. Therefore, we can promote the $U(1)$ global symmetry to a local one, and then replace the zero component of the gauge field with $\mu$. We are allowed to consider $\mu>m$ since the theory is an interacting one for $\lambda\neq 0$.

In order to study the ground state of the system we can study the static part
of of eq.(\ref{efflagra}) 

\begin{equation}\label{effpot0}
V(\Phi)=(\mu^2-m^2)\Phi^*\Phi+\lambda(\Phi^*\Phi)^2
\end{equation}

We notice that, by expanding eq.(\ref{efflagra}) and neglecting derivative terms, the chemical potential enters as a negative square mass term, thus it tends to favour the spontaneous breaking of the $U(1)$ symmetry. In fact, for $\mu>m$ ($\mu$ is chosen positive without loss of generality) the minimum moves from $\langle\Phi\rangle=0$ to $\langle\Phi\rangle^2=\frac{\mu^2-m^2}{2\lambda}$. This state is a Bose condensate, with a charge density given by

\begin{equation}\label{chargedensity}
\rho=\frac{\mu}{\lambda}(\mu^2-m^2)
\end{equation}

If we expand the action up to the second order with respect to the fields, we can compute the mass spectrum. 
We can do this by decomposing the complex field $\Phi$ as 
\begin{equation}\label{eigenmodeenergy}
\Phi=\langle\Phi\rangle+\chi_1+i\chi_2
\end{equation}
and searching the energy eingenmodes as a function of $\chi_1,\chi_2$ 
For $\mu<m$ we have two modes with effective masses $m\pm\mu$.
At the critical value $\mu=m$ one mode becomes massless; this mode represents the  Goldstone Boson corresponding to the spontaneous breaking of the $U(1)$ symmetry in the condensed phase.  
In the condensed regime $\mu>m$, there is of course a massless mode.
The study of the effective action in the condensed phase allows to describe the system as a superfluid. 

Actually, in the language of Quantum Field Theory, the condensation of a field is reached when the effective mass of the field goes to zero. When the effective mass is zero, the energy cost for exciting a particle with zero spatial momentum is vanishing. Therefore, in the zero temperature limit, only the $\vec{p}=0$ modes contribute to the ground state; all of the charge is carried by the ground state.

Here we have considered the zero temperature, finite chemical potential regime of the theory; one can introduce the temperature by means of standard techniques, see e.g. ref. \cite{Kapusta:1989kp}.
Actually, this can be done quite easily in the path integral formulation of the euclidean theory.
It is possible to study the distribution of charge density in momentum space as a function of temperature. In any case when we are in the condensed regime, the condensate, which is the zero momentum mode of the field, carries a finite fraction of the charge density of the system. As the temperature is raised up, some of the charge is excited out of the condensate.
Eventually, the temperature becomes high enough to completely melt, or thermally disorder, the condensate; this happens at some critical temperature $T_c$.
However, in the non-relativistic limit, one recovers the same critical temperature as in the first-quantization formalism. 
The ground state can be interpreted as a coherent state which well reproduces a classical state with definite number of particles ($\Delta N\ll N$) in the thermodynamic limit.

In the following of this work we will be interested in the condensation of composite particles, like mesons, which cannot be described by a simple bosonic Lagrangian as in eq.(\ref{eqcharged}). To this end, will we deal with a class of QCD effective models. The condensation of pions and/or kaons is described 
by the non zero expectation value of the relative quantum field, with the consequent breaking of a phase symmetry.

\section{Isospin chemical potential and pion condensation}
\label{ischempot}

In recent years, the study of the phase diagram of 2 flavor QCD in the space of quark $\mu_q=(\mu_u+\mu_d)/2=\mu_B/3$ (where $\mu_B$ is the baryon chemical chemical potential) and isospin $\mu_I=(\mu_u-\mu_d)/2$ chemical potentials has been the subject of several papers.
In 2000, Son and Stephanov 
\cite{Son:2000xc} have shown that a high enough value of $\mu_I$ may cause the condensation of charged pions.
By using a chiral model they showed that $\mu_I$ 
is basically half of the charged pions chemical potentials; since pions are zero spin particles we expect a phenomenon of condensation when the pion chemical potential is equal or greater than the pion mass (we are using the relativistic notations for chemical potentials).

Chiral models are field theories built in terms of the lighter degrees of freedom  present in the spectrum, namely the pseudo-Goldstone bosons corresponding to the breaking of the approximate symmetry $SU(2)_L\otimes SU(2)_R$ (or the larger and less valid $SU(3)_L\otimes SU(3)_R$). The constraint of invariance under the residual $SU(2)_V$ of QCD produces an interaction between those light degrees of freedom, as in a non-linear sigma model. This interaction allows, differently from the case of free bosons, to have chemical potentials higher than the mass. 
The formation of a pseudoscalar condensate leads to the spontaneous breaking of parity and of a $U(1)$ isospin symmetry: therefore the existence of a Goldstone mode is expected and the system behaves like a supefluid. \\
The chiral Lagrangian describing the low-energy dynamics is

\begin{equation}\label{chirallag}
\mathcal{L}=\frac{1}{4}f_{\pi}^2\mbox{Tr}[\partial_{\mu}\Sigma\partial_{\mu}\Sigma^{\dagger}-2m_{\pi}^2\mbox{Re}\Sigma]
\end{equation}
where $\Sigma=e^{i\vec{\pi}\cdot\vec{\tau}/f_{\pi}}\in SU(2)$ is the matrix pion field ($\vec{\pi}$ are the real components of pion field), and $f_{\pi}$ the pion decay constant. 
The theory is invariant under $SU(2)_V$ transformations. The introduction of an isospin chemical potential explicitly breaks the symmetry down to $U(1)_V$. In order to introduce a chemical potential a possible way is obtained by noticing that the chemical potential enters in the Lagrangian as the zero-component of a gauge field. Then one can promote the global $SU(2)_L \otimes SU(2)_R$ symmetry to be a local one, and gauge invariance fixes the way the chemical potential enters \cite{Kogut:2000ek}:

\begin{equation}\label{lagchirchempot}
\mathcal{L}_{eff}=\frac{f_{\pi}^2}{4}\mbox{Tr}\nabla_{\nu}\Sigma\nabla_{\nu}\Sigma^{\dagger}-\frac{m_{\pi}^2~f_{\pi}^2}{2}\mbox{ReTr}\Sigma
\end{equation}
The covariant derivative is defined as

\begin{equation}\label{chempot}
\nabla_0=\partial_o\Sigma-{\mu_I}(\tau_3\Sigma-\Sigma\tau_3),~~~
~~\nabla_i\Sigma=\partial_i
\end{equation}
We can see that the chemical potential breaks Lorentz invariance. If we consider now the static part of (\ref{lagchirchempot}) we can study the behaviour of $\Sigma$ as a function of $\mu_I$. One can prove that, by introducing the following parametrization for $\Sigma$ on the minimum of the effective potential

\begin{equation}\label{sigmaminimum}
\bar{\Sigma}=\mbox{cos}\alpha+i(\tau_1~\mbox{cos}\phi+\tau_2~\mbox{sin}\phi)~\mbox{sin}\alpha
\end{equation}
the effective potential depends on $\alpha$ only \cite{Son:2000by}

\begin{equation}\label{Veffchirlag}
V_{eff}(\alpha)={f_{\pi}^2\mu_I^2}(\mbox{cos}2\alpha-1)-f_{\pi}^2m_{\pi}^2\mbox{cos}\alpha
\end{equation}
For $|\mu_I|$ lower than $m_{\pi}/2$ the system lies in the ground state with $\alpha=0$, $\bar{\Sigma}=1$. When $|\mu_I|$ exceeds the critical value $m_{\pi}/2$, the minimum of $\bar{\Sigma}$ occurs for 

\begin{equation}\label{minimumsigmaalpha}
\mbox{cos}\alpha=\frac{m_{\pi}^2}{4\mu_I^2}
\end{equation}
For $\alpha\neq 0$ we have the spontaneous breaking of the symmetry associated with rotations in the $\phi$ coordinate, and consequently one Goldstone boson. We can interpret this phenomenon as a Bose Einstein condensation of charged pions (positive or negative depending on the sign of $\mu_I$).\\
From (\ref{minimumsigmaalpha}) the chiral $\langle\bar{u}u+\bar{d}d\rangle$ and the pion $\langle\bar{u}\gamma_5 d\rangle$ condensates take the following values
\begin{equation}\label{chircondensa}
\langle\bar{u}u+\bar{d}d\rangle=2\langle\bar{\Psi}\Psi\rangle_{vac}~\mbox{cos}\alpha~~~~~~~~~~\langle\bar{u}\gamma_5 d\rangle+\mbox{h.c.}=2\langle\bar{\Psi}\Psi\rangle_{vac}~\mbox{sin}\alpha
\end{equation}
The chiral condensate rotates into a pion one as $\mu_I$ grows. In correspondence of the superfluid phase the pion field must be redefined with respect to the vacuum case: both the superfluid mode and the 
massive excitations are suitable linear combinations of the original $\pi^+,\pi^-$ fields, and can be interpreted as Bogoliubov quasi-particles of the condensed phase.\\
It is worth to stress that the ground state at $|\mu_I|>m_{\pi}$ is a Bose Einstein condensate of charged pions, which are composite particles.
We recall here that this theory is reliable as long as pions are the only particles excited by the chemical potential. Therefore, an upper limit for using such a theory can be identified as $\mu_{\pi}=2\mu_I\simeq m_{\rho}=770~\mbox{MeV}$.
On the other hand, in the regime of asymptotically high
values of $\mu_I$ the description in terms of chiral models breaks down; a perturbative calculation is sensible, and the result is that the ground state should be characterized by a non vanishing value of the field  

\begin{equation}\label{asymptotmui}
\langle\bar{d}\gamma_5u\rangle\neq 0
\end{equation}

This condensate carries the same quantum numbers as that at smaller chemical potentials, but the physical interpretation is quite different. In the regime $|\mu_I|\gg \Lambda_{QCD}$ the condensate is a BCS condensate of $\bar{d},u$ quarks (for positive $\mu_I$).
Therefore, at some intermediate value of $\mu_I$ there should be a transition BEC-BCS, but since the condensate has the same quantum numbers in the two regimes the transition should be continuous.

In heavy ion collision we do not expect values of $\mu_I$ sufficiently high to cause pion condensation; the attainable values of $\mu_I$ could cause a separation of the critical lines relative to the two light quarks, being different the respective chemical potentials $\mu_u=\mu_q+\mu_I$, $\mu_d=\mu_q-\mu_I$. This aspect can have a strong phenomenological significance.\\
There is another reason why QCD at finite isospin densities has been widely studied in recent years.
Actually, QCD at $\mu_I\neq0$ (and $\mu_q=0$) can be simulated on the lattice \cite{Alford:1998sd}, whereas the general and more physical case of $\mu_q\neq0$ suffers from the so-called sign problem (the fermion determinant is complex in euclidean space). 
However recently some approximate solutions have been proposed to overcome this difficulty. 
When integrating out the fermionic degrees of freedom, one is left with the following effective action

\begin{equation}\label{pathinteffact}
Z=\int \mathcal{D} \phi~\mbox{det}~M(\phi)~e^{-S_{YM}(\phi)}
\end{equation}
where $\phi$ are bosonic variables, and $\mbox{det} M(\phi)$ is the fermion functional determinant, acting in flavor, Dirac, color and momentum-frequency spaces, stemming from a Lagrangian density $\bar{\Psi}M(\phi)\Psi$ after the integration over fermion fields. If $det M(\phi)$ is a real-positive
definite quantity, we can treat this term as a modification of the functional measure
$\mathcal{D} \phi$, and standard Monte-Carlo simulations are allowed. To guarantee that the measure is positive definite, generally one must have an even number of flavors, with $det(M)$ real for each of them. One particular case is when there exists an operator $P$ such that for each flavor

\begin{equation}\label{chempotdet}
M^{\dagger}=PMP^{-1}
\end{equation}
In this case $det~M^{\dagger}=det~M$, and therefore $(det~M)^2=det~|M|^2$ which is positive, and thus for each flavor $det~M$ is real. The full determinant is positive for an even number of flavors. In the case of a chemical potential $\mu$ real and equal for two flavors, relation (\ref{chempotdet}) does not hold, and this brings to the well known sign problem.

One possible way to avoid this problem is to consider an imaginary chemical potential instead of a real one \cite{Alford:1998sd}. In this case the fermion determinant is positive definite, but after Monte-Carlo simulations an analytical continuation is needed to come back to the physical case of a real $\mu$. Some caution in this procedure is needed.

A case where numerical simulations are possible is that of QCD at finite density of isospin. In that case the operator $M(\mu)$ has the following structure

\begin{equation}\label{detchempotisosp}
M(\mu)=\left( 
\begin{array}{cc}
L(\mu) & 0\\
0 & L(-\mu)
\end{array}
\right)
\end{equation}
$L(\mu)$ is the Dirac operator for one flavor with chemical potential $\mu$. $L(\mu)$ satisfies $L(\mu)^{\dagger}=\gamma_5 L(-\mu)\gamma_5$, and eq. (\ref{chempotdet}) is satisfied by setting 

\begin{equation}\label{detgamma5chempotisosp}
P=\left( 
\begin{array}{cc}
0 & \gamma_5\\
\gamma_5 & 0
\end{array}
\right)
\end{equation}

The relation $det~M(\mu)=|det~L(\mu)|^2$ follows straightforwardly.
Therefore the study of QCD at finite isospin density allows to establish a direct connection between analytical calculations, performed within effective models, and numerical lattice simulations of the basic theory; this is a reason why this argument became recently rather popular.

Another case where $det~M$ is real with a chemical potential is when there exists an invertible operator $Q$ such that:

\begin{equation}\label{aggiuntochempot}
M^{*}=QMQ^{-1}
\end{equation}
Examples of this sort are afforded by models with four-fermion interaction, like the NJL model.

\section{General remarks on QCD at $\mu\neq 0$}
\label{Genremneq0mu}

The insertion of a chemical potential breaks explicitly charge conjugation (and also Lorentz invariance).
Futhermore, the condensation of a pseudoscalar field at $\mu_I\neq 0$ leads to the spontaneous breaking of parity; also the combined symmetry CP is broken.
The breaking of P is not in contradiction with the Vafa-Witten theorem \cite{Vafa:1983tf,Vafa:1984xg}, which states that, in vector-like gauge theories like QCD (i.e. theories where gauge invariant bare masses are possible for all fermions, which implies that both states of chirality appear in the theory), one cannot have the spontaneous breaking of parity and of vector-like global symmetries (like isospin and baryon number). The theorem holds at zero temperature and chemical potentials only, and in the case of vanishing theta term, $\theta=0$.
The $\theta$ term, see eq.(\ref{topologicalterm}), is related to the so-called strong CP problem in QCD. A $\theta\neq 0$ would break CP, due to instantonic effects, but experimental analyses put a very low upper bound on $\theta$ ($\theta\lesssim10^{-9}$). This problem could be avoided considering a dynamical degree of freedom associated with $\theta$ \cite{Peccei:1977hh,Weinberg:1977ma}; the axion would result as a pseudo-Goldstone boson corresponding to a Peccei-Quinn symmetry.\\
The previous arguments may have a strong cosmological significance. The condensation of a pseudoscalar field can be used as a toy model description for breaking CP, in the attempt of giving a unified explanation of cosmological issues by means of particle physics \cite{Kobsarev:1974hf}.

The physics of QCD at finite chemical potential presents many other interesting features.
One is related to the counting of Goldstone bosons. There is a theorem due to Nielsen and Chadha \cite{Nielsen:1975hm} stating that, in relativistically non-covariant theories, there is a possible mismatch between the number of broken generators and number of massless particles. One can distinguish two types of Goldstone bosons: those with the energy proportional to an even power of the momentum and those showing a dispersion relation which is an odd power of the momentum. The particles of the first type must be counted twice. QCD with a non zero chemical potential is just an example of a Lorentz non-invariant theory. \\ This theorem can be applied for instance to the case of non zero chemical potential for strangeness \cite{Schafer:2001bq}. 
Let us consider a theory of interacting bosons:
\begin{equation}\label{efflagrakaon}
\mathcal{L}=(\partial_0+i\mu)\Phi^*(\partial_0-i\mu)\Phi-(\vec{\nabla}\Phi^*)(\vec{\nabla}\Phi)-m^2\Phi^*\Phi-\lambda(\Phi^*\Phi)^2
\end{equation}
where $\Phi$ is a doublet of complex fields and $m$ is the mass term. The theory is invariant under $SU(2)\otimes U(1)$ transformations, and the chemical potential $\mu$
is introduced with respect to the $U(1)$ charge. 
This simple model can mimic the behaviour of the four kaons, being $\mu$ a chemical potential associated with strangeness; in a realistic model the interaction term should be better specified. 
However, as far as the problem of kaon condensation is concerned, this model encompasses much of the relevant features.
If the chemical potential is higher than the mass $m$, the potential energy has a non trivial minimum with $\Phi\neq 0$. This causes the breaking of $SU(2)\otimes U(1)$ down to $U(1)$. A naive analysis would suggest the existence of three Goldstone bosons, corresponding to the three broken generators. Conversely, if one expands the action around the minimum of the potential, and studies the particle spectrum, one finds only two Goldstone bosons; this is possible because one of them has a quadratic dispersion relation, and the other a linear one. The counting of massless particles stems form the Nielsen-Chadha theorem.

Another interesting argument concerns the behaviour of the QCD partition function at zero temperature and finite chemical potential 

\begin{equation}\label{effactchempotq}
Z_q(T,\mu_q)=\int \mathcal{D} A (det(\displaystyle{\not}{D}+m-\mu_q\gamma_0))^2~e^{-S_{YM}}
\end{equation}
where $S_{YM}$ is the euclidean finite temperature Yang-Mills action, and we are here considering
two degenerate flavors with equal chemical potential $\mu_q$. The fermion determinant is taken for each quark flavor, therefore it appears to the second power.
The only place where the chemical potential enters in is the fermion determinant. One would expect that rising up $\mu_q$ from zero, at $T=0$, the fermion determinant is modified and the system is excited from its ground state. Instead, until a critical value of $\mu_q$ is reached, the system still lies in the ground state, and the partition function remains unchanged. 
The critical value $\mu_q=(m_N-\epsilon)/3$, where $m_N$ is the nucleon mass and $\epsilon$ is the nuclear binding energy, must be associated with the transition to the nuclear matter regime.
This problem (called {\it Silver Blaze} problem from a Conan Doyle story \cite{Cohen:2003kd}) appears hard to be faced by means of a functional integral approach. 
As a first step, we can consider an isospin chemical potential.
The partition function can be written as

\begin{equation}\label{effactchempotI}
Z_I(T,\mu_I)=\int \mathcal{D} A|det(\displaystyle{\not}{D}+m-\mu_I\gamma_0)|^2~e^{-S_{YM}}
\end{equation}
because the fermion determinant is real and positive. An explicit calculation of the eigenvalues of the Dirac operator at finite $\mu_I$ ($det(\displaystyle{\not}{D}+m-\mu_I\gamma_0)=\prod_j\lambda_j$) can be done \cite{Cohen:2004qp} : the result is that, even if the eigenvalues and consequently the full fermion determinant are modified by $\mu_I$, 
the average over gauge configurations of
the fermion determinant remains unchanged until $\mu_I<m_{\pi}/2$; for higher values of $\mu_I$ pion condensation occurs. Actually, the gauge configurations which contribute to the partition function have a functional determinant identical to that of the vacuum.

As far as the quark chemical potential is concerned, a similar calculation shows that the partition function at $T=0$ is independent on $\mu_q$ for low chemical potentials \cite{Cohen:2004qp,Osborn:2005ss}, but the determination of the critical value for $\mu_q$ is still lacking.

However, the fermion determinant at $\mu_q\neq 0$ (and $\mu_I=0$) differs from that at $\mu_I\neq 0$ (and $\mu_q\neq0$) only by a phase; 
the so-called Phase-Quenched approximation consists in taking the absolute value of the determinant, and thus removing the complex phase.
Rigorous QCD inequalitites can be used to put some bound on the value of fermion determinant at finite quark chemical potential, which is the most interesting phenomenological situation.
A $1/N_c$ 't Hooft expansion \cite{'tHooft:1973jz} can be applied to shed some light on the problem \cite{Cohen:2004mw,Toublan:2005rq}.
A numerical study of the dependence of the critical curve on $\mu_q$ and $\mu_I$ in the NJL model will be presented in the sixth Chapter of this work.

\chapter{Ladder-QCD at finite isospin chemical potential}
\label{cap:ladderQCD}

In this Chapter, we present a calculation, done by using the ladder-QCD model, to explore the phase
diagram of QCD for chiral symmetry breaking and restoration, and pion condensation, at finite
temperature with different $u,d$ quark chemical potentials.
The steps for deriving the mean-field effective potential of the model are elucidated.
In agreement with previous investigations, we find that a finite pion condensate
takes place for high enough isospin chemical potential
$\mu_{I}$, with a critical value at $T=\mu_q=0$ which is just half of the pion mass. For $\mu_{I}$ smaller than ${m_{\pi}}/{2}$ the phase
diagram in the $(\mu_q,T)$ plane exhibits two first order transition
lines and two critical ending points. The latter fact could be a relevant consequence of a non-vanishing $\mu_I$ in heavy ion collision experiments.
We refer here to the paper \cite{Barducci:2003un}.

\section{Introduction}
\label{intcap1}
As said in the Introduction, according to
different models
\cite{Hatsuda:1994pi,Barducci:1989wi,Halasz:1998qr,Berges:1998rc} the
QCD phase diagram in the plane $(\mu_q,T)$ should exhibit a
tricritical point. Since this point should be located at moderate
density and temperature there is some possibility of observation
in heavy ion experiments. An important point in heavy ion
physics is the fact that in the experimental setting there is a
non zero isospin chemical potential $\mu_I$. Studies at finite
$\mu_I$ have been the object of several papers in the last few years
\cite{Alford:2000ze,Bedaque:1999nu,Buballa:1998pr,Steiner:2000bi,
Neumann:2002jm,Steiner:2002gx} but mainly in the regime of low
temperature and high baryon chemical potential. The first
study of the phase diagram in the space of parameters
$(\mu_q,\mu_I,T)$ by employing a microscopic model was done with a Random Matrix
model \cite{Klein:2003fy}. It has been found that the first order
transition line ending at the tricritical point of the case
$\mu_I=0$ actually splits in two first order transition lines and
correspondingly two crossover regions are present at low values
of baryon chemical potential. Shortly thereafter, the existence of this splitting has
also been shown in the context of a Nambu-Jona-Lasinio model in
\cite{Toublan:2003tt}. 

In this Chapter we will consider the effect of a finite isospin
chemical potential in a model (ladder-QCD) where the existence of
a tricritical point was shown several years ago
\cite{Barducci:1989wi}. 
The reason of doing this analysis in a
model different from the Random Matrix or the NJL ones is due to the fact that QCD at
finite baryon density is difficult to be studied on the lattice
(however it should be noticed that recently a new technique has
been proposed \cite{Fodor:2001au} and a first evaluation of the
tricritical point has been given in \cite {Fodor:2001pe}). It is
therefore important to study certain features in different models
in order to have a feeling about their universality. For instance
the existence of a tricritical point seems to enjoy such a
characteristic.
Moreover, the interest for this topic is due to
results from lattice simulations and effective theories which
forecast the existence of a phase transition toward a regime of condensed pions at finite $\mu_I$
\cite{Son:2000xc,Kogut:2002tm,Kogut:2002zg}.

\section{The model}
\label{sec:physicscap1}
In this section we will review a model used several
years ago to describe  the chiral phase of QCD both at zero
temperature \cite{Barducci:1984pr,Barducci:1987gn} and at finite
temperature and density \cite{Barducci:1989wi,Barducci:1989eu,Barducci:1993bh}.
This model is an approximation of QCD based on the evaluation of
the effective potential at two-loop level and on a
parametrization of the self-energy consistent with the OPE
results. 
The most striking features of the model are the dynamical breaking of chiral symmetry and the presence of a critical point in the ($T,\mu$) phase diagram, which are features generally assumed to hold in QCD.

The effective action we evaluate is a slight
modification (see for instance \cite{Barducci:1989wi}) of the
Cornwall-Jackiw-Tomboulis action for composite operators
\cite{Cornwall:vz,Jackiw:cv}. We recall here the major steps of
this calculation. We start from the Cornwall-Jackiw-Tomboulis
formula for the effective euclidean action $\Gamma$:

\begin{equation}
\Gamma [S]~=~-~\Gamma_2[S]~+~{\rm
Tr}\left[S\frac{\delta\Gamma_2}{\delta S}\right]~-~{\rm Tr}~{\rm
ln}\left[S_0^{-1}+\frac{\delta\Gamma_2}{\delta S}\right]+ {\rm
counterterms} \label{eq:gammacjt}\end{equation} where the logarithm term is the standard one loop fermionic term. The free
fermion inverse propagator is

\begin{equation}
S_0^{-1}(p)~=~i{\hat p}~-~m \label{eq:freeprop}
\end{equation}
being $m$ a matrix in $SU(3)$ flavor space; $\Gamma_{2}$ is the sum of all 2PI vacuum diagrams with a propagator
$S$ which has to coincide with the exact fermion propagator at
the absolute minimum of $\Gamma$. Thus $S$ is the dynamical
variable in this variational approach. 
However it turns out useful
to trade $S$ for
\begin{equation}
\Sigma~=~-\frac{\delta\Gamma_2}{\delta S} \label{eq:sdeq}
\end{equation}
which  coincides with the fermion self-energy at the minimum of
$\Gamma$.

\begin{figure}[htbp]
\begin{center}
\includegraphics[width=3cm]{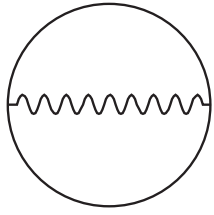}
\caption{\it The two-loop diagram needed to evaluate $\Gamma_2$.}
\label{fig:gam2}
\end{center}
\end{figure}

In the present model (ladder-QCD) we will make the very rough
approximation of evaluating $\Gamma_2$ at the lowest order. That
is, we evaluate $\Gamma_2$ at two-loops with one gluon exchange.
The relevant Feynman diagram is given in Fig. \ref{fig:gam2}. 
Of course, in the regime where the coupling gets small this is a good approximation, since other graphs are suppressed by a power of $\alpha_s$.
On the other hand, in the low-energy regime this procedure could appear as rather
uncertain.
However, it turns out that this approximation works rather well
phenomenologically (see for instance \cite{Barducci:1987gn}); 
the intrinsecally non-perturbative nature of this approach is able to display the phenomenon of spontaneous mass generation.
Therefore, at this order the effective action is simply given by
\cite{Barducci:1987gn}

\begin{eqnarray}
\Gamma[\Sigma]~&=&~-{\rm Tr}~{\rm ln}\left(S_0^{-1}-\Sigma\right)
-\frac 1 2~{\rm Tr}\left(S~\Sigma\right)+ {\rm c.t.}\nonumber\\
&=&~-{\rm Tr}~{\rm
ln}\left(S_0^{-1}-\Sigma\right)~+~\Gamma_{2}\left[\Sigma\right]+{\rm
c.t.} \label{eq:simplesp}\end{eqnarray}

As in refs. \cite{Barducci:1984pr,Barducci:1987gn}, we use the
following parametrization for $S$

\begin{equation}
S(p)~=~iA(p){\hat p}~+~B(p)~+~i\gamma_{5} C(p) \label{eq:parprop}
\end{equation}
Then, by working in the Landau gauge, it is possible to show that
no renormalization of the wave function is required and also that
the  Ward identity at this order  is satisfied by taking the free
quark-gluon vertex and the free gluon propagator.
Also the latter assumptions are in agreement with the behaviour of the theory in the UV regime.

We also consider the so-called rigid case, where the strong coupling $g$
is considered fixed at $p^2=M^2$, where $M$ is a convenient  mass
scale to be determined in the fit procedure. 
In the case we would consider a behaviour of coupling, masses and self-energies in accordance with the renormalization group,
we would automatically take into account the vertex perturbative corrections, i.e. ladder-diagrams, in the evaluation of the effective action. This the reason of the name of the model. However, we will neglect the renormalization group corrections since they make much harder the evaluation of the effective potential.

In this way the relation between the
scalar and pseudoscalar contribution to $\Sigma$ and the terms
$B,~C$ in Eq. (\ref{eq:parprop}) can be easily inverted, leading
to the following expression for the effective action, fully
expressed in terms of $\Sigma$ (\cite{Barducci:1987gn})
\begin{equation}
\Gamma[\Sigma]=\Gamma_{2}[\Sigma]+\Gamma_{\rm log}[\Sigma]
\end{equation}
where, separating $\Sigma=\Sigma_s+i\gamma_5\Sigma_p$

\begin{equation}
\Gamma_2[\Sigma]=-\frac{8\Omega_4 N_{c}\pi^2} {3g^2C_2}\int
\frac{d^4q} {(2\pi)^4}{\rm tr}~[\Sigma_s(q^2)\square~\Sigma_s(q^2)+
\Sigma_p(q^2)\square~\Sigma_p(q^2)]\label{eq:7}
\end{equation}
with $\Omega_{4}$ being the four-volume;  $C_{2}=4/3$ for
$N_{c}=3$ is the quadratic Casimir of $SU(3)_c$. Besides $M$, also
$g$ is a  parameter of the model. The one-loop term is
\begin{equation}
\Gamma_{\rm log}[\Sigma]=-{\rm Tr}~{\rm
log}\left(S_0^{-1}-\Sigma\right)= -{\rm Tr}~{\rm log}~\left[i{\hat
p}- m -\Sigma_s(p^2)-i\gamma_5\Sigma_p(p^2)\right]
\label{eq:unloop}
\end{equation}
where the scalar and pseudoscalar parts of the dynamical variable
are matrices in $SU(3)$ flavor space, related to
the scalar and pseudoscalar quark condensates through the
following equation

\begin{equation}
\Sigma_s(p^2)+i\gamma_5\Sigma_p(p^2) =({s}+i\gamma_5 {p})~f(p^2)
\label{eq:ope}
\end{equation}
The function $f(p^2)$ which contains the momentum dependence of
the self-energy will be discussed later. The fields

\begin{eqnarray}
&\langle s_{ab}\rangle&=-\frac{3C_2g^2} {4NM^3}\langle
\bar{\Psi}_{a}\Psi_b\rangle \label{eq:10}\\
&\langle p_{ab}\rangle&=-\frac{3C_2g^2} {4NM^3}\langle \bar{\Psi}_a
i\gamma_5\Psi_b\rangle\label{condpseud}
\end{eqnarray}
will be determined by minimizing the effective action.  In Eq.
(\ref{eq:unloop}), $m$ is the mass matrix in flavor space which is
taken diagonal
$$
 m=\left(
\begin{array}{ccc}
m_u & 0 & 0 \\
0  & m_d & 0\\
0 & 0 & m_s
\end{array}
\right)
$$
In the spirit of a variational approach, we must make an {\it Ansatz} on the behaviour of $f$ in momentum space. Of course, this would not be necessary if we were able to solve exactly the theory.
The function $f(p^2)$ will be chosen requiring that it goes to a
constant for $p\to 0$ and behaves as $1/p^2$ (apart from logarithmic terms) for large euclidean
values of $p$.
The former constraint derives from the comparison with four-fermion interaction models, where the self energy is constant up to a cut-off scale; actually, those models give a reasonable description of chiral symmetry breaking in the IR regime.
The latter constraint stems from the operator product expansion (OPE) results, in the deep euclidean ultraviolet limit. 
In practice, we want obtain an effective potential which traces that of the NJL model in the IR regime, plus UV corrections which are in agreement with the perturbative asymptotic behaviour of the theory. The merit of this approach is that we have started with the fundamental theory itself, by operating suitable approximations.

By introducing a
dimensionless variable $x^2=p^2/M^2$ we will consider the
following family of functions

\begin{equation}
\frac{f_N(x^2)}{M}= \displaystyle{\frac{1+x^2+x^4+...+x^{2N-2}}
{1+x^2+x^4+...+x^{2N-2}+x^{2N}}}
\label{eq:smoothnew}\end{equation} In the limit $N\to\infty$ we
get the ``step'' function used in \cite{Barducci:1984pr,Barducci:1987gn}
\begin{equation}
{f(x^2)\over
M}=\theta\left(1-x^2\right)~+~\theta\left(x^2-1\right)~
\displaystyle{{1\over x^2}} \label{eq:stepold}\end{equation}
The reason of choosing a test function which is the ratio of polynomial expressions lies in the fact that the finite temperature extension of the effective potential will result straightforward.

Notice that for $x\to 0$ we get
\begin{equation}
{f_{N}(x^2)\over M}\sim 1 - x^{2N}+.. \label{eq:expandf}
\end{equation}
Now, let us consider for simplicity the chiral limit and zero
chemical potential. In this case it is simple to get the
mass-shell condition from the one loop term in
Eq.~(\ref{eq:unloop}) (see for instance \cite{Barducci:1989wi})
\begin{equation}
p^2+\langle s\rangle ^2 f^2_N=0 \label{eq:mashell}
\end{equation}
where $\langle s\rangle$ is the field proportional to the scalar
condensate (see Eq.~(\ref{eq:10})). If we want to recover, at
least in the infrared regime, a free particle-like dispersion
relation (as for instance happens in theories with four-fermion interaction), we
see from Eqs.~(\ref{eq:expandf}) and (\ref{eq:mashell}) that we
need $N\geq 2$. In this work we will choose $N=2$. Notice that
the choice $N=1$ would lead to the following dispersion relation
in the limit of small momenta
\begin{equation}
p^2(1-2\langle s\rangle ^2)+M^2\langle s\rangle ^2+...=0
\end{equation}
This might give rise to some problems in the broken phase where the
coefficient of $p^2$ can become negative. However no
difficulties arise for the determination of the critical points
where $\langle s\rangle\simeq 0$. 

We can thus evaluate explicitly the effective potential
\begin{equation}
V={\Gamma\over\Omega_{4}}
\end{equation}
which is UV-finite in the chiral limit, whereas it needs to be
properly renormalized in the massive case. We have employed the
following normalization condition
\begin{equation}\label{eq:condnorm}
\frac{\partial V}{\partial (m_a \langle
\bar{\Psi}_a\Psi_a\rangle)}\Big|_{min}=1
\end{equation}
In the chiral limit, this requirement is equivalent to the
Adler-Dashen relation (see for instance \cite{Barducci:1987gn})
\begin{equation}\label{AdlerDashen}
m^2_{\pi}f^2_{\pi}=-2m~\langle\bar{\Psi}\Psi\rangle_0
\end{equation}
where $\langle\bar{\Psi}\Psi\rangle_0$ is the condensate for one massless flavor, and $m$ is a current mass term. At leading order $m_{\pi}^2$ scales as the explicit chiral violating term $m$.
Here we will require the validity of this renormalization condition at the values
of the quark current masses.\\
By defining $\alpha_{a}=m_{a}/M$ and
\begin{equation}{\chi}_{a}=-
\frac{g^2}{3M^3}\langle\bar\Psi_a\Psi_a\rangle \label{eq:hatchi}
\end{equation}
the normalization condition, using
Eq. (\ref{eq:10}), can be written as
\begin{equation}
\frac{\partial V_a}{\partial (\alpha_a
\chi_a)}\Big|_{min}=-{3M^{4}\over 2\pi^{2}} c = -{3 M^{4}\over
g^2(M)}\label{eq:condorm1}
\end{equation}
where we have introduced the parameter
\begin{equation}c\equiv\frac{2\pi^2}{g^2(M)}\end{equation}
The chemical potential and the temperature dependence are
introduced  following standard methods \cite{Dolan:qd} (see for
example \cite{Barducci:1989wi}). In particular the chemical
potential is introduced from the very beginning  via the usual
substitution $p^{\nu}\rightarrow (p_0+i\mu,{\vec{p}})$ (we are working in the euclidean space) in ${\hat
{p}}$ appearing in the Dirac operator in Eq.~ (\ref{eq:unloop}).
On the other hand the temperature dependence is introduced by
substituting to $p_0$ the Matsubara frequency $\omega_n=(2n+1)\pi
T$ in all the $p_0$ dependent terms appearing in the effective
action. The reason for this asymmetric treatment is that the
$p$ dependence in the self-energies becomes relevant only at
$p\gg M$ whereas, as we shall see, we will be interested in
chemical potentials lower than $M$. We will not review in detail this procedure, since it is a standard technique of thermal field theory. However, 
we have to decompose the structure $(p^2+(m+\chi~f(p^2))^2)$ of the one-loop term into a product $\prod_i(p^2+z_i^2)$, since the latter terms can be trivially extended to finite temperature.
Of course $\mbox{log}(\prod_i(p^2+z_i^2))=\sum_i~\mbox{log}(p^2+z_i^2)$, and therefore
the roots $z_i$ play the role of the effective mass of some quasi-particle mode. The fact that some of the $z_i$ can be complex leads to instabilities in the equation of state in the regime of low temperatures. This was the case corresponding to the choice of the $N=1$ parameter in the test function. In particular, for $N=1$ the pressure (defined form the effective potential as $p=-(V(T,\mu)-B)$, with $B=V(T=0,\mu=0)$ being the bag consant) was not a monotonic, increasing function of the temperature. This caused a non positive-definite energy density in the regime of low temperatures ($\epsilon=-p+Ts+\mu\frac{\partial ~p}{\partial~\mu}$, being the entropy density $s$ defined as $\frac{\partial~p}{\partial~T}$), and a corresponding negative specific heat $C_V=\frac{\partial\epsilon}{\partial~T}$ in the same regime.
In order to avoid these instabilities, we have chosen the test function corresponding to $N=2$.

\section{Results at finite temperature and density}

In order to obtain the effective action we need to calculate the
determinant of the operator appearing in Eq. (\ref{eq:unloop}). We
set to zero the strange quark chemical potential $\mu_{s}=0$ and
define $\mu_{I}=(\mu_{u}-\mu_{d})/2$ and
$\mu_q=(\mu_{u}+\mu_{d})/2$. The operator is given by the
following  $3\otimes 3$ matrix in flavor space:

\begin{equation}
\begin{pmatrix}
i(\omega_{n}+i\mu_{u})\gamma_{0}+i{\vec p}\cdot{\vec\gamma} -
F_{u}  &~~~~ \rho f_2\gamma_{5}& 0 \cr  -\rho f_2\gamma_{5}&
i(\omega_{n}+i\mu_{d})\gamma_{0}+i{\vec p}\cdot{\vec\gamma} -
F_{d} &0 \cr 0 & 0 &i\omega_{n}\gamma_{0}+i{\vec
p}\cdot{\vec\gamma} - F_{s} \label{determinant}
\end{pmatrix}
\end{equation}
where we have defined $F_{a}=m_{a}+ f_{2}\chi_{a}$ ($a=u,d,s$); $f_{2}$ 
is defined in Eq. (\ref{eq:smoothnew}), with $N=2$. Also,
$p^2=\omega_n^2+|\vec p\,|^2$. Here $\rho$ is related to the
charged pion condensate
\begin{equation}\label{pionfield}\rho=-\frac{g^2}{6M^3}
(\langle \bar u\gamma_5 d\rangle-\langle\bar d\gamma_5 u\rangle)
\end{equation}

We have directly set to zero the hyper-charged condensates in the
strange sector, since we do not consider here the formation of such
condensates for $\mu_{s}=0$. 
This point will be the subject of a further study in Chapter five.
The strange sector thus factorizes
and the determination of the phase diagram for (approximate) chiral
symmetry restoration is performed by studying the behavior of the
light $\langle {\bar u}u\rangle$ and $\langle {\bar d}d\rangle$
quark condensates, independently on the strange quark condensate.
To evaluate the effective action, we perform the sum over the
Matsubara frequencies using standard methods \cite{Dolan:qd}. We 
have to find numerically the roots $z_i$ which solve the mass-shell condition given
by the vanishing of the determinant in Eq.~(\ref{determinant}). Although formally
straightforward, the calculation is a hard computational task.
Actually, at each integration step on $|{\vec p}|$, we have to
solve a twentieth order algebraic equation in $\omega_{n}$ in the
$u-d$ sector. The part relative to the strange quark is obviously
easier to deal with. The coefficients of this equation  depend
also on the parameters
$\chi_u,\chi_d,\rho,\mu_u,\mu_d,m_{u},m_{d}$.

Before discussing the results at finite density and temperature
let us review how we fix the parameters appearing in the effective
action. This is done by looking at $T=\mu=0$. The parameters that
we have to fit are $c$, $M$, $m_u$, $m_d$, $m_s$. These are
obtained by using as input parameters the following physical
quantities: $m_{\pi^\pm}$, $m_{K^\pm}$, $m_{K^0}$, $f_\pi$ and
$f_K$. Here we do not show the explicit expressions, but we limit ourselves to recall the major steps in the calculation.\\
The static meson masses have been computed as second derivatives of the effective potential with respect to the corresponding canonical meson field. For instance for $m_{\pi^{\pm}}$ we have
\begin{equation}\label{mpailadderr}
m^2_{\pi^{\pm}}=\frac{d^2~V}{d\pi^2}
\end{equation}
The relation between the canonical pion field $\pi$ and the field $\rho$ of eq.(\ref{pionfield}) can be derived by studying the effective action for meson fields. The two fields are related to each other by a constant multiplicative factor.
\\The decay constants are derived by evaluating the couplings of the canonical meson fields to the axial-vector currents.
This can be done by considering the effective action as a function of space-time dependent scalar and pseudoscalar fields.
Here we just write the expression for $f_{\pi}$ in the chiral case, the so-called Pagels Stokar relation
\begin{equation}\label{Pagelsstokar}
f^2_{\pi}=\frac{3}{(2\pi)^2}\int_0^{\infty}dp^2~p^2~\left[\bar{\Sigma}^2-\frac{1}{2}p^2\bar{\Sigma}\frac{d\bar{\Sigma}}{d~p^2} \right]/\left[ p^2+\bar{\Sigma}^2\right]^2
\end{equation}
where $\bar{\Sigma}$ is the self-energy $\Sigma$ evaluated at the minimum of the effective potential.
A similar expression will be obtained in Chapter IV in the framework of the NJL model.
In any case, at leading order in the current quark mass, the expressions we find for $m_{\pi},f_{\pi}$ statisfy the Adler-Dashen relation eq.(\ref{AdlerDashen}).

The results of the fit are given in Table I; in Table II we show the
values of the input experimental quantities, together with the
result we get from the fit procedure.
\begin{table}[htbp]
\begin{center}
$$
\begin{array}{|c|c|}
\hline
{\rm {\bf Parameters}} & {\rm {\bf Fitted ~ Values}}\\
\hline
M & 529 ~ {\rm MeV}\\
\hline
c & 1.0\\
\hline
m_u & 4.4 ~{\rm MeV}\\
\hline
m_d & 6.2 ~{\rm MeV}\\
\hline
m_s & 110 ~{\rm MeV}\\
\hline
\end{array}
$$
\end{center}
\caption{{\it Fit of the parameters.}} \label{Fitpar}
\end{table}

\begin{table}[htbp]
\begin{center}
$$
\begin{array}{|c|c|c|}
\hline
{\rm{\bf Input~parameters}}&{\rm{\bf Fitted~Values}}&{\rm{\bf Experimental~Values}}\\
\hline
m_{\pi^{\pm}} & 139~{\rm MeV} & 139.6~{\rm MeV}\\
\hline
m_{K^{\pm}} &494~{\rm MeV} & 493.7~{\rm MeV}\\
\hline
m_{K^{0}} &499~{\rm MeV} & 497.7~{\rm MeV}\\
\hline
f_{\pi} & 92 ~{\rm MeV} & 92.4~{\rm MeV}\\
\hline
f_{K} & 105 ~{\rm MeV} & 113~{\rm MeV}\\
\hline
\end{array}
$$
\end{center}
\caption{{\it Comparison between the values of the input
parameters as obtained from the fit  and the experimental
results.}} \label{Valexp}
\end{table}

With these values of the parameters, we find that at $T=\mu=0$ the
quark condensate for one flavor in the chiral limit has the value
\begin{equation}\langle{\bar\psi}\psi\rangle_{0}=-(248~{\rm MeV})^3\end{equation}
 whereas in
the massive case \begin{equation}\langle{\bar
u}u\rangle=-(251~{\rm MeV})^3,~~~\langle{\bar
d}d\rangle=-(253~{\rm MeV})^3,~~~\langle{\bar
s}s\rangle=-(305~{\rm MeV})^3\end{equation}  

The values we find for the light quark condensates are in agreement with the commonly accepted value $\langle\bar{\Psi}\Psi\rangle\sim-(250~\mbox{MeV})^3$.
By defining the
constituent quark masses a la Politzer \cite{Politzer:1976tv}
\begin{equation}
M_{const}=\bar{\Sigma}(p^2=4M_{const}^2)
\end{equation}
we get
\begin{equation}\label{Masseff2}
M_s=385~{\rm MeV},\ \ \ \ \ \ \  M_{u,d}=256\ \rm{MeV}
\end{equation}
where, here and in the following, the light quarks have been
taken degenerate with average mass ${\hat m}=(m_{u}+m_{d})/2=5.3~{\rm MeV}$.
For the ratio $m_s/{\hat m}$ we find a value of $\sim 21$ instead of the generally assumed value $\sim 25$. The values we find for both the fitted quantities and the output parameters motivate the statement that this model reproduces a reasonable phenomenology.\\
In the case of $T=\mu=0$ and in the chiral limit, one also
finds that in order to break the chiral symmetry we must
have
\begin{equation}
c<1.37
\end{equation}
which corresponds to
\begin{equation}\label{alphas}
\alpha_s(M)=\frac{g^2}{4\pi}\gtrsim1.1
\end{equation}
This condition is satisfied by our choice of parameters.

\begin{figure}[htbp]
\begin{center}
\includegraphics[width=16cm]{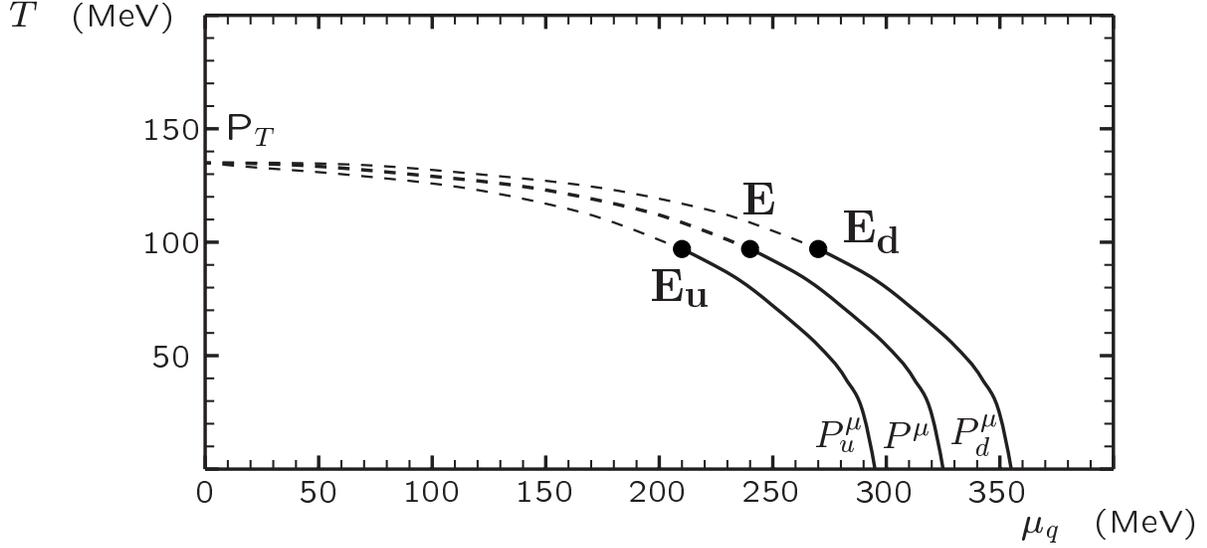}
\end{center}
\caption{\it Phase diagram for chiral symmetry in the
$(\mu_{q},T)$ plane. For $\mu_{I}=0$ (central line), the
cross-over transition line starts from the point $P_{T}=(0,135)$
and ends at the point $E=(240,97)$. The line between $E$ and the
point $P^{\mu}=(325,0)$ is the line for the first order transition
with discontinuities in the $\langle{\bar u}u\rangle$ and
$\langle{\bar d}d\rangle$ condensates. For $\mu_{I}=30~{\rm MeV}$
(side lines), the two cross-over transition lines start from the
point $P_{T}=(0,135)$ and end at the points $E_{u}=(210,97)$ and
$E_{d}=(270,97)$. The lines between $E_{u}$ and the point
$P^{\mu}_{u}=(295,0)$ and  between $E_{d}$ and the point
$P^{\mu}_{d}=(355,0)$ are the lines for the first order
transitions with discontinuities in the $\langle{\bar u}u\rangle$
and $\langle{\bar d}d\rangle$ condensates respectively.}
\label{fig:diafaison2}
\end{figure}

When $\mu_I=0$ it was shown that in the version of the model discussed in
\cite{Barducci:1989wi} there is a tricritical point in the chiral
limit. The model we are presenting here is essentially the same as \cite{Barducci:1989wi},
apart from some slight modifications, as for instance, the choice
of the function $f_2(x)$. However the tricritical point is still
present as it can be seen from Fig. \ref{fig:diafaison2} (central
line), with the choice of parameters of Table I.

A complete analysis of the full three parameter space
$(\mu_q,\mu_I,T)$ has not been completed yet especially in
relation with the possible appearance of a hyper-charged condensate. 
A carefully study will be done in Chapter five in the framework of the NJL model.
We have
examined here the case $T=\mu_q=0$ and we have found that there is a
phase transition indicated by a finite pion condensate starting
at $\mu_I=70$ MeV. Since in our model $m_\pi\approx 140$ MeV we
agree with the results found in the literature
\cite{Son:2000xc,Klein:2003fy,Kogut:2002tm,Kogut:2002zg}.
We also found that for $m=0$ the critical $\mu_I$ is zero; actually in that case the pion mass is exactly zero because the pion is a rigorous Goldstone boson.
In the following we will limit our studies to the regme of small isospin
chemical potential, say $\mu_I=30$ MeV, where we expect the pion
condensate $\rho$ to vanish. In this situation the determinant in
 (\ref{determinant}) factorizes and the effective action is
given by the sum of three independent terms, one for each  flavor.
Therefore the structure of the action is analogous to the case of $\mu_I=0$, with each
flavor evaluated at its own chemical potential. It follows that
the two light flavor critical lines show the same tricritical structure
exhibited from the central line in Fig. \ref{fig:diafaison2} for
$\mu_I=0$. Consequently the phase diagram we obtain for a small,
fixed isospin chemical potential ($\mu_{I}=30~{\rm MeV}$) is
described by the two side lines in Fig. \ref{fig:diafaison2}; each flavor $u$ and $d$ shows the same structure as the central
line but with a splitted chemical potential
$\mu_{u,d}=\mu_q\pm\mu_I$. Notice also that although the Figure \ref{fig:diafaison2}
extends up to zero temperature, its low temperature, high chemical potential part should not be taken too
seriously due to the existence of color superconductivity (which in the present study is not considered).

It should also be noticed that in 
\cite{Frank:2003ve} a study of the $(\mu_q,T)$ phase diagram with non zero $\mu_I$ has been performed within a NJL model augmented by the
four-fermi instanton interaction relevant in the case of two
flavors. These authors have found that the coupling induced by
the instanton interaction between the two flavors might wash
completely the splitting of the first order transition line. This
happens for values of the ratio of the instanton coupling to the
NJL coupling of order 0.1-0.15.
In our model all the three flavors are present, and the relevant
instanton effects would give rise to a six-fermion contact
interaction. 
However, the `t Hooft term has not be considered in the present version of the model. The six fermion term would give rise to a non renormalizable term, whereas
the present model is expected to be renormalizable, being derived form QCD theory. Therefore, that term would require some caution to be included in the model.
Due to the lack of this flavor-mixing term, it is still an open question whether these effects would wash
out the splitting as in the two flavor case \cite{Frank:2003ve}.
We
also ignore the effects from color superconductivity, since these
are present only for temperatures lower than some tens of ${\rm
MeV}$, and they would not regard the physics of heavy ions.
 What we are presenting here is a limited
study of the phase diagram, and we restrict the analysis at small
isospin chemical potential. A complete study of the phase diagram should take into account instanton effects and di-quark condensation too.
However, a more detailed analysis will be done in the next Chapter, by using the NJL model, since it is simpler to deal with.

\section{Conclusions}
\label{sec:conclucap1} 
In this Chapter we have considered an approximate
model of QCD (ladder-QCD) at finite temperature and densities. In
particular we have studied the experimentally important
situation of a non-vanishing isospin chemical potential. This
situation has been already explored previously by various authors
and we confirm, in particular, the results found in
\cite{Klein:2003fy} and \cite{Toublan:2003tt} about the splitting
of the first order transition line in the plane $(\mu_q,T)$, for
small $\mu_I$. As pointed out in \cite{Toublan:2003tt} this
result could be relevant for heavy ion physics since the first order
transition line is split symmetrically with respect to the
original line at $\mu_I=0$. This implies a reduction of the value
of the quark chemical potential at the tricritical point of an
amount given by $\mu_I$, making easier the possibility of
discovering it experimentally. Since it is very difficult to
perform first principle analysis of QCD at finite baryon density,
we think that it is important to show that certain features as
the existence of the tricritical point and the possible splitting
of the first order transition line are common to different models.
This suggests that these features might possess some universal
character, valid also for QCD. However we should stress that the result of the
splitting of the first order transition line is strictly related
to the factorization in flavor space. For instance, in the two
flavor case, the four-fermi interaction due to the instanton
effects leads to a mixing of the flavors that, if sufficiently
large, might wash out the mixing. This point needs
further analysis in the more complete scheme with three flavors and instanton-induced interaction. In the sixth Chapter we will perform a detailed study of the effect of instanton-induced interaction on the behaviour of the critical line, in the NJL model. We will show that, in the three flavor chiral case, this coupling can drive the zero chemical potential transition from second order to first order; therefore the 't Hooft coupling is crucial for defining the properties of the chiral transition. 
\newpage

\chapter{Pion condensation in a 2-flavor NJL model}
\label{NJL2}

We study the phases of a two-flavor Nambu-Jona-Lasinio model at
finite temperature $T$, quark and isospin chemical potentials $\mu_q,\mu_I$.
This study, related to that of the paper \cite{Barducci:2004tt}, completes the analysis performed in the previous Chapter, in the ladder-QCD model, 
where only small values of $\mu_{I}$ were considered. The paper \cite{Barducci:2004tt} was the first in literature to study pion condensation and the restoration of chiral symmetry in the scalar sector in the {\it whole} space of parameters $(\mu_q,\mu_I,T)$, after the analysis performed within the random matrix model of \cite{Klein:2003fy}.

\section{Introduction}
\label{intcap2}
The possible formation of a pion condensate due to a finite
isospin chemical potential $\mu_{I}$ has been, in recent years,
the subject of several papers
\cite{Son:2000xc,Barducci:2003un,
Klein:2003fy,Toublan:2003tt}.
Pion condensation has been so far primarily
investigated by means of low energy models based on chiral
lagrangians
\cite{Son:2000xc,Splittorff:2000mm,Kogut:2001id,Jakovac:2003ar}.
Although these models are well suited to study the phases of QCD
as they have the right symmetry properties, they do not include
the combined effect of the isospin chemical potential $\mu_{I}$
with a finite quark chemical potential $\mu_{q}$ in order to
study the pattern of chiral symmetry breaking and restoration as
well. 
In order to consider the effect of both $\mu_{q}$ and $\mu_{I}$ on the behaviour of scalar and pseudoscalar condensates, we need a model
with quarks as microscopic degrees of freedom, and to study chiral symmetry breaking as a dynamical effect.

 The effect of a small
$\mu_{I}$ has been investigated in
ref. \cite{Toublan:2003tt} in the context of the
Nambu-Jona-Lasinio model and in ladder-QCD
\cite{Barducci:2003un}. The result is the splitting of the
critical curves for chiral symmetry restoration for the two light
flavors, whereas a full study for arbitrary $\mu_{I}$ has only
been done in the context of a random matrix model
\cite{Klein:2003fy}.
Studies on the lattice have been performed at finite $\mu_{I}$ and
$\mu_{q}=0$ in refs.
\cite{Kogut:2002tm,Kogut:2002zg,Gupta:2002kp,deForcrand:2002ci,Sinclair:2003rm}
and with a finite $\mu_q$ and $\mu_I=0$ in refs.
\cite{Fodor:2001au,Fodor:2001pe,Sinclair:2003rm,Allton:2002zi,D'Elia:2002gd}.
In the work \cite{Nishida:2003fb} the effect of both $\mu_{q}$ and a
small $\mu_{I}$ has also been considered. The case of high $\mu_q$
and small $\mu_I$ has been taken into account in \cite{Hands:2004uv}.

In this Chapter we extend the analysis of
\cite{Barducci:2003un,Toublan:2003tt} where it was found that the
first order transition line ending at the tricritical point of the
case $\mu_I=0$ actually splits into two first order transition
lines and correspondingly two crossover regions are present at low
values of baryon chemical potential. In particular we will be
working in the context of a NJL model with a form factor included
such as to imply a UV decrease of the fermion self-energy
compatible with the operator product expansion.
In the present version of the model, we do not consider instanton effects.
As mentioned in the previous Chapter, it should also be noticed
that the coupling induced by the
instanton interaction between the light flavors might completely
wash out the splitting of the first order transition line. 
This effect was studied in ref. \cite{Frank:2003ve} in a NJL model.
To shed some light on this controversial point it should be also important to consider
 the explicit $SU(2)_V$ breaking due to the mass asymmetry between $u$ and $d$ quarks, and the possible restoration of $U(1)_A$ symmetry, due to the damping of instanton effects in the regime of high temperatures and/or densities. These aspects will be discussed, although partially, in the sixth Chapter of this work.

In section II we summarize the relevant features of the NJL model
we have here considered, with an isospin charge included. The expression for the one-loop
effective potential and the values of the fit parameters are
reported. 
In section III we describe the bosonization procedure, which is necessary to express the effective action in terms of the composite fields.
In section IV we discuss the several phases
of the model, together with the corresponding symmetries, by
studying the behavior of the scalar and pion condensates with
respect to the thermodynamic parameters 
$T,\mu_q,\mu_I$ (or $\mu_u,\mu_d$). Results are shown for growing
temperatures, starting from zero up to temperatures above that of
the critical ending point. 
In section V we present a calculation which confirms the analytical agreement between $\mu_I^C$ and half of the pion mass, in the framework of the NJL model. 
Section VI is devoted to
conclusions.

\section{The model}
\label{sec:physicscap2}

Our purpose is to explore the structure of the phase diagram for
chiral symmetry breaking and pion condensation in QCD at finite temperature
and quark densities, by using a microscopic model with quark
degrees of freedom. This task has been accomplished, up to now, in
the context of a random matrix model simulating QCD with two
flavors \cite{Klein:2003fy} and, in the case of small differences
between the $u$ and $d$ quark chemical potentials, also in the
Nambu-Jona-Lasinio model (NJL) \cite{Toublan:2003tt} and in
ladder-QCD
\cite{Barducci:2003un}.\\
One motivation for using a model with quarks as microscopic degrees of
freedom is that it gives us the possibility of studying chiral
symmetry breaking and pion condensation at finite isospin and
baryon chemical potentials; this study is not possible within effective
chiral models.\\ In ref. \cite{Toublan:2003tt}, the authors made
use of the NJL model with a suitable form factor included in the
quark self-energy to mimic asymptotic freedom
\cite{Alford:1997zt,Berges:1998rc}. This version of the NJL model
turns out to be very close to ladder-QCD as developed in refs.
\cite{Barducci:1989eu,Barducci:1987gn} where the momentum
dependence of the quark self-energy is consistently dictated by
the study of the Schwinger-Dyson equation within a variational
approach (see the previous references for details). However,
although ladder-QCD is a covariant and self-consistent approach,
the dependence on the four-momentum of the quark self-energy makes
the numerical computation of the one-loop effective potential with
finite quark densities much more onerous than in the NJL
case, where the quark self-energies depend only on the
three-momentum. For this reason, in the present work we study the
NJL model. It is reasonable to expect that when employing
ladder-QCD, the resulting physical picture does not considerably
differ from that of the NJL model. This has been the case in
previous applications too
\cite{Hatsuda:1994pi,Barducci:1989eu,Klevansky:1992qe}.

As already said in the Introduction we are not going to consider
the effects of di-fermion condensation. Therefore our results can
be considered valid only outside the region of the color
superconductive phase, which roughly should take place for
$\mu_{q}\gtrsim 400\sim500~\mbox{MeV}$ and $T\lesssim 50~\mbox{MeV}$. For the same reason we will not consider values of $\mu_{u}$ or
$\mu_{d}$ higher than $400\sim500~\mbox{MeV}$, where other di-fermion
condensates might arise (see for instance ref.
\cite{Alford:2002rz}).

Let us now consider the Lagrangian of the NJL model, in Minkowski space, for two
flavors $u,d$ with the same mass $m$ but with different chemical
potentials $\mu_{u}$ and $\mu_{d}$
\begin{eqnarray}
{\cal {L}}&=& {\cal {L}}_{0}+{\cal {L}}_{m}+{\cal {L}}_{\mu}+{\cal {L}}_{int}\nonumber\\
&=&{\bar{\Psi}} i{\hat{\partial}}\Psi ~-~ m~{\bar{\Psi}}\Psi~+~
\Psi^{\dagger}~A~\Psi~+~{G\over 2}\sum_{a=0}^{3}\left[\left(
{\bar{\Psi}}\tau_{a}\Psi
\right)^{2}+\left({\bar{\Psi}}i\gamma_{5}\tau_{a}\Psi\right)^{2}
\right] \label{eq:njlagr}
\end{eqnarray}
where $\Psi=\left(
\begin{array}{c} u\\ d
\end{array} \right)$, $~~A=\left(
\begin{array}{c} \mu_u\\0
\end{array} \begin{array}{c} 0\\\mu_d
\end{array}\right)$ is the matrix of chemical potentials and
$~~\tau_{a}$, $a=0,1,2,3$, is the set of the three Pauli matrices plus the identity.
The interaction term of eq.(\ref{eq:njlagr}) is invariant under 
$SU(2)_L\otimes SU(2)_R\otimes U(1)_V$, since we want to mimic the spontaneous breaking of chiral symmetry. In eq.(\ref{eq:njlagr}) we recover all the $SU(2)_A$ invariant combinations, for instance
\begin{equation}\label{chiralmultip}
(\bar{\Psi}\Psi)^2+(\bar{\Psi}i\gamma_5\vec{\tau}\Psi)^2
\end{equation}
Actually the $\sigma$ particle (scalar with isospin $I=0$) is the chiral partner of the three pions (pseudoscalar with $I=1$).
\\
The interaction term of eq.(\ref{eq:njlagr}) is also invariant under $U(1)_A$ transformations since we are not considering here the instanton-induced interaction which would explicitly break $U(1)_A$. Therefore, in eq.(\ref{eq:njlagr}) we find terms like
\begin{equation}\label{chiralmultippseudo}
(\bar{\Psi}\Psi)^2+(\bar{\Psi}i\gamma_5\Psi)^2
\end{equation}
which are invariant under $U(1)_A$. Actually, if $U(1)_A$ symmetry was observed in nature, there would be a neutral, I=0 pseudoscalar meson with the same mass as the $\sigma$. Phenomenologically, this turns not to be the case.\\
We note that we can express ${\cal {L}}_{\mu}$ either by using the
variables $\mu_{u},\mu_{d}$ or the two combinations
$\mu_{q}=\mu_B/3=\displaystyle{{\mu_{u}+\mu_{d}\over 2}}$ and
$\mu_{I}=\displaystyle{{\mu_{u}-\mu_{d}\over 2}}$, which couple to
one third of the baryon charge density and to the third component of isospin
respectively
\begin{equation}
{\cal {L}}_{\mu}=\mu_{q}~\Psi^{\dagger}\Psi
~+~\mu_{I}~\Psi^{\dagger}\tau_{3}\Psi
\label{eq:lmuiso}
\end{equation}
To study whether a pion condensate shows up, we need to evaluate
the effective potential. This is obtained by using the standard
technique to introduce bosonic (collective) variables through the
Hubbard-Stratonovich transformation and by integrating out the
fermion fields in the generating functional. This procedure will be discussed in the next section. However, the
effective potential that we have considered is not directly
obtained from the Lagrangian in Eq. (\ref{eq:njlagr}). To mimic
asymptotic freedom we have to include ``by hand'' a form factor $F(\vec{p})$ coupled to each fermionic leg as in ref.
\cite{Alford:1997zt}, and thus we follow the same procedure as in
refs.  \cite{Berges:1998rc,Toublan:2003tt}. We could derive this momentum dependence of the fermionic fields by starting from a non-local action \cite{Berges:1998rc}.
Each condensate involving a quark and an anti-quark field couples to $F(\vec{p})^2$. 
The result is a
one-loop effective potential which generalizes that of the theory
described by the Lagrangian in Eq. (\ref{eq:njlagr}), and which
reduces to it in the limit of a constant form factor $F({\vec
p})=1$ (in that case, a cut-off scale on UV momenta has to be introduced):

\begin{equation}
\label{Poteff} V=\frac{\Lambda^2}{8G}
(\chi_u^2+\chi_d^2+2\rho^2)+V_{\mbox{log}}
\end{equation}

\begin{eqnarray} \label{Vlog}
V_{\mbox{log}}=-\mbox{Tr log} \left(
\begin{array}{cc}
h_u & -F^2(\vec{p})~\Lambda~\rho~\gamma_5\\
F^2(\vec{p})~\Lambda~\rho~\gamma_5 & h_d
\end{array}
\right)
\nonumber\\
\\
h_f=(i\omega_n+\mu_f)\gamma_0~-~\vec{p}\cdot\vec{\gamma}~-~
\left(m~+~F^2(\vec{p})~\Lambda~\chi_f\right)\nonumber
\end{eqnarray}
where $\omega_{n}$ are the Matsubara frequencies and the
dimensionless fields $\chi_{f}$ and $\rho$ are connected to the
condensates by the following relations
\begin{eqnarray} \label{eq:fields}
\chi_f &=& - ~2G~{\langle{\bar{\Psi}}_f\Psi_f\rangle\over \Lambda}\nonumber\\
\\
\rho &=& - ~G~{\langle\bar{u}\gamma_5d-\bar{d}\gamma_5u\rangle\over
\Lambda}\nonumber
\end{eqnarray}
and are variationally determined at the absolute minimum of the
effective potential. In the previous equations, $\Lambda$ is a
mass scale appearing in the form factor $F({\bf
p}^2)=\displaystyle{{\Lambda^2\over \Lambda^2+{\bf p}^2}}$
\cite{Alford:1997zt}.\\
It is worth noticing that the one-loop effective potential in Eq.
(\ref{Vlog}) has the same expression as the one derived in the previous Chapter within ladder-QCD, where the form factor, in place of $F^2(\vec{p})$ in Eq.
(\ref{Vlog}), was a function
introduced after the study of the one-loop Schwinger-Dyson equation
for the quark self-energy. The only difference is that $F^2$ depends on the
three-momentum whereas the quoted test function depends on the
four-momentum; in addition the two asymptotic behaviors are different
($\sim 1/p^2$ in the test function of ref. \cite{Barducci:2003un}
and $\sim 1/{\vec p}~^4$ in $F^2$ of Eq. (\ref{Vlog})). Otherwise
the two effective potentials would be identical. This observation
also explains the reason why we have adopted the NJL model instead
of ladder-QCD to generalize the analysis of ref.
\cite{Barducci:2003un} at high isospin chemical potentials: in the NJL model the
numerical analysis is much simpler.

To fix the free parameters of the model, which are $\Lambda$, the
average current quarks mass $m=(m_{u}+m_{d})/2$ and the coupling
$G=g/\Lambda^2$, we work at zero temperature and quark densities.
We first choose the mass scale $\Lambda$ within the range $\Lambda
\sim 500\div600~ \mbox{MeV}$. Then we determine the strength of the coupling
$g$ and the mass parameter $m$ by requiring a light quark
condensate of the order $\langle {\bar \Psi}_{f}\Psi_{f}\rangle
\simeq -(200~ \mbox{MeV})^3$ and a static pion mass $m_{\pi}\simeq 140~\mbox{MeV}$. The
latter quantity is evaluated, as in Chapter three, through the curvature of the effective potential
in the direction of the canonical pion field  \cite{Barducci:1987gn}
(having fixed $f_{\pi}$ at
its experimental value). Actually, the relation between canonical pion field and $\rho$ involves a decay constant $f_{\pi}$.

The output parameters are the following
\begin{equation}
\Lambda=580~\mbox{MeV};~~~~~~~~~~g=7;~~~~~~~~~~~m=4.5~\mbox{MeV}
 \label{eq:parameters}
\end{equation}
With these values we obtain a condensate $\langle{\bar
\Psi}_{f}\Psi_{f}\rangle= -(172~\mbox{MeV})^3$ and a constituent quark
mass $M_{f}=428~\mbox{MeV}$ (defined as in
\cite{Hatsuda:1994pi,Berges:1998rc}). The critical isospin chemical
potential at zero temperature turns out to be $\mu_I^C=89~\mbox{MeV}$.
The discrepancy of about $25\%$ (we recall that the expected value
of $\mu_I^C$ would be $m_{\pi}/2\simeq 70~\mbox{MeV}$) is entirely due to the
approximate fit procedure; the fact is that the inclusion of a non-covariant form factor $F({\vec p}^{~2})$ complicates remarkably the calculation of $f_{\pi}$.
Actually, in our previous work based on
ladder-QCD \cite{Barducci:2003un}, where $f_{\pi}$ was
consistently calculated within the model,
the result we obtained was $\mu_I^C=m_{\pi}/2$. 
However, in section \ref{Analyticalmuic} we will show that the equivalence between $m_{\pi}/2$ and $\mu_I^C$ is an analytical result in the framework of the ``standard'' NJL model, i.e. without the form factor.

\section{The procedure of Bosonization}
\label{sec:Bosonization}

In this section we review the standard procedure of bosonization, which is useful for introducing bosonic degrees of freedom in the effective potential, by integrating out the fermionic variables. It is particularly important for describing the condensation of fermionic couples, which behave as a composite bosonic field. We will refer here to the $SU(2)$ flavor case. The generalization to $SU(3)$ is straightforward, and will not be explicitly described in the next Chapter. We consider here the case of the standard NJL model, i.e. without form factor and with a cut-off $\Lambda$ on spatial momentum.

We start from the Lagrangian density eq.(\ref{eq:njlagr}), and we write the generating functional for the Green's functions in the path-integral formulation:

\begin{eqnarray}\label{gfenerfunct1}
Z[\eta,\bar{\eta}]&=&\mathcal{A}\int\mathcal{D}(\Psi,\bar{\Psi})exp\{i\int d^4 x\left[\bar{\Psi}(i\hat{\partial}-m)\Psi+\Psi^{\dagger}A\Psi+\right.\nonumber\\
&+&\left.{G\over 2}\sum_{a=0}^{3}\left[\left(
{\bar{\Psi}}\tau_{a}\Psi
\right)^{2}+\left({\bar{\Psi}}i\gamma_{5}\tau_{a}\Psi\right)^{2}
\right]+\bar{\eta}\Psi+\bar{\Psi}\eta\right]\}=\\
&=&\mathcal{A'}\int\mathcal{D}(\Psi,\bar{\Psi})\mathcal{D}(\sigma_a,\pi_a)exp\{i\int d^4 x\left[\bar{\Psi}(i\hat{\partial}-m)\Psi+\Psi^{\dagger}A\Psi+\right.\nonumber\\
&-&\left.\frac{1}{2G}\sum_a(\sigma_a^2+\pi_a^2)-\bar{\Psi}\tau_a\Psi\sigma_a-(\bar{\Psi}\tau_ai\gamma_5\Psi)\pi_a+\bar{\eta}\Psi+\bar{\Psi}\eta\right]\}\label{gfenerfunct2}
\end{eqnarray}
where $\eta,\bar{\eta}$ are fermionic sources coupled to $\bar{\Psi},\Psi$; we can pass from eq.(\ref{gfenerfunct2}) to eq.(\ref{gfenerfunct1}) by performing the gaussian functional integral over $\sigma_a,\pi_a$ fields. 

We have started from a four-fermion interaction and we obtain a Yukawa interaction with auxiliary fields $\sigma_a,~\pi_a$.
If we study the equations of motions for the auxiliary fields deriving from the action in eq.(\ref{gfenerfunct2}), we obtain the following relations:

\begin{eqnarray}
\sigma_a&=&-G\bar{\Psi}\tau_a\Psi\label{equatofmotion1}\\
\pi_a&=&-G\bar{\Psi}\tau_ai\gamma_5\label{equatofmotion2}\Psi
\end{eqnarray}
When dealing with eq.(\ref{gfenerfunct2}), it is possible to integrate over fermionic fields; by
setting to zero the fermionic sources, and by introducing bosonic sources $J_{\sigma},J_{\pi}$ coupled to auxiliary fields,
we obtain a generating functional expressed in terms of $\sigma_a,\pi_a$ fields:

\begin{eqnarray}
Z[J_{\sigma},J_{\pi}]&=&\mathcal{N}\int\mathcal{D}(\sigma_a,\pi_a)exp\{i\int d^4 x\left[-\frac{1}{2G}\sum_a(\sigma_a^2+\pi_a^2)+J_{\sigma}^a\sigma_a+J_{\pi}^a\pi_a\right]\cdot\nonumber\\
&\cdot&\mbox{Det}\left[i\hat{\partial}-m+\gamma_0 A-\tau_a\sigma_a-\tau_ai\gamma_5\pi_a\right]=\label{gfenerfunctaux1}\\
&=&\mathcal{N'}\int\mathcal{D}(\sigma_a,\pi_a)exp\{i\int d^4 x\left[-\frac{1}{2G}\sum_a(\sigma_a^2+\pi_a^2)+J_{\sigma}^a\sigma_a+J_{\pi}^a\pi_a\right]\cdot\nonumber\\
&\cdot& exp\{\int i~d^4 x\{-i~\mbox{tr~ln}\left[i\hat{\partial}-m+\gamma_0 A-\tau_a\sigma_a-\tau_ai\gamma_5\pi_a\right]\}\}\label{gfenerfunctaux2}
\end{eqnarray}
the functional determinant $\mbox{Det}$ in eq.(\ref{gfenerfunctaux1}) arises after the functional integration over fermions, and we have used the identity $\mbox{ln~Det}=\mbox{Tr~ln}$ to pass from eq.(\ref{gfenerfunctaux1}) to eq.(\ref{gfenerfunctaux2}). The operators $\mbox{Det}$ and $\mbox{Tr}$ act on Dirac, flavor, colour and space-time spaces; $\mbox{tr}$ acts on Dirac, flavor and colour spaces only.\\
For vanishing sources we obtain the effective action for auxiliary fields in the approximation of one fermionic loop:

\begin{equation}\label{effactaux}
\mbox{S}_{eff}=\int d^4 x \{ -\frac{1}{2G}\sum_a(\sigma_a^2+\pi_a^2)-i~\mbox{tr~ln}\left[i\hat{\partial}-m+\gamma_0 A-(\sigma_a+i\gamma_5\pi_a)\tau_a\right]\}
\end{equation}

When we are interested in the properties of the ground state only, in particular in the dynamical breaking of the chiral symmetry, we can consider
constant fields $\sigma_a(x)=\sigma_a$,~$\pi_a(x)=\pi_a$. Actually the minima of the effective action correspond to constant fields, in order to ensure the invariance of the vacuum under rotations and translations.
In this case we get the one loop effective potential:

\begin{equation}\label{effpotaux}
\mbox{V}=\frac{1}{2G}\sum_a(\sigma_a^2+\pi_a^2)+i~\mbox{tr~ln}\left[i\hat{\partial}-m+\gamma_0 A-(\sigma_a+i\gamma_5\pi_a)\tau_a\right]
\end{equation}
In order to obtain the final expression (\ref{Poteff}) of the effective potential, it is necessary to replace $\sigma_a,\pi_a$ fields with $\bar{u}u,~\bar{d}d,~\bar{u}\gamma_5d$. This can be done by expanding the relations (\ref{equatofmotion1}),(\ref{equatofmotion2}) in flavor space.
Charged scalar ($\sim \bar{\Psi}\tau_1\Psi$) and neutral pseudoscalar ($\sim \bar{\Psi}i\gamma_5\Psi$) combinations can be set to zero from the beginning, whenever they are not expected to get a non-vanishing expectation value.
 The  expectation values $\langle\bar{u}u\rangle,\langle\bar{d}d\rangle,\langle\bar{u}\gamma_5d\rangle$ are given by the location of minima of the effective potential in the space of corresponding fields.

The approximation of taking constant fields is valid at the mean-field level; we consider the propagation of constituent quarks as quasi-particles with an effective mass
\begin{equation}\label{massquarknjl}
M_f=m-2G\langle\bar{\Psi}_f\Psi_f\rangle
\end{equation}
where $f$ is a flavor index. In this model without flavor mixing, for each flavor the contribution to the constituent mass arising from interactions is proportional to the corresponding scalar condensate. In the chiral case (or for current masses $\sim10~\mbox{MeV}$), a constituent mass $\sim350~\mbox{MeV}$ is provided by means of interactions only. This is the phenomenon of mass generation.\\
In the following of this work we will perform calculations at the mean-field level, and therefore we study the effective potential as the fundamental tool for defining the vacuum properties.
However, we could go beyond mean-field and consider space-time dependent fields

\begin{eqnarray}
\sigma_a(x) &=& \langle\sigma_a\rangle+s_a(x)\label{fieldsxdep1}\\
\pi_a(x) &=& \langle\pi_a\rangle+p_a(x)\label{fieldsxdep2}
\end{eqnarray}

The inclusion of fluctuations $s,p$ is important for defining the properties of bound states, both in the vacuum and at finite temperature and densities. For instance, if we want to compute the dynamical mass of pions
we have just to expand the effective action up to the second order for the fields $\pi_a(x)$, then operate the functional derivative

\begin{equation}
\frac{\delta^2S_{eff}^{(2)}}{\delta p_a(x)\delta p_b(y)}=G_{ab}^{-1}(x-y)
\end{equation}
where $G_{ab}(x)$ is the boson propagator; obviously $G_{\pi}(p)^{-1}=0$ defines the on-shell condition. By doing this, we can prove that an isospin chemical potential enters in the effective pion mass just like as the half of the charged pion chemical potential, see e.g. \cite{He:2005nk}; 
\begin{equation}
m_{\pi^{+}}=m_{\pi}-2\mu_I~~;~~m_{\pi^{-}}=m_{\pi}+2\mu_I
\end{equation}
The latter expressions hold when we are in the phase with zero pion condensates, namely for $|\mu_I|<m_{\pi}/2$.
Actually, if we go in momentum space and consider momentum-dependent fields, we are 
taking into account their kinetic terms. Therefore, we are considering the fields $\sigma_a,\pi_a$ as dynamical, propagating degrees of freedom, which can give a sensible contribution to the thermodynamic functions. However, as for $f_{\pi}$, the calculation of the effective action as a function of the fluctuations can be done by considering the NJL model without the form factor.

\section{Phase diagram for chiral symmetry breaking and pion condensation}
\label{sec:physics2}

In order to discuss the structure of the phase diagram, it is
worth summarizing the symmetries of the Lagrangian density in Eq.
(\ref{eq:njlagr}). Both ${\cal {L}}_{0}$ and ${\cal {L}}_{int}$
are $SU_{L}(2)\otimes SU_{R}(2)\otimes U_{V}^{B}(1)\otimes
U^{B}_{A}(1)$ invariant. The symmetry is reduced by the mass term
${\cal {L}}_{m}$ to $SU_{V}(2)\otimes U_{V}^{B}(1)$ and further
reduced from the term ${\cal {L}}_{\mu}$ which selects a direction
in the isospin space, as is evident from Eq. (\ref{eq:lmuiso}),
unless $\mu_{u}=\mu_{d}$ and thus $\mu_{I}=0$. The remaining
symmetry can be expressed either as $U_{V}^{u}(1)\otimes
U_{V}^{d}(1)$ or $U_{V}^{B}(1)\otimes U_{V}^{I}(1)$, depending on
the basis of the fields that we are choosing.\\
The baryon number symmetry $U_{V}^{B}(1)$ is dynamically
respected, whereas a non vanishing v.e.v. of the $\rho$ field
defined in Eq. (\ref{eq:fields}) may appear, which dynamically
breaks $U_{V}^{I}(1)$. This implies the appearance of a Goldstone
mode, which is either the charged $\pi^{+}$ or $\pi^{-}$ at the
threshold, depending on the sign of $\mu_{I}$,
whereas the other two pions are massive modes. The nice feature is that 
this is a {\it genuine} Goldstone boson, rigorously massless even in presence of a finite current mass.
\\
As far as the scalar condensates $\chi_{u},\chi_{d}$ are concerned (see Eq.(\ref{eq:fields})), they do not break any symmetry when a non zero current mass is considered.
However, since the mass term is small, their value is almost
entirely due to the approximate spontaneous breaking of chiral
symmetry. Consequently we distinguish regions where the dynamical
effect is relevant, from regions where the scalar condensates are
of order $\sim m/\Lambda$, namely where only the effect of the
explicit breaking of chiral
symmetry survives.\\
The study of the various phases has been performed
numerically by minimizing the one-loop effective potential. We
start by showing the results in the $(\mu_{u},\mu_{d})$ plane, for
fixed values of the temperature. Different regions are labelled,
as in ref. \cite{Klein:2003fy}, by the symbol of the field which
acquires a non vanishing v.e.v. due to dynamical effects, whereas
the other fields are vanishing ($\rho$), or of the order $\sim
m/\Lambda$
($\chi_{u}$ and/or $\chi_{d}$).\\
Solid lines refer to discontinuous transitions and dashed lines to
continuous ones. However, we recall that strictly speaking only
the lines surrounding regions with a non vanishing field $\rho$
refer to genuine phase transitions, associated with the breaking
and restoration of the $U_{V}^{I}(1)$ symmetry.

\subsection{Zero temperature}

\begin{figure}[htbp]
\begin{center}
\includegraphics[width=12cm]{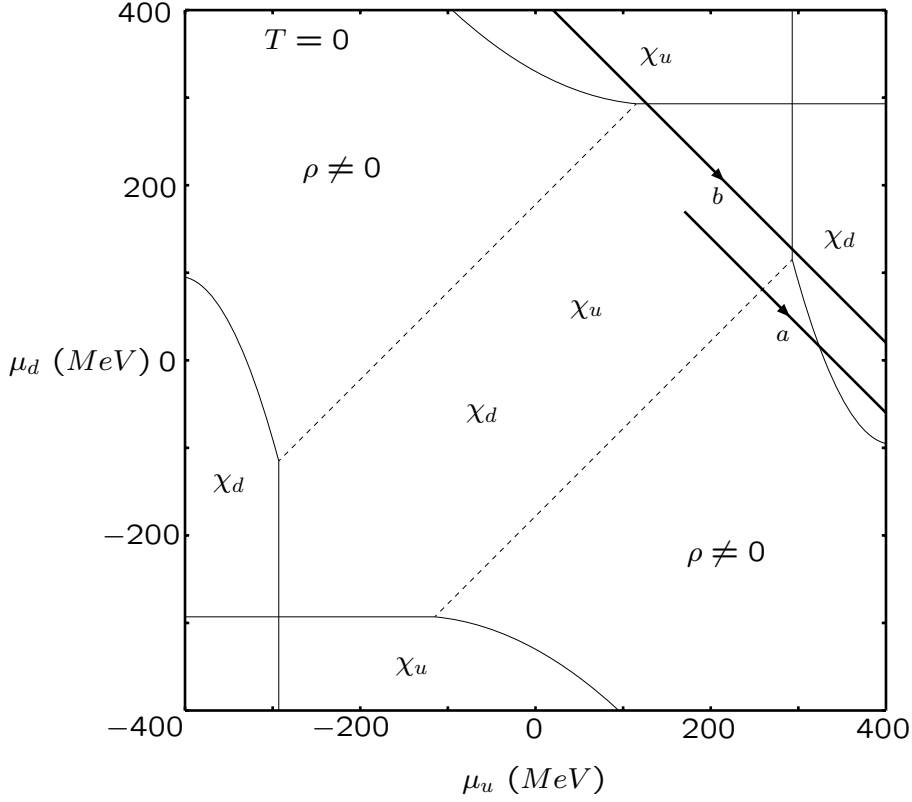}
\end{center}
\caption{\it Phase diagram for chiral symmetry restoration in the
plane $(\mu_{u},\mu_{d})$ of chemical potentials of quarks, at $T=0$.
Different regions are specified by the non vanishing of a given
condensate, whereas the others are vanishing ($\rho$) or order
$\sim m/\Lambda$ ($\chi_{u}$ and $\chi_{d}$). Dashed lines are for
the continuous vanishing of $\rho$ or for cross-over phase
transitions for $\chi_{u}$ or $\chi_{d}$, whereas solid lines are
for discontinuous behaviors. The solid lines $a$ and $b$ refer to
specific paths at fixed values of $\mu_{q}$, with
$\mu_{q}=170~MeV$ (line $a$) relative to Fig. \ref{fig:mufetta100}
and $\mu_{q}=210~MeV$ (line $b$) relative to Fig.
\ref{fig:mufetta250}.} \label{fig:totdiamuftz1}
\end{figure}

\begin{figure}[htbp]
\begin{center}
\includegraphics[width=11cm]{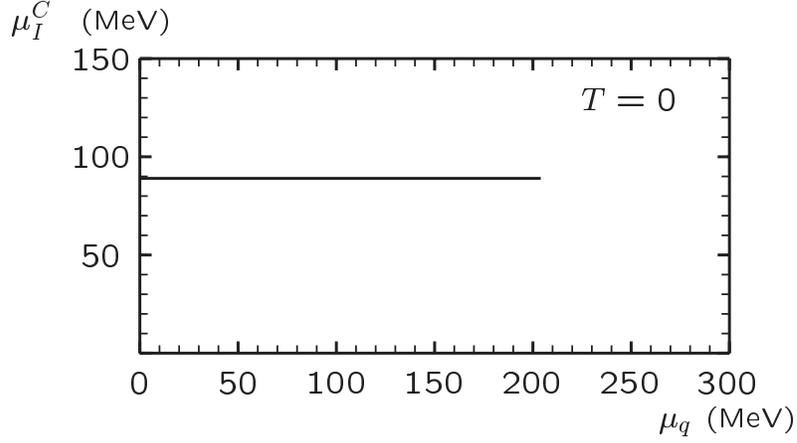}
\end{center}
\caption{\it Critical value of the isospin chemical potential,
beyond which a pseudoscalar condensate forms, vs. quark chemical
potential, at zero temperature. $\mu_{q}=204~MeV$ is the highest
allowed value for pion condensation to occur. The path followed in
the phase diagram of Fig. \ref{fig:totdiamuftz1} is along the
upper-half of the dashed line in the lower half-plane.}
\label{fig:critisovsmub}
\end{figure}

\begin{figure}[htbp]
\begin{center}
\includegraphics[width=11cm]{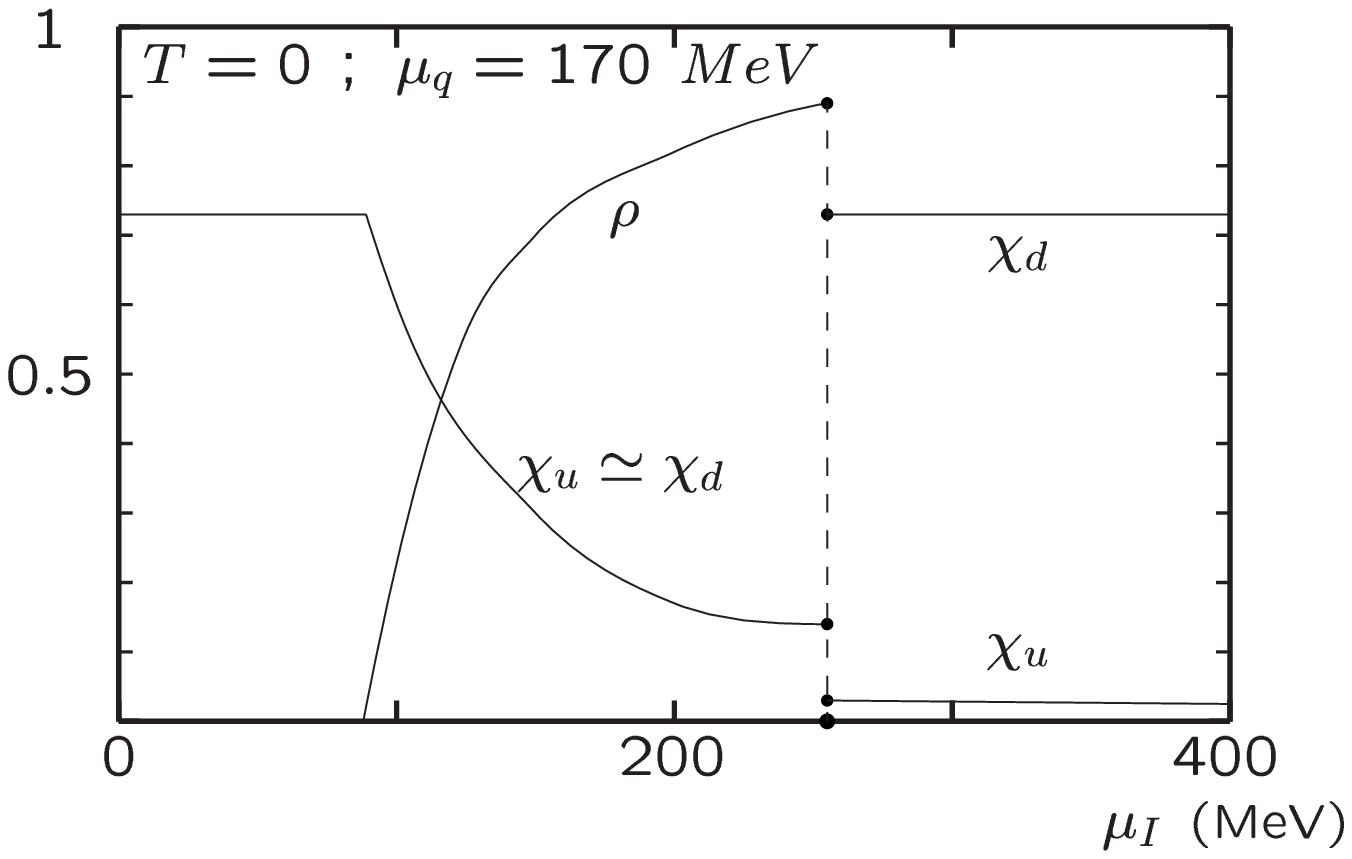}
\end{center}\caption{\it Scalar and pseudoscalar
condensates vs. $\mu_I$, for $\mu_q=170~MeV$ and $T=0$. The path
followed in the phase diagram of Fig. \ref{fig:totdiamuftz1} is
that of the solid line $a$.} \label{fig:mufetta100}
\end{figure}

\begin{figure}[htbp]
\begin{center}
\includegraphics[width=11cm]{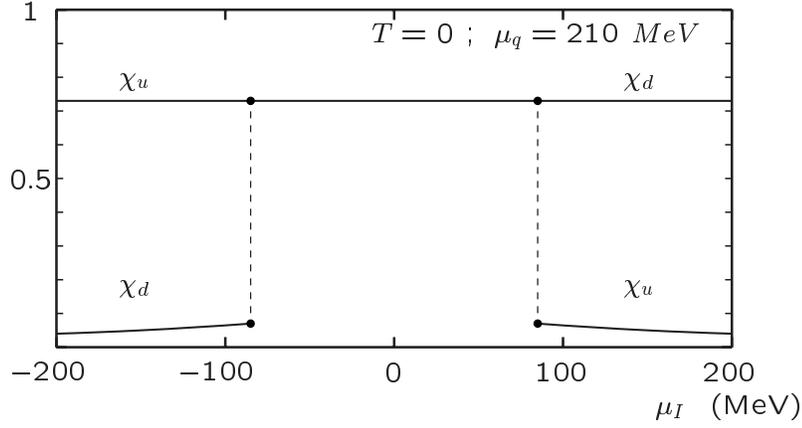}
\end{center}
\caption{\it
Scalar condensates vs. $\mu_{I}$ for $\mu_{q}=210~MeV$ and $T=0$.
The pseudoscalar condensate is zero. The path followed in the
phase diagram of Fig. \ref{fig:totdiamuftz1} is that of the solid
line $b$.} \label{fig:mufetta250}
\end{figure}

In Fig. \ref{fig:totdiamuftz1} we show the phase diagram in the
$(\mu_{u},\mu_{d})$ plane at zero temperature. Let us start from
the vacuum at $T=\mu_{u}=\mu_{d}=0$, at the center of the picture.
Here, in the chiral limit, the pions are the Goldstone bosons
associated with the spontaneous breaking of $SU(2)_L\otimes
SU(2)_R$. We have chosen these variables in order to compare the
structure of the phase diagram with that obtained in ref.
\cite{Klein:2003fy}. However, if we want to recover known results
given in terms of the quark chemical potential, we have to move
from the center along the diagonal at $\mu_u=\mu_d$ and thus at
$\mu_{I}=0$, and increase the absolute value of $\mu_{q}$. At
$\displaystyle{{|\mu_{u}+\mu_{d}|\over 2}=|\mu_{q}|}=293~\mbox{MeV}$ we
meet the approximate restoration of chiral symmetry due to the
sudden jump of the condensates of the two (degenerate) quarks to
values of order $\sim m/\Lambda$, which is a discontinuous
transition. The same thing happens by moving along lines parallel
to the main diagonal in the region labelled by
$\chi_{u},~\chi_{d}$, enclosed between the two dashed lines at
$\displaystyle{{|\mu_{u}-\mu_{d}|\over 2}}=|\mu_I|= 89~\mbox{MeV}$ (see
also Fig. \ref{fig:critisovsmub} where it is shown that the
critical value of $\mu_{I}$ at $T=0$ is independent on $\mu_{q}$)
and by varying $|\mu_{q}|$. The regions in the top-right and
bottom-left corners of Fig. \ref{fig:totdiamuftz1} thus have the
$U_{V}^{u}(1)\otimes U_{V}^{d}(1)$ symmetry of ${\cal {L}}$ with
$\rho=0$ and $\chi_{u}$,$\chi_{d}$ of order $\sim m/\Lambda$.\\
By moving from the center along the diagonal at $\mu_{u}=-\mu_{d}$
(and thus $\mu_{q}=0$) or parallel to it, when crossing one of the
two dashed lines at $|\mu_{I}|=89~\mbox{MeV}$ (we have already discussed
the origin of the discrepancy between this value and half of the
pion mass in the model), the absolute minimum of the effective
potential starts to rotate along the $\rho$ direction. We thus
have a continuous breaking of $U_{V}^{I}(1)$ and a second order
phase transition with one Goldstone mode which is at the threshold, right along the
dashed line, either the $\pi^{+}$ (in the upper part of the
diagram) or the $\pi^{-}$ (in the lower part). In the chiral limit
these two dashed lines merge together in coincidence with the
diagonal at $\mu_{I}=0$ as the pion becomes massless in this limit
and the rotation is sudden, giving first order phase transitions
for pion condensation. In this case there are two Goldstone bosons
associated with the spontaneous breaking of two $U(1)$ symmetry
groups
($U_{A}^{B}(1)\otimes U_{V}^{I}(1)$) \cite{Kogut:2002zg}.
We recall here that in the present application we are not considering the explicit $U(1)_A$ violation; therefore, in a more realistic NJL model, with 't Hooft determinant and $U(1)_A$ breaking, only one Goldstone boson, corresponding to the spontaneous $U_{V}^{I}(1)$ breaking, would exist.
\\
Coming back to the massive case, and still with reference to Fig.
\ref{fig:totdiamuftz1}, we conclude that by considering
$|\mu_{q}|$ not too large and by growing $|\mu_{I}|$, we find a
second order phase transition with the rotation of the scalar
condensates into the pseudoscalar, namely we are faced with pion
condensation in a relatively simple picture. A difference with
ref. \cite{Klein:2003fy} is that we do not find the vanishing of
$\rho$ for values of $|\mu_{I}|$ high with respect to the pion
mass, but still sufficiently low to avoid considering
superconductive phases (actually, for very low $|\mu_{q}|$ this
transition would occur in the present model for $|\mu_{I}|\sim 1~\mbox{GeV}$).\\
To explore the possibility of multiple phase transitions and thus
of a richer phenomenology, we need to grow $|\mu_{q}|$ as for
instance we do in the case described in Fig. \ref{fig:mufetta100}
where we follow the path of the solid line $a$ in Fig.
\ref{fig:totdiamuftz1} at $\mu_{q}=170~\mbox{MeV}$ for growing
$\mu_{I}\geq 0$. The fields $\chi_{u}$ and $\chi_{d}$ are almost
degenerate, both in the region of the approximate dynamical
breaking of chiral symmetry (below $\mu_{I}=89~\mbox{MeV}$) and in the
region of spontaneous breaking of $U_{V}^{I}(1)$, where they
rotate into the $\rho$ field. Then, when the line $a$ in Fig.
\ref{fig:totdiamuftz1} crosses the solid line surrounding the
region labelled by $\chi_{d}$, we see that $\rho$ suddenly jumps
to zero with the restoration of the $U_{V}^{I}(1)$ group and that
$\chi_{u}$ and $\chi_{d}$ split. Actually the latter suddenly
acquires a value due to the dynamical breaking of chiral symmetry
whereas $\chi_{u}$ undergoes a further decrease and remains of
order $\sim m/\Lambda$.\\
In Fig. \ref{fig:mufetta250} we plot the behavior of the scalar
condensates $\chi_{u}$ and $\chi_{d}$ vs. $\mu_{I}$ at
$\mu_{q}=210~\mbox{MeV}$, namely by following the path described by the
solid line $b$ in Fig. \ref{fig:totdiamuftz1}. We see that we
never cross the region with $\rho\neq 0$ and that we simply pass
from a region where the dynamical effect of the breaking of chiral
symmetry is entirely due to a large value of $\chi_{u}$ and
$\chi_{d}$ of order $\sim m/\Lambda$ (at large negative $\mu_{I}$
and small $\mu_{u}$), to a region where this effect manifests
itself with a large value of $\chi_{d}$ and $\chi_{u}$ of order
$\sim m/\Lambda$ (at large positive $\mu_{I}$ and small
$\mu_{d}$). The region in between has almost degenerate and both
large $\chi_{u}$ and $\chi_{d}$. Pion condensation does not occur
for this value of $\mu_q$
(see also Fig. \ref{fig:critisovsmub}).\\
Finally, in Fig. \ref{fig:tfetta2}, we plot the behavior of the
condensates at fixed $\mu_u=200~\mbox{MeV}$ vs. $\mu_d$ (see again Fig.
\ref{fig:totdiamuftz1}). The rotation of the pion condensate into
the scalar ones occurs when the vertical line at $\mu_u=200~\mbox{MeV}$
meets the dashed line at $\mu_I=89~\mbox{MeV}$, which happens for $\mu_d$
of few $\mbox{MeV}$. Then, when $\mu_d$ has sufficiently increased,
$\chi_d$ falls to a small value of order $\sim m/\Lambda$ with a
discontinuous transition, whereas $\chi_u$ remains constant at its
large value.

\subsection{Finite temperature}
\label{Fintempcap2}

\begin{figure}[htbp]
\begin{center}
\includegraphics[width=11cm]{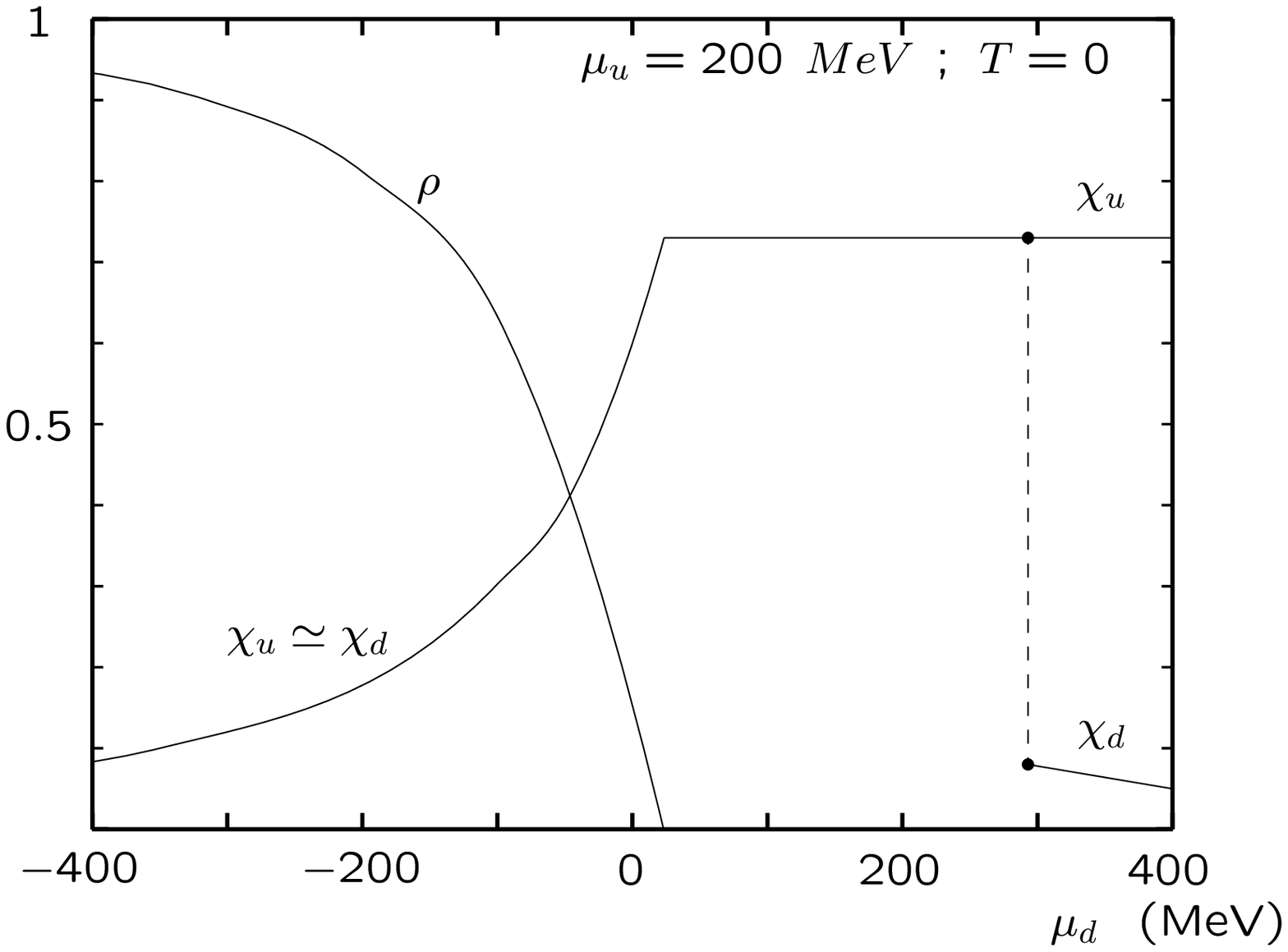}
\end{center}
\caption{\it Scalar and pseudoscalar condensates vs. $\mu_d$, for
$\mu_u=200~MeV$, $T=0$ (see Fig. \ref{fig:totdiamuftz1}).}
\label{fig:tfetta2}
\end{figure}

\begin{figure}[htbp]
\begin{center}
\includegraphics[width=15cm]{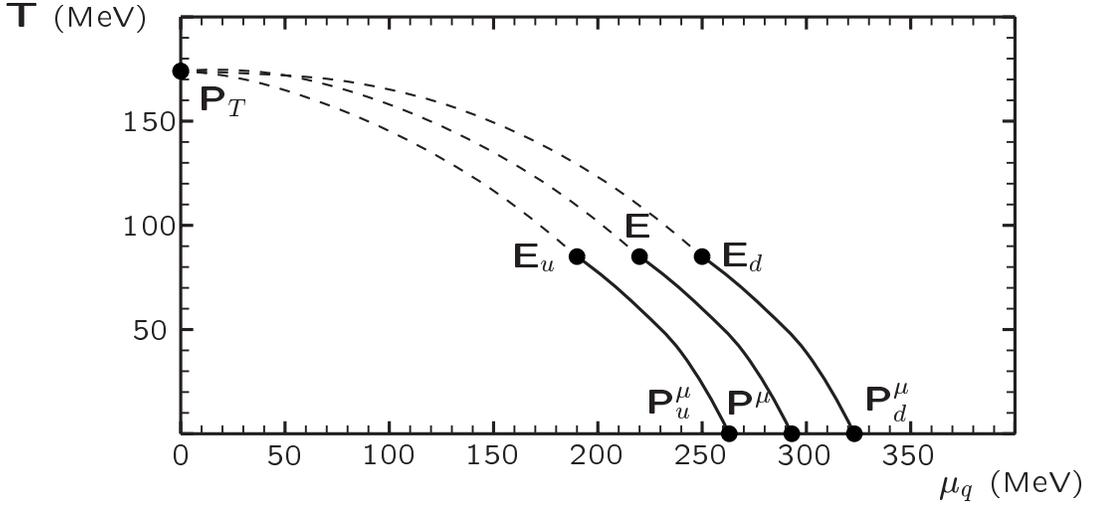}
\end{center}
\caption{\it{Phase diagram for chiral symmetry in the
$(\mu_{q},T)$ plane for zero or small isospin chemical potential
$\mu_{I}$. For $\mu_{I}=0$ (central line), the cross-over
transition line starts from the point $P_{T}=(0,174)$ and ends at
the point $E=(220,85)$. The line between $E$ and the point
$P^{\mu}=(293,0)$ is the line for the first order transition with
discontinuities in the $\langle{\bar u}u\rangle$ and $\langle{\bar
d}d\rangle$ condensates. For $\mu_{I}=30~{\rm MeV}$ (side lines),
the two cross-over transition lines start from the point
$P_{T}=(0,174)$ and end at the points $E_{u}=(190,85)$ and
$E_{d}=(250,85)$. The lines between $E_{u}$ and the point
$P^{\mu}_{u}=(263,0)$ and between $E_{d}$ and the point
$P^{\mu}_{d}=(323,0)$ are the lines for the first order
transitions with discontinuities in the $\langle{\bar u}u\rangle$
and $\langle{\bar d}d\rangle$ condensates respectively.}}
\label{fig:diafaisopic}
\end{figure}

\begin{figure}[htbp]
\begin{center}
\includegraphics[width=12cm]{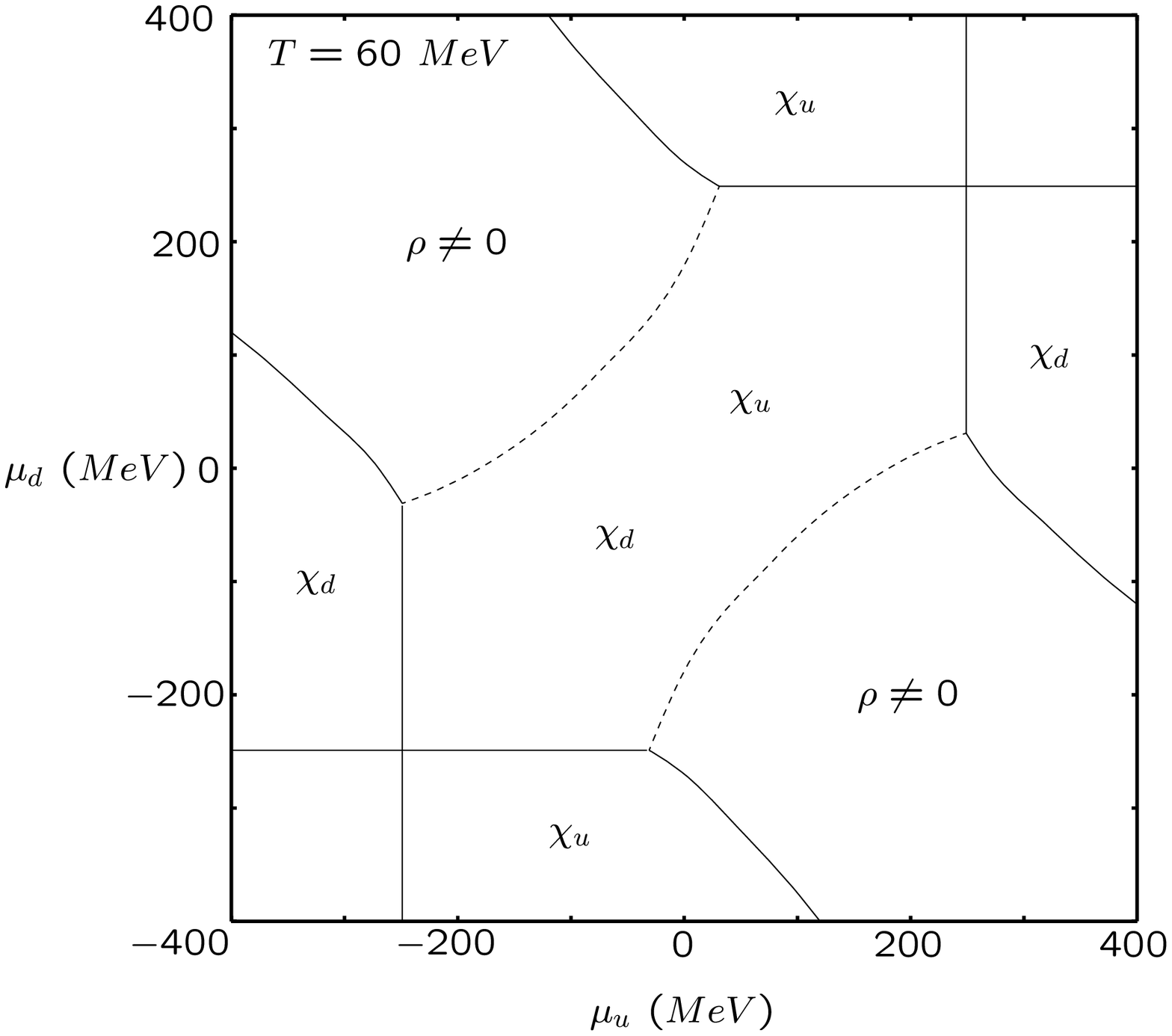}
\end{center}
\caption{\it Phase diagram for chiral symmetry restoration in the
plane $(\mu_{u},\mu_{d})$ of quark chemical potentials, at
$T=60~MeV$, which is below the temperature of the critical ending
point (see Fig. \ref{fig:diafaisopic}). Different regions are
specified by the non vanishing of a given condensate, whereas the
others are vanishing ($\rho$) or $\sim m/\Lambda$ ($\chi_{u}$ and
$\chi_{d}$). Dashed lines are lines for the continuous vanishing
of $\rho$ or for cross-over phase transitions for $\chi_{u}$ or
$\chi_{d}$, whereas solid lines are for discontinuous behaviors.}
\label{fig:totdiamuftz2}
\end{figure}

\begin{figure}[htbp]
\begin{center}
\includegraphics[width=12cm]{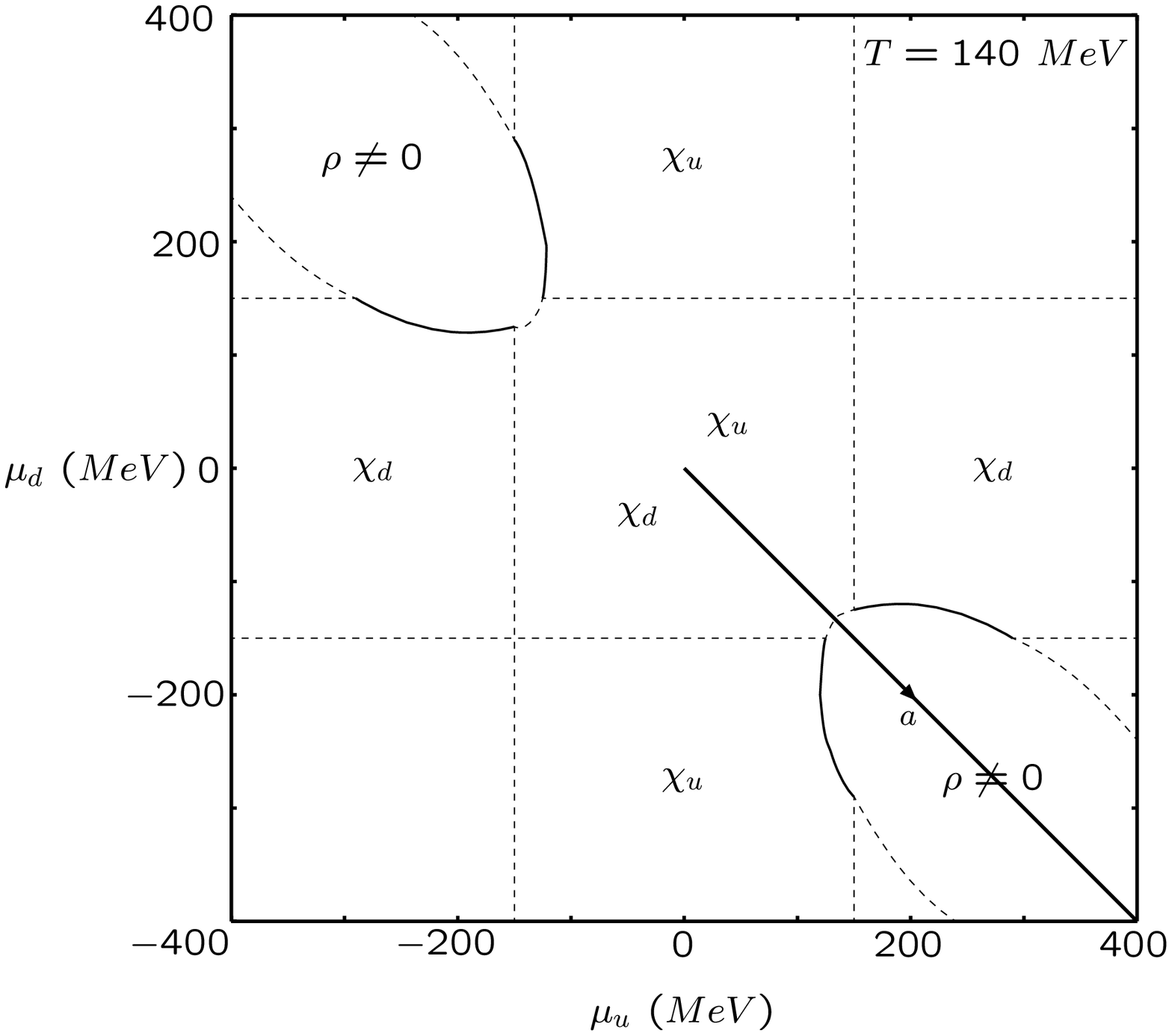}
\end{center}
\caption{\it  Phase diagram for chiral symmetry restoration in the
plane $(\mu_{u},\mu_{d})$ of chemical potentials of quarks, at
$T=140~MeV$, which is above the temperature of the critical ending
point (see Fig. \ref{fig:diafaisopic}). Different regions are
specified by the non vanishing of a given condensate, whereas the
others are vanishing ($\rho$) or $\sim m/\Lambda$ ($\chi_{u}$ and
$\chi_{d}$). Dashed lines are lines for the continuous vanishing
of $\rho$ or for cross-over phase transitions for $\chi_{u}$ or
$\chi_{d}$, whereas solid lines are for discontinuous behaviors.
The solid line $a$ refer to the path at $\mu_{q}=0$ vs.
$\mu_{I}\geq 0$ followed in Fig. \ref{fig:tfetta150}.}
\label{fig:totdiamuftz3}
\end{figure}

\begin{figure}[htbp]
\begin{center}
\includegraphics[width=11cm]{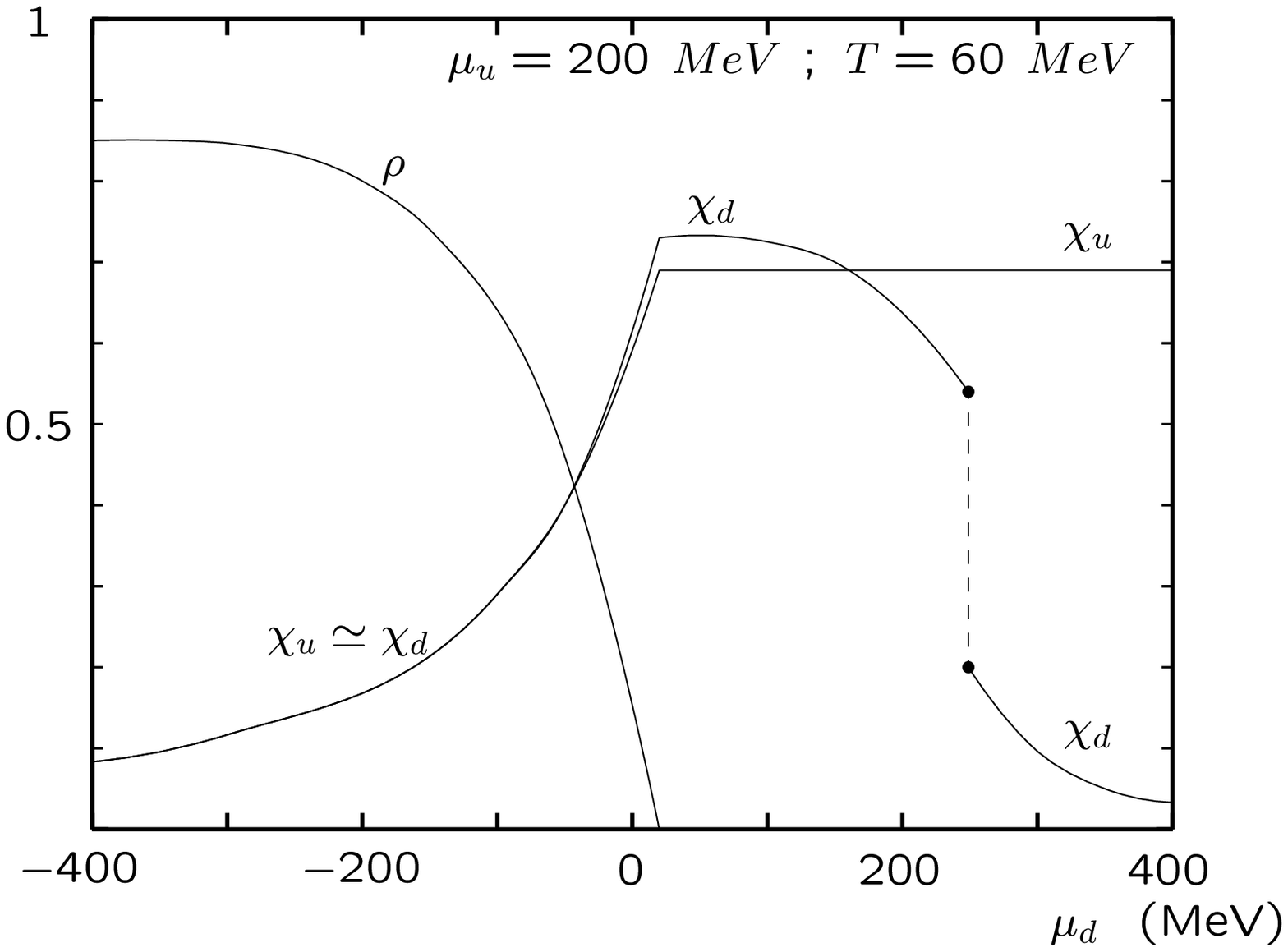}
\end{center}
\caption{\it Scalar and pseudoscalar condensates vs. $\mu_d$, for
$\mu_u=200~MeV,~T=60~MeV$ (see Fig. \ref{fig:totdiamuftz2}).}
\label{fig:tfetta5}
\end{figure}

\begin{figure}[htbp]
\begin{center}
\includegraphics[width=11cm]{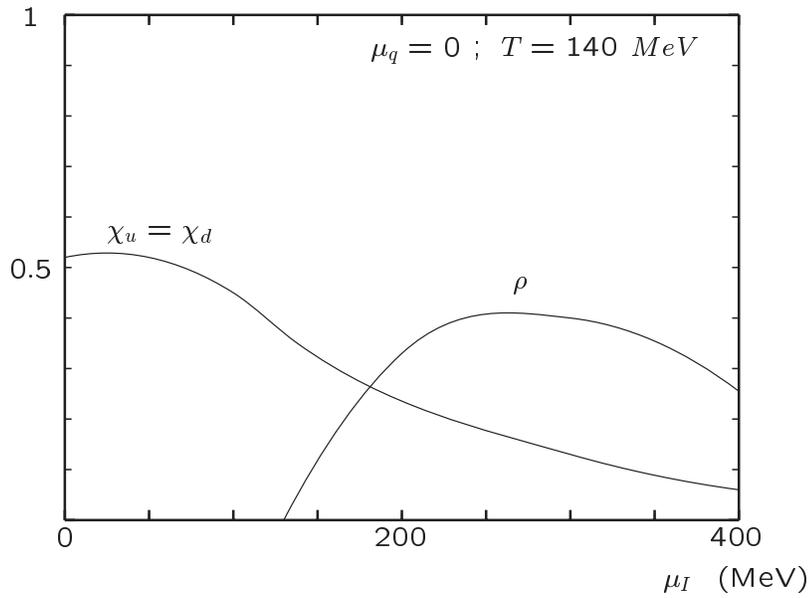}
\caption{\it Scalar and pseudoscalar condensates vs. $\mu_I$, for
$\mu_q=0$ and $T=140~MeV$. The figure is obtained following the
path $a$ in the phase diagram of Fig. \ref{fig:totdiamuftz3}.}
\label{fig:tfetta150}
\end{center}
\end{figure}

\begin{figure}[htbp]
\begin{center}
\includegraphics[width=11cm]{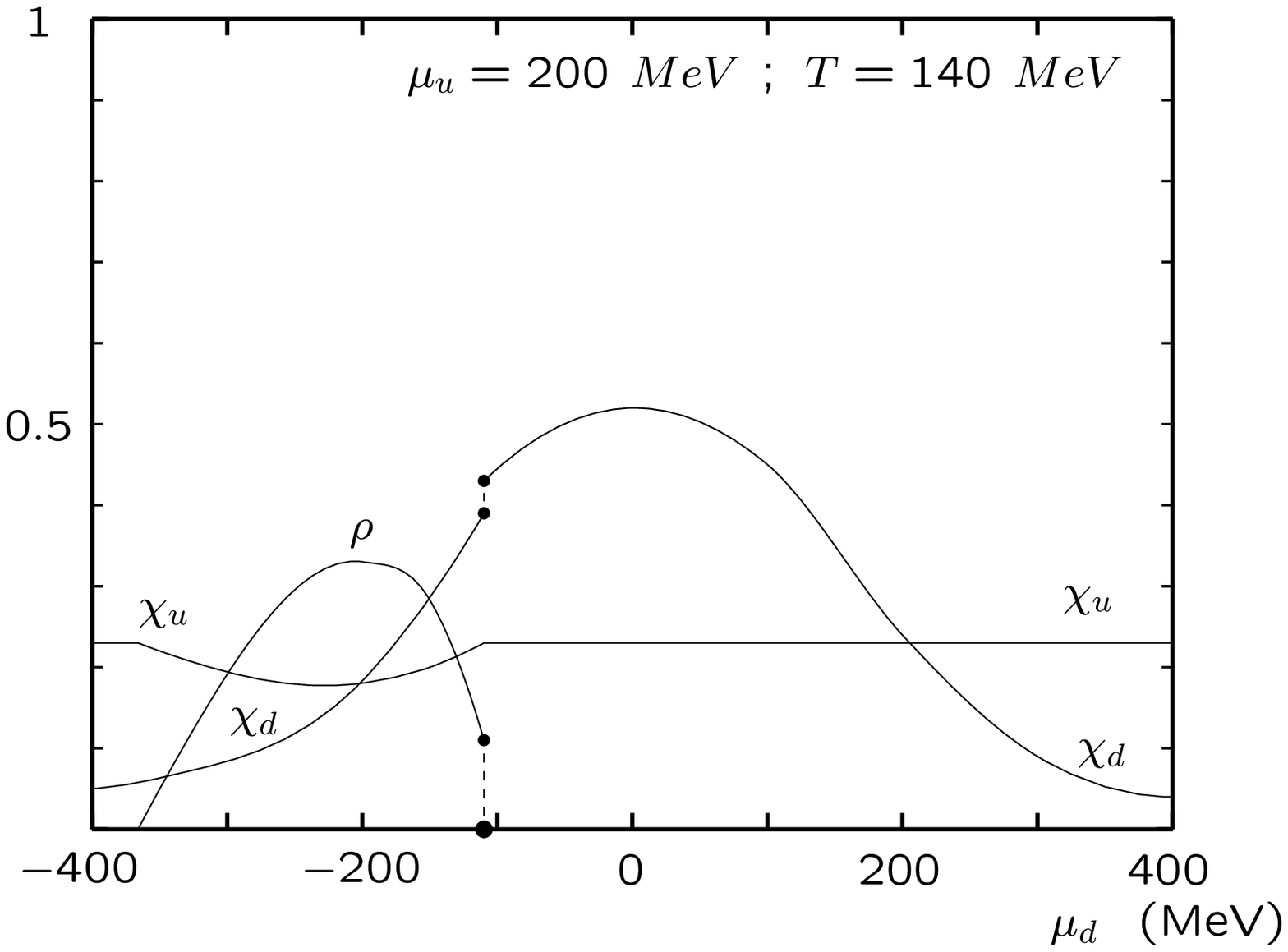}
\end{center}
\caption{\it Scalar and pseudoscalar condensates vs. $\mu_d$, for
$\mu_u=200~MeV,~T=140~MeV$ (see Fig. \ref{fig:totdiamuftz3}). }
\label{fig:tfetta33}
\end{figure}

\begin{figure}[htbp]
\begin{center}
\includegraphics[width=11cm]{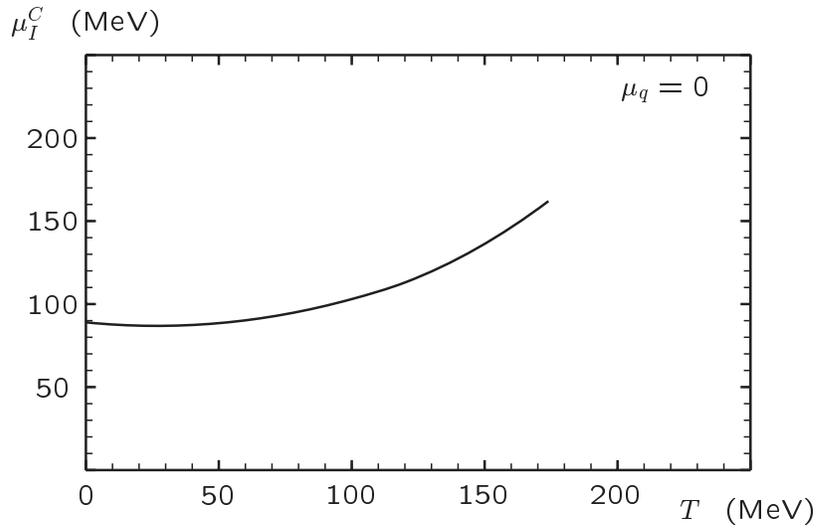}
\end{center}
\caption{\it Critical value of the isospin chemical potential,
beyond which a pseudoscalar condensate forms, vs. temperature, at
$\mu_q=0$.} \label{fig:critisovst}
\end{figure}

\begin{figure}[htbp]
\begin{center}
\includegraphics[width=12cm]{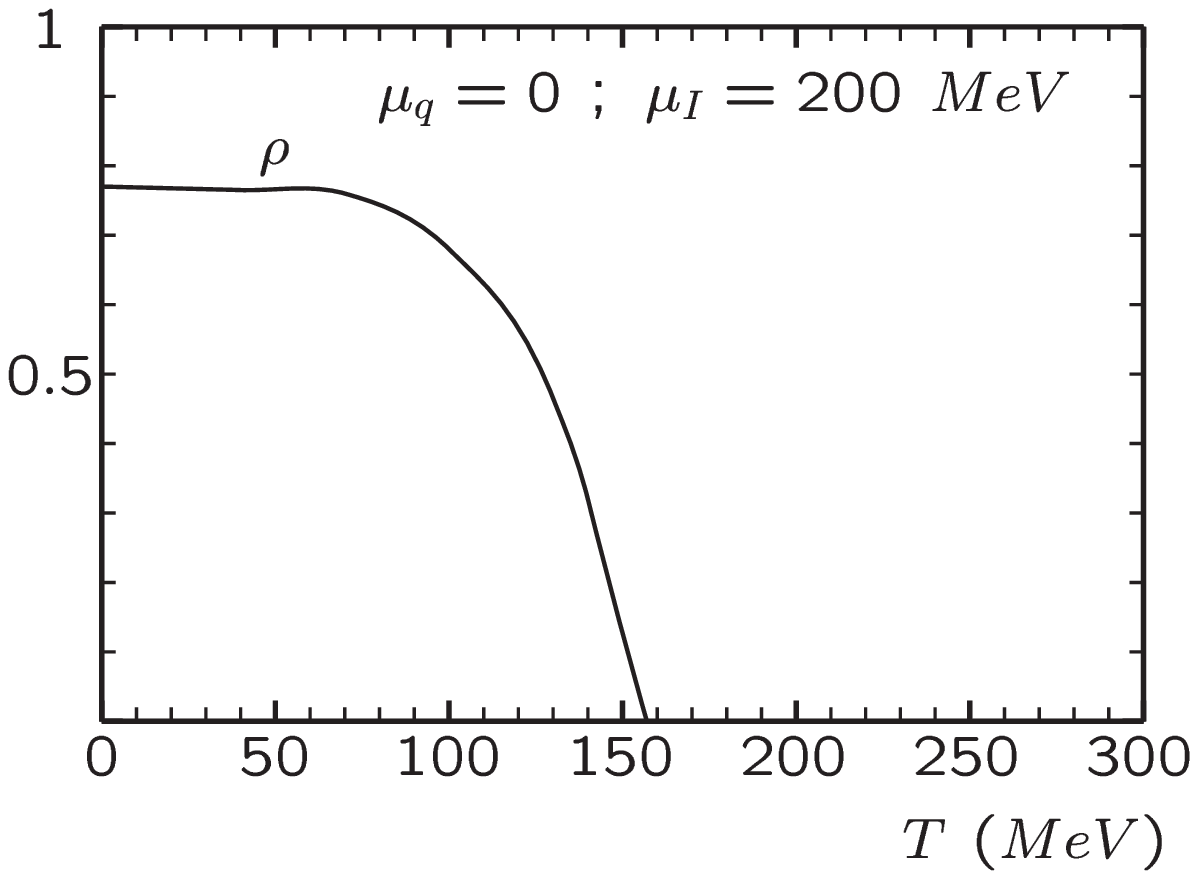}
\end{center}
\caption{\it Pion condensates vs. $T$, for $\mu_q=0$ and
$\mu_{I}=200~MeV$. The scalar condensates $\chi_{u}$ and
$\chi_{d}$ are of order $\sim m/\Lambda$.}
\label{fig:ifetta200vst}
\end{figure}

The evolution of the phase diagram for growing temperatures is
easily understood as far as the regions with $\rho=0$ are
concerned. Actually, in this case the effective potential at the
minimum is the sum of two independent terms, one for each flavor,
and the results are straightforwardly given through the analysis
of chiral symmetry breaking and restoration for a single flavor at
finite temperature and chemical potential (see for instance refs.
\cite{Barducci:1989eu,Barducci:1993bh}). In Fig.
\ref{fig:diafaisopic} we show the phase diagram at zero, or small
isospin chemical potential (see also refs.
\cite{Barducci:2003un,Toublan:2003tt}). From this picture we see
that moving along any of the critical lines of chiral symmetry
restoration at fixed $\mu_{I}$, the critical value of the quark
chemical potential $\mu_{q}$ decreases for growing temperatures.
Furthermore, for temperatures below that of the critical ending
point $E$,  $T<T(E)=85~\mbox{MeV}$, the transitions are always
discontinuous whereas they become cross-over transitions for
$T>T(E)$. Consequently the regions labelled by $\chi_{u}$ and/or
$\chi_{d}$ in Fig. \ref{fig:totdiamuftz1} shrink when growing $T$
and their rectilinear sides become lines of cross-over transitions
for $T> T(E)$ (see Fig. \ref{fig:totdiamuftz1} and Figs.
\ref{fig:totdiamuftz2}, \ref{fig:totdiamuftz3} where we plot the
phase diagrams in the $(\mu_{u},\mu_{d})$ plane at $T=60~\mbox{MeV}$ and
$T=140~\mbox{MeV}$, which is respectively below and above $T(E)$). The
new feature concerns the regions with $\rho\neq 0$, which also
reduce their size for growing $T$, whereas the order of the
transitions starts to change from first to second, beginning from
the critical points at highest $|\mu_{I}|$, until they reach the
points of the boundaries which coincide with those of the regions
labelled by $\chi_{u}$ or $\chi_{d}$ (see Fig.
\ref{fig:totdiamuftz3}). Also the length of the curves of second
order phase transitions to pion condensation at fixed values of
$|\mu_{q}|$ sensibly reduces from low temperatures to high
temperatures (see again Fig. \ref{fig:totdiamuftz1}, and Figs.
\ref{fig:totdiamuftz2}, \ref{fig:totdiamuftz3}).  A similar
behavior, for high $T$, is found in ref. \cite{Klein:2003fy}. For
$T> T(P_T)=174~\mbox{MeV}$, which is the cross-over temperature at zero
chemical potential (see Fig. \ref{fig:diafaisopic}), all these
regions disappear from the phase diagram, which is thus
characterized by $\rho=0$ and
$\chi_{u},\chi_{d}~\sim ~ m/\Lambda$.\\
The behavior of the scalar and pseudoscalar condensates at
$T=60~\mbox{MeV}$ are much similar to those at $T=0$. As an example we
plot, in Fig. \ref{fig:tfetta5}, the condensates at
$\mu_u=200~\mbox{MeV}$ and $T=60~\mbox{MeV}$ vs. $\mu_{d}$ (compare with Fig.
\ref{fig:tfetta2}). The situation is different if we consider
temperatures above $T(E)=85~\mbox{MeV}$. For instance, at $T=140~\mbox{MeV}$, we
see from Fig. \ref{fig:totdiamuftz3}, that the structure of the
phase diagram is only slightly modified with respect to the case
of two independent flavors which undergo cross-over phase
transitions at sufficiently high values of their own chemical
potentials (actually the region of pion condensation has sensibly
reduced with respect to Figs. \ref{fig:totdiamuftz1},
\ref{fig:totdiamuftz2}). The phase transition associated with the
spontaneous breaking of $U_{V}^{I}(1)$ can be both second or first
order, depending on the path followed. In Fig.
\ref{fig:totdiamuftz3}, the solid line $a$ refers to a path at
$\mu_{q}=0$ vs. $\mu_{I}\geq 0$ where the transition to pion
condensation is continuous. The behavior of the condensates
relative to this path is plotted in Fig. \ref{fig:tfetta150}. In
Fig. \ref{fig:tfetta33} we plot the scalar and pseudoscalar
condensates
at $\mu_{u}=200~\mbox{MeV}$ and $T=140~\mbox{MeV}$ vs. $\mu_{d}$. \\
In Fig. \ref{fig:critisovst} we plot the value of the critical
isospin chemical potential $\mu_{I}^C$ beyond which a pion
condensate forms vs. temperature $T$ at zero quark chemical
potential $\mu_{q}$. The growth of $\mu_{I}^{C}$ is easily
understood since the pion mass (which should be twice
$\mu_{I}^{C}$) is expected to grow near the critical temperature
for chiral symmetry restoration \cite{Barducci:1991rh}; actually, in the chirally-restored phase pion cannot be interpreted anymore as a pseudo-Goldstone boson.
On the
other hand, no phase transition to pion condensation is found
above the cross-over temperature for chiral symmetry restoration.
Thus the line of
critical values ends at $T=174~\mbox{MeV}$.
This is a somehow empirical result. There must be a deeper connection 
between the critical temperature $T_c$ for an interacting Bose-Einstein condensate of pions and the critical temperature for chiral symmetry restoration.
\\
The pion condensate is also expected to decrease for growing
temperatures; in Fig. \ref{fig:ifetta200vst} we show $\rho$ vs.
$T$ at $\mu_{q}=0$ and $\mu_{I}=200~\mbox{MeV}$. Similar behaviors are
obtained for fixed values of $\mu_{q}$ and $\mu_{I}$.\\

\section{Analytical equivalence between $\mu_I^C$ and $m_{\pi}/2$}
\label{Analyticalmuic}

In this section we present a calculation which shows the analytical agreement between $\mu_I^C$ and $m_{\pi}/2$, at $\mu_q=T=0$, by making a $(1/m)$ expansion in the NJL model. We will consider the case of two light quarks with the same mass $m$.
Let us first consider the $T=\mu=0$ effective potential for one flavor, with vanishing pseudoscalar condensate

\begin{equation}\label{effpotoneflav}
V=\frac{\Lambda^2}{8G}\chi^2-\frac{3}{\pi^2}\int_0^{\infty}dp~p^2\sqrt{p^2+(m+F^2\Lambda\chi)^2}
\end{equation}
Here $\chi$ is related to the scalar condensate through eq.(\ref{eq:fields}), and $F^2=F(\vec{p}^{~2})^{2}$. In the chiral case we have
\begin{equation}\label{effpotoneflavmasszero}
V=\frac{\Lambda^2}{8G}\chi^2-\frac{3}{\pi^2}\int_0^{\infty}dp~p^2\sqrt{p^2+F^4\Lambda^2\chi_0^2}
\end{equation}
$\chi_0$ standing for the dimensionless field in the chiral case. We find the gap equation by imposing the stationarity condition ${\partial V}/{\partial \chi}\left|\right._{\chi_0}=0$

\begin{equation}\label{gapequationchiral}
\frac{\chi_0}{4G}-\frac{3}{\pi^2}\int_0^{\infty}dp~p^2\frac{F^4\chi_0}{\sqrt{p^2+F^4\Lambda^2\chi_0^2}}=0
\end{equation}
Here we are interested in minima of the effective potential characterized by $\chi_0\neq0$, namely in solutions of 

\begin{equation}\label{gapequationchiral2}
\frac{1}{4G}-\frac{3}{\pi^2}\int_0^{\infty}dp~p^2\frac{F^4}{\sqrt{p^2+\Lambda^2 F^4\chi_0^2}}=0
\end{equation}
In the massive case, the gap equation becomes

\begin{equation}\label{gapequationmassive}
\frac{\Lambda\chi}{4G}-\frac{3}{\pi^2}\int_0^{\infty}dp~p^2\frac{(m+\Lambda F^2\chi)F^2}{\sqrt{p^2+(m+\Lambda F^2\chi)^2}}=0
\end{equation}
If we consider low enough values of masses, we can separate the scalar field into the chiral part plus a mass correction

\begin{equation}\label{chiralmass}
\chi\approx \chi_0+\chi_1~~~~~~~~\Lambda\chi_1=\alpha~m
\end{equation}
$\alpha$ plays the role of chiral susceptibility $\alpha=\partial(\Lambda\chi)/\partial m|_{m=0}$.. After some tedious but trivial calculations which follow from eqs. (\ref{gapequationchiral2}),(\ref{gapequationmassive}) we find for $\alpha$

\begin{equation}\label{alphavalue}
\alpha=\left[\frac{3}{\pi^2}\int_0^{\infty}dp~\frac{F^2p^4}{(p^2+\Lambda^2 F^4\chi_0^2)^{3/2}}\right]/\left[\frac{1}{4G}-\frac{3}{\pi^2}\int_0^{\infty}dp\frac{p^4F^4}{(p^2+\Lambda^2 F^4\chi_0^2)^{3/2}}
\right]\end{equation}

We will use this result later on. Now, we can study the effective potential for two degenerate flavors with mass $m$, with a non-vanishing pion condensate, at $\mu_q=T=0$ and $\mu_I\neq0$ (we are therefore considering the regime $\mu_u=-\mu_d$). In this particular case the expression for the effective potential turns out to be quite compact:

\begin{equation}\label{effpotonpioncondens}
V=\frac{\Lambda^2}{4G}(\chi^2+\rho^2)-\frac{3}{\pi^2}\int_0^{\infty}dp~p^2\left[\sqrt{(E+\mu_I)^2+F^4\Lambda^2\rho^2}+\sqrt{(E-\mu_I)^2+F^4\Lambda^2 \rho^2}\right]
\end{equation}
where $\chi_u=\chi_d\equiv\chi$ (the two fields are equal for opposite values of the chemical potential), $E=\sqrt{p^2+(m+\Lambda F^2\chi)^2}$. Now we should find the minimum of the effective potential with respect to the variables $\chi,\rho$; in order to do this, we will make the reasonable assumption of taking the scalar condensate fixed at the $\mu=0$ value. We are here interested in values of $\mu_I\sim m_{\pi}/2\sim70~\mbox{MeV}$, and it is a well-known fact that the physics at $T=0$ is independent on the chemical potential until a threshold value. We assume that, at $\mu_q=0$ and $\mu_I \neq 0$, this critical value is such that a non zero $\rho$ forms with a $\chi$ fixed to its $\mu=0$ value. In analogy with chiral models, we also assume that the transition is continuous; therefore, the critical value $\mu_I^C$ must verify

\begin{equation}\label{uiccritvalue}
\frac{\partial^2V}{\partial\rho^2}\left|\right._{\rho=0}(\mu_I^C)=0
\end{equation}
After some algebra, we have

\begin{equation}\label{uiccritcondition1}
\frac{1}{4G}-\frac{3}{\pi^2}\int_0^{\infty}dp~p^2\frac{F^4}{2}\left[\frac{1}{|E+\mu_I^C|}+\frac{1}{|E-\mu_I^C|}\right]=0
\end{equation}

Now, we consider without loss of generality positive values for $\mu_I$; we notice that the term $(E-\mu_I)$ is positive, since $E$ is in any case greater than the constituent mass, which is of the order of $\sim350\mbox{MeV}$, and we consider here $\mu_I\sim70\mbox{MeV}$. Therefore, we can safely remove the absolute values 
in the latter equation.\\
At the end of the calculation, we find that $\mu_I^C$ must satisfy

\begin{equation}\label{uiccritcondition2}
\frac{1}{4G}=\frac{3}{\pi^2}\int_0^{\infty}dp~p^2{F^4}\left[\frac{E}{E^2-(\mu_I^C)^2}\right]
\end{equation}
As a first check, we want verify that in the chiral case, when the pion mass is exactly zero, the critical value $\mu_I^C$ vanishes. By substituting $m=\mu_I^C=0$ in the last equation, we find back the gap equation in the chiral case. This check works out. Now, we want to consider the massive case; in the spirit of a $1/m$ expansion, we will consider values of $m$ such that $m\ll\Lambda\chi_0$ in eq.(\ref{chiralmass}), and expand all of the previous expressions in powers of $m$. At the end of this procedure, only the most significant term in $m$ will be kept. We skip the intermediate calculations and show the final expression only

\begin{equation}\label{uiccritconditionmassive}
(\mu_I^C)^2=(\Lambda\chi_0m)\left[\int_0^{\infty}dp~p^2\frac{F^4}{(p^2+\Lambda^2 \chi_0^2F^4)^{3/2}}(F^2+\alpha F^4)\right]/\left[\int_0^{\infty}dp~p^2\frac{F^4}{(p^2+\Lambda^2 \chi_0^2F^4)^{3/2}}\right]
\end{equation}
Now, we must compute the pion mass and compare it with the latter expression.
The static pion mass is related to the second derivative of the effective potential with respect to the pion field, apart a normalization factor which depends on the relation between canonical pion field and $\rho$ field. We have

\begin{equation}\label{pionmass1}
m_{\pi}^2=\frac{\chi^2}{f_{\pi}^2}\frac{\partial^2V}{\partial\rho^2}|_{min}
\end{equation}
where $f_{\pi}$ is the pion decay constant.\\
The expression for $f_{\pi}$ in a NJL model without form factor and with a cut-off $\Lambda$ is \cite{Klevansky:1992qe}:

\begin{equation}\label{fpai1}
f_{\pi}^2=\frac{N_c}{(2\pi)^3}\int_0^{\Lambda}d^3p\frac{M^2}{(p^2+M^2)^{3/2}}
\end{equation}
 where $M$ is the constituent mass. One can think of removing the cut-off and introducing the form factor in the expression eq.(\ref{fpai1}) by substituting 
$M$ with $(m+\Lambda\chi F^2)$. This procedure could appear as rather heuristic, and will be discussed later. By doing this one finds for $m_{\pi}$, computed at the leading order in $m$, the following expression:

\begin{equation}\label{mpaicrit}
m_{\pi}^2=(4\Lambda\chi_0m)\left[\int_0^{\infty}dp~p^2\frac{F^4}{(p^2+\Lambda^2 \chi_0^2F^4)^{3/2}}(F^2+\alpha F^4)\right]/\left[\int_0^{\infty}dp~p^2\frac{F^4}{(p^2+\Lambda^2 \chi_0^2F^4)^{3/2}}\right]
\end{equation}
At the leading order in $m$ the equivalence 

\begin{equation}\label{equiv}
\mu_I^C\equiv m_{\pi}/2
\end{equation}
is therefore proved. As far as the calculation of $f_{\pi}$ is concerned, if we consider the standard NJL model with hard cut-off on the three-momentum, the described procedure is rigorous. In the less-used version of the model with the form factor some difficulties arise when computing $f_{\pi}$; the chosen dependence of $F$ on the three-momentum only breaks covariance and a closed formula for $f_{\pi}$ has not been found. This is the reason why we did not fixed $f_{\pi}$ in the fit procedure of the paper \cite{Barducci:2004tt}, and why there was a little numerical mismatch between $\mu_I^C$ and $m_{\pi}/2$ in that paper. 

With a self-consistent treatment of quarks in the mean fied approximation and of mesons in the Random Phase Approximation, the authors of \cite{He:2005sp}, by using a NJL model with $U(1)_A$ breaking and without the form factor, have shown that
the result of eq.(\ref{equiv}) is exact at all orders. Moreover, the result is independent on the size of the $U(1)_A$ breaking coupling.

\section{Conclusions}\label{sec:conclucap2}

In this Chapter we have continued the study of pion condensation at
finite quark and isospin density in the NJL model that we had started
in the previous Chapter in the case of ladder-QCD for
small isospin chemical potentials. The extension to higher isospin
chemical potentials confirms the structure predicted in ref.
\cite{Klein:2003fy}, where two-flavor QCD was simulated in the
context of a random matrix model. Some differences between the two
analyses are present, at low temperatures, in the region of high
isospin chemical potentials, at the boundary of the region where
color superconductivity should take place. Actually in this region
we find that pion condensation is still active, whereas in ref.
\cite{Klein:2003fy} the authors find that the pion condensate
vanishes. We have also shown the expected behavior of scalar and
pion condensates by following different paths, for growing
temperatures, both in the plane of quark chemical potentials
$(\mu_{u},\mu_{d})$ and in that of isospin and quark chemical
potentials $(\mu_{I},\mu_{q}$). The analysis that we have
performed should also be confirmed, with only small quantitative
differences, within ladder-QCD.

\eject

~
\newpage
~

~
\newpage
~

\chapter{Pion and kaon condensation in a 3-flavor NJL model}

We analyze the phase diagram of a three-flavor Nambu-Jona-Lasinio
model at finite temperature $T$ and chemical potentials $\mu_u,
\mu_d, \mu_s$. We study the competition of pion and kaon condensation and we propose a physical situation where kaon condensation could be led by light quark finite densities only. We refer to the paper \cite{Barducci:2004nc}.

\section{Introduction}
\label{int3}

In the Chapters III and IV we have studied pion condensation by using two effective models, in order to mimic 2-flavors QCD.
The existence of phases with condensed pion is confirmed by lattice analyses \cite{Kogut:2002tm,Walters:2004ya}, within chiral models \cite{Son:2000xc,Splittorff:2000mm,Loewe:2002tw,Loewe:2004mu} and
within microscopic models \cite{Klein:2003fy,Toublan:2003tt,Frank:2003ve,He:2005sp}. 

In this Chapter we extend our previous analyses to the strange flavor. It has been suggested a kaon condensed phase in the context of high density nuclear matter \cite{Kaplan:1986yq,Brown:1992ib,Brown:1993yv,Glendenning:1998zx} and, more recently, in particular in the CFL regime \cite{Schafer:1999fe,Bedaque:2001je,Kryjevski:2004cw}: in this context, the mass of the strange quark plays the role of distinguishing the pattern of symmetry breaking between a two flavor superconducting (2SC) and a three flavor color flavor locking (CFL) phase. For this reason, an exhaustive study of these phenomena must take into account the strange quark sector. It is also worth to mention that taking into account weak equilibrium and color and electric neutrality many other interesting superconducting phases could exist \cite{Alford:2000ze,Neumann:2002jm,Alford:2003fq,Shovkovy:2003uu}.

These study could regard the physics of compact stars, more than that of colliders. The existence of stars whose core is so dense that quarks can be treated as fundamental degrees of freedom (i.e. quasi-particles) has been considered for instance in \cite{Steiner:2000bi,Buballa:2003qv,Weber:2004kj,Xu:2004zf}. 
Actually, compact stars can be thought as the only hypothetical experimental settings for color superconductivity to occur.

We will propose a physical scenario in which kaon condensation could be driven solely by a light quark finite density. The possibility of a kaon condensation at $\mu_s=0$ could be important in the context of neutron stars, where, by neglecting electroweak effects, the densities are associated with light quarks only.

This has been the first study in literature dealing with kaon condensation in a microscopic model\footnote{In Ref.\cite{Costa:2003uu} a study of meson properties at finite quark densities within the NJL model is performed. However, they do not deal with meson condensation.}.

\section{The model}
\label{sec:physics3}
Our purpose is to explore the general structure of the phase
diagram for chiral symmetry and pion and/or kaon condensation in
three-flavor QCD at non zero quark densities and its evolution in
temperature, by using the Nambu-Jona Lasinio model
(NJL). In this way we generalize, to
three flavors, previous works concerning the simpler two-flavor
case \cite{Barducci:2003un,Barducci:2004tt}. 
To our knowledge, so far the only study of meson condensation in
the general case of three-flavor QCD  is based on a chiral
Lagrangian \cite{Brown:1992ib,Brown:1993yv,Kogut:2001id}; nevertheless the use of a model such as the NJL
model or ladder-QCD \cite{Barducci:1987gn} is necessary if we want to
include the effects of a net baryon charge and also study
chiral symmetry restoration through the behaviour of scalar
condensates, besides the pseudoscalar ones. 

Let us thus start with the Lagrangian of the NJL model with three
flavors $u,d,s$, with current masses $m_u=m_d\equiv m$ and $m_s$
and chemical potentials $\mu_u,\mu_d,\mu_s$ respectively

\begin{eqnarray}
{\cal {L}}&=& {\cal {L}}_{0}+{\cal {L}}_{m}+{\cal {L}}_{\mu}+{\cal {L}}_{int}\nonumber\\
&=&{\bar{\Psi}}i{\hat{\partial}}\Psi-{\bar{\Psi}}M\Psi~+~
\Psi^{\dagger}~A~\Psi~+~{G}\sum_{a=0}^{8}\left[\left(
{\bar{\Psi}}\lambda_{a}\Psi
\right)^{2}+\left({\bar{\Psi}}i\gamma_{5}\lambda_{a}\Psi\right)^{2}
\right] \label{eq:njlagr3}
\end{eqnarray}
where
\begin{equation} \Psi=\left(
\begin{array}{c} u\\ d\\s
\end{array} \right),~~~~~A=\left(
\begin{array}{c} \mu_u\\0\\0
\end{array} \begin{array}{c} 0\\ \mu_d\\0
\end{array}\begin{array}{c} 0 \\ 0\\ \mu_s
\end{array}\right),
~~~~M=\left(
\begin{array}{c} m\\0\\0
\end{array} \begin{array}{c} 0\\ m\\0
\end{array}\begin{array}{c} 0 \\ 0\\ m_s
\end{array}\right)
\nonumber \end{equation}

\bigskip\noindent $M$ is the quark current mass matrix which is
taken diagonal and $A$ is the matrix of the quark chemical
potentials. As usual $\lambda_0=\sqrt{\displaystyle{{2\over 3}}}~
\mbox{\bf{I}}$ and
$\lambda_{a}~$, $~a=1,...,8~$ are the Gell-Mann matrices.\\
For $M=0$ and $A=0$ or $A\propto \mbox{\bf{I}}$, the Lagrangian is
$U(3)_L\otimes U(3)_R$ invariant. The chiral symmetry is broken by
$M\neq 0$ which also breaks $SU(3)_{V}$ down to $SU(2)_{V}$ as
$m\neq m_s$. However, this symmetry is also lost, since we
generally consider $\mu_u\neq\mu_d$. The remaining symmetry is
thus the product of three independent phase transformations
$U_u(1)\otimes U_d(1)\otimes U_s(1)$.
For simplicity, in this application we do not consider the 't Hooft determinant, which explicitly breaks $U(1)_A$. 
Also, we turn off the electroweak effects and we do not consider the possibility of di-quark condensation; therefore, our results will be reliable only in the regime of intermediate densities (with chemical potentials lower than $400\div500~\mbox{MeV}$).

We note that we can express  ${\cal {L}}_{\mu}$ either by using
the variables $\mu_{u},\mu_{d},\mu_{s}$ or, by introducing an
averaged light quark chemical potential
\begin{equation}
\mu_{q}=(\mu_u+\mu_d)/2 \label{eq:muq}
\end{equation}
and the three combinations

\begin{equation}
\hat{\mu}=(\mu_u+\mu_d+\mu_s)/3;~~~\mu_I=(\mu_u-\mu_d)/2;~~~\mu_Y=(\mu_q-\mu_s)/2
\label{eq:defmu}
\end{equation}
which couple to charge densities proportional to the baryon
number, to the third component of isospin and to hypercharge
respectively,

\begin{equation}
{\cal {L}}_{\mu}=\hat{\mu}~\Psi^{\dagger}\Psi
~+~\mu_{I}~\Psi^{\dagger}\lambda_{3}\Psi+{2\over\sqrt{3}}~\mu_{Y}~\Psi^{\dagger}\lambda_8\Psi
\label{eq:lmuiso3}
\end{equation}

Otherwise we could decide to express ${\cal {L}}_{\mu}$ in terms of $\mu_q,\mu_I,\mu_S$, where $3\mu_q$ would couple to the baryon number and $\mu_S$ would be the chemical potential associated with strangeness.
Obviously, the two choices are equally legitimate, and we will adopt (\ref{eq:lmuiso3}).

To study whether a pion and/or kaon condensate is formed, we need
to calculate the effective potential. To do this we generalize to $SU(3)_F$ the bosonization procedure exposed in the previous Chapter.
The one-loop effective potential we get is:
\begin{equation}
\label{eq:poteff3} V=\frac{\Lambda^2}{8G}
(\chi_u^2+\chi_d^2+\chi_s^2+2\rho_{ud}^2+2\rho_{us}^2+2\rho_{ds}^2)+V_{\mbox{log}}
\end{equation}

\begin{eqnarray} \label{eq:vlog3}
V_{\mbox{log}}=-{1\over\beta}\sum_{n=-\infty}
^{n=+\infty}\int{d^{3}p\over (2\pi)^{3}} ~\mbox{tr log} \left(
\begin{array}{ccc}
h_u & -\gamma_5 F^2(\vec{p})~\Lambda~\rho_{ud}~&-\gamma_5 F^2(\vec{p})~\Lambda~\rho_{us}~\\
\gamma_5 F^2(\vec{p})~\Lambda~\rho_{ud}~ & h_d & -\gamma_5 F^2(\vec{p})~\Lambda~\rho_{ds}~\\
\gamma_5 F^2(\vec{p})~\Lambda~\rho_{us}~& \gamma_5
F^2(\vec{p})~\Lambda~\rho_{ds}~ & h_s
\end{array}
\right)\nonumber\label{eq:vlog3b}
\end{eqnarray}
\begin{eqnarray}
\end{eqnarray}
\begin{eqnarray}
&&~~~~~~~~~h_f=(i\omega_n+\mu_f)\gamma_0~-~\vec{p}\cdot\vec{\gamma}~-~
\left(m_f~+~F^2(\vec{p})~\Lambda~\chi_f\right)\nonumber
\end{eqnarray}
where the form factor $F({\bf
p}^2)=\displaystyle{{\Lambda^2\over \Lambda^2+{\bf p}^2}}$ ($\Lambda$ is a mass scale) has been coupled to fermionic legs to mimic asymptotic freedom \cite{Alford:1997zt}. In eq. (\ref{eq:vlog3b}) $\mbox{tr}$ means sum over Dirac, flavor and color indices
and $\omega_{n}$ are the Matsubara frequencies.
The dimensionless fields $\chi_{f}$ and $\rho_{ff^{'}}$ are
connected to the condensates by the following relations

\begin{eqnarray} \label{eq:fields3}
\chi_f &=& - ~4G~{\langle{\bar{\Psi}}_f\Psi_f\rangle\over \Lambda}\nonumber\\
\\
\rho_{ff^{'}} &=& -~2G
~{\langle{\bar{\Psi}}_f\gamma_{5}\Psi_{f^{'}}-{\bar{\Psi}}_{f^{'}}\gamma_{5}\Psi_{f}\rangle\over
\Lambda}\nonumber
\end{eqnarray}
and are variationally determined at the absolute minimum of the
effective potential. 
We expect a superfluid phase with condensed pions when
the isospin chemical potential exceeds a critical value $\mu_I^C$ ($\mu_I^C=m_{\pi}/2$ at $T=\mu_q=0$); analogously a kaon 
condensation phase is expected when $\mu_K$ is higher than the kaon mass $m_K$. Since we are neglecting electroweak effects, the four kaons have all the same mass $m_K$.
We can express the kaon chemical potentials as a function of quark chemical potentials, by studying the effective action in the kaon fields sector. 
In practice, we have to study the effective kaon masses when chemical potentials are included.
In the normal phase (i.e. without meson condensates) we find: 
\begin{equation}\label{kaonchempot}
\mu_{K^+}=2\mu_Y+\mu_I~~~;~~~\mu_{K^0}=2\mu_Y-\mu_I
\end{equation}
For $\mu_I=0$, at $T=0$ the critical value for kaon condensation should be $\mu_Y^C=m_K/2$. 

To fix the free parameters of the model, which are $\Lambda$, the
average light quark and the strange quark masses $m$, $m_s$ and
the coupling $G=g/\Lambda^2$, we calculate the charged pion and
kaon masses (by fixing  their decay constants to the experimental
values) and the light quark scalar condensate, in the vacuum. The fit procedure is analogous to that of Chapter IV.

By choosing the free parameters as follows
\begin{equation}
\Lambda=1000~\mbox{MeV};~~~~~~~g=G~\Lambda^2=6;~~~~~~~~m=1.7~\mbox{MeV};~~~~m_s=42~\mbox{MeV}~~~~
 \label{eq:parameters3}
\end{equation}
we get for the light quark condensate, the
constituent light quark mass, the  pion mass  and the  kaon mass
respectively the results $\langle{\bar
\Psi}_{f}\Psi_{f}\rangle=-(252~\mbox{MeV})^3$, $M_{q}=385~\mbox{MeV}$,
$m_{\pi}=142~\mbox{MeV}$, $m_K=500~\mbox{MeV}$. Furthermore, we find the expected
agreement between $m_{\pi}/2$ and $\mu_I^C$ (within  1$\%$), and also between
$\mu_Y^C$ and $m{_K}/2$ (within  4$\%$).

The very low value we find from our fit for the quark masses is entirely due to
the choice of form factors (instantaneous approximation), see \cite{Schmidt:1994di}, but their ratio $m_s/m\simeq 25$ is in agreement with the value given by the ratio of the pseudoscalar masses.

\section{Phase diagram for chiral symmetry and meson condensation}
\label{sec:physics3b}
As usual, we determine the phases of the model by looking at the
absolute minimum of the effective potential, given in
Eqs. (\ref{eq:poteff3}), (\ref{eq:vlog3}). 

The effective potential depends on three scalar fields and three pseudo-scalar fields; moreover, a non zero value of a field $\rho_{ff'}$ induces
an effective coupling between the flavors $ff'$. Therefore, in general, the minimum has to be searched in a six-dimensional space, and the analysis turns out to be very hard from the numerical side.
\\
Since there are four
thermodynamic parameters, namely $T,\mu_u,\mu_d,\mu_s$ (or
$T,\hat{\mu},\mu_I,\mu_Y$), we start from $T=0$ and then follow the
evolution in temperature of two-dimensional slices
in the space of chemical potentials. We recall that the global symmetry in the model is given by the product of three $U(1)$ groups. The possible formation of a non-zero value of one of the pseudoscalar fields $\rho_{ff^{'}}$ implies the spontaneous breaking of a $U(1)$ symmetry with the appearance of a Goldstone boson. The transition to this superfluid phase leaves two $U(1)$ groups unbroken and between them there is always the $U(1)^B_V$. For instance, the formation of a $\rho_{ud}$ condensate implies the breaking of the $U(1)_V^I$ symmetry, whereas $U(1)_V^Y \otimes U(1)_V^B$ are left unbroken. At the threshold one of the charged pions, depending on the sign of $\mu_I$, is the Goldstone boson, whereas the other mesons of the octet remain massive (see also {\cite{Kogut:2001id}).

Thus we characterize different regions of pion or kaon
condensation with the pseudoscalar field which acquires a non
vanishing vacuum expectation value (v.e.v.), whereas the other
v.e.v.'s are vanishing. Then, to facilitate the
physical interpretation and to directly compare it with the results of ref.
\cite{Kogut:2001id}, in the pictures we label these regions
directly with the symbol of the particle that condenses, 
rather than with that of the corresponding field $\rho_{ff^{'}}$.\\
The scalar fields $\chi_{f}$ do not break any symmetry as chiral
symmetry is explicitly broken by the current quark masses. However
these fields undergo cross-over or discontinuous
transitions for given values of the thermodynamic parameters.\\
Thus we distinguish between regions where a non zero v.e.v. of a
scalar field is primarily due to dynamical effects, and we
indicate them with the relative symbol $\chi_{f}$, from regions
where their values are of order $\sim m_{f}/\Lambda$, where we put
no symbol. We also avoid putting the symbol $\chi_f$ in a region where the pseudoscalar field associated with that flavor starts to form, and $\chi_f$ is not yet of order $\sim m_f/\Lambda$.
Furthermore, both for pseudoscalar meson condensation and for the transitions
associated with scalar fields, the boundaries of the regions
separated by discontinuous transitions are represented by solid
lines whereas dashed lines indicate continuous transitions.\\
To start with, we set $\mu_{u}=-\mu_{d}$, which is the case studied by
Toublan and Kogut in ref. \cite{Kogut:2001id}. However, to compare
our results with theirs, we have to remark that we adopt
different combinations of the microscopic chemical potentials
$\mu_{u},\mu_{d},\mu_{s}$. In this case, where $\mu_{q}=0$ (see
Eq. (\ref{eq:muq})), we plot the phase diagram in the plane of
$\mu_{Y}$ and $\mu_{I}$, which are proportional respectively to
$\mu_{s}$ and to $\mu_{u}-\mu_{d}$ and which correspond to half of
the chemical potentials employed in ref. \cite{Kogut:2001id}.\\
In Fig. \ref{fig:Diagmuimus} we plot the quadrant with
$\mu_{Y}>0,\mu_{I}>0$ of the phase diagram at zero temperature.
Let us start by looking at what happens by moving along the vertical
line at $\mu_{I}=0^+$, where $0^+$ stands for an infinitesimally small positive value, in order to remove the degeneracy between $K^+$ and $K^0$.

 In this case one starts from a ``normal''
phase (see also ref. \cite{Kogut:2001id}) characterized by chiral
symmetry breaking by large scalar condensates and then, for
$\mu_{Y}\simeq m_{K}/2\simeq 250~ \mbox{MeV}$, the scalar $\chi_{u}$ and
$\chi_{s}$ condensates smoothly rotate into a $\rho_{us}$
condensate, whereas the $\chi_{d}$ field remains unchanged across
the transition. We have indicated in the picture this new region
with the symbol $\langle K^{+}\rangle$ (rather than with
$\rho_{us}$) as with these signs of the chemical potentials it is
the positively charged kaon to condense. By further increasing
$\mu_{Y}$ the dynamical effect associated with the strange quark
condensation is weakened by a high strangeness density and the
favored phase becomes that with $\chi_{s}\sim m_{s}/\Lambda$ and
large $\chi_{u},\chi_{d}$. The kaon condensate vanishes in this
case through a first order transition. The same situation hereby
described also holds when we move along vertical lines with
$\mu_{I}\lesssim m_{\pi}/2$, with a linear decrease of the
critical $\mu_{Y}$ for kaon condensation and a weak linear
increase for the second transition at higher $\mu_{Y}$.\\
On the other hand, when we move along the horizontal line at
$\mu_{Y}=0$ and we increase $\mu_{I}$, we pass from the ``normal''
phase to a pion condensation phase (in this case the $\pi^{+}$)
described by a finite $\rho_{ud}$, whereas $\chi_{s}$ remains
large. This situation was already known by the previous studies of
the two-flavor NJL model \cite{Barducci:2004tt}. This
transition is also second order. This situation persists at larger
$\mu_{I}$ too. In the intermediate regions there is a competition
between pion and kaon condensation. For instance, by fixing
$\mu_{I}\gtrsim m_{\pi}/2$, starting from low $\mu_{Y}$ 
 in the $\pi^{+}$-condensed phase with a large $\chi_{s}$
and increasing $\mu_{Y}$\footnote{In this case,  working at $\mu_{q}=0$, it simply means to increase $|\mu_{s}|$.}, we find that the system
goes to the phase where the $K^{+}$ condenses when $\mu_{Y}\gtrsim m_K/2$.
Consequently $\chi_s$ becomes $\sim m_{s}/\Lambda$ whereas
$\chi_d$ becomes large (if $\mu_{I}<325~\mbox{MeV}$) and the full
transition is discontinuous. When $\mu_{Y}$ is about 300
$\mbox{MeV}$ (with a linear weak increase for growing $\mu_{I}$), the
kaon condensate vanishes and a pion condensed phase shows up,
with all the scalar condensates $\chi_{f}\sim m_{f}/\Lambda$. The
transition is first order. Finally, for high $\mu_{I}$ and $\mu_{Y}$
in the range of values corresponding to kaon condensation, there is a phase analogous to
the previous one, with a kaon condensate instead of a pion condensate
and $\chi_{f}\sim m_{f}/\Lambda$. This region is
connected to the others through discontinuous transitions
too.

Along the line $\mu_I=0$ the Nielsen-Chadha theorem \cite{Nielsen:1975hm} would apply. The condensation of a kaon field would cause the following pattern of symmetry breaking \cite{Schafer:2001bq}
\begin{equation}\label{Kaonpatternsym}
SU(2)\otimes U(1)\rightarrow U(1)
\end{equation}

with three broken generators. Due to their dispersion relations, there are only two corresponding Goldstone bosons \cite{Schafer:2001bq}. 

In Fig. \ref{fig:CondT0mui230} we plot the behaviour of the scalar
and pseudoscalar condensates vs. $\mu_{I}$ at $\mu_{q}=0$,
$\mu_{Y}=230~ \mbox{MeV}$ and $T=0$ 
corresponding in Fig. \ref{fig:Diagmuimus} to the path marked by the solid line $a$. It is worth remarking the rotation of the $\chi_{u}$ and $\chi_{s}$
fields into a $\rho_{us}$ field at the continuous transition for
kaon condensation (at $\mu_{I}\simeq 50~\mbox{MeV}$) and the further ``exchange'' of the two pseudoscalar fields
$\rho_{us}$ and $\rho_{ud}$ at the pion condensation transition
(at $\mu_{I}\simeq 125~\mbox{MeV}$).

In Fig. \ref{fig:condvert} we plot the behaviour of the scalar and
pseudoscalar condensates vs. $\mu_{Y}$ at $\mu_{q}=0$, $\mu_I=200
~\mbox{MeV}$ and $T=0$ (following the path of the solid line $b$ in Fig.
\ref{fig:Diagmuimus}).

In Fig. \ref{fig:Diagallmuimus} the phase diagram of Fig.
\ref{fig:Diagmuimus} is extended to negative values of $\mu_{I}$
and $\mu_{Y}$ too. This picture is thus simply obtained by
reflection of Fig. \ref{fig:Diagmuimus} around its axes at
$\mu_{Y}=0$ and $\mu_{I}=0$. The only difference is that in the
other three quadrants different pseudoscalar mesons of the octet condense
(see also ref. \cite{Kogut:2001id}).

The evolution of the scalar fields for growing temperatures is
well-known. In particular, as expected the growth of
temperature fights against pion and kaon condensation. We find
that the kaon condensed phase disappears at $\mu_{I}=0$ for
$T>85~\mbox{MeV}$. In Fig. \ref{fig:DiagmuimusT100} we show the evolution
of the phase diagram of Fig. \ref{fig:Diagmuimus} at $T=100~\mbox{MeV}$.
The region for kaon condensation has shrunk and the transitions
associated with chiral symmetry approximate restoration have become
continuous. Above $T\simeq 110~\mbox{MeV}$ the regions of kaon
condensation disappear. In Fig. \ref{fig:DiagmuimusT140} we plot
the further evolution of the phase diagram of Fig.
\ref{fig:Diagmuimus} at $T=140~\mbox{MeV}$. In the range of values of
$\mu_{I},\mu_{Y}$ that we have considered, only the
regions of pion condensation and the usual regions with chiral
symmetry breaking separated by continuous transitions from
analogous phases with $\chi_{s}\sim m_{s}/\Lambda$ for large
$\mu_{Y}$ remain. Finally, pion condensation disappears above the
cross-over critical temperature $T=192~\mbox{MeV}$.\\

\section{Kaon condensation in the region of high $\mu_q$}\label{Kaoncond}

So far we have considered the limit $\mu_{u}=-\mu_{d}$ (and thus $\mu_q=0$). A more general picture of the full phase diagram can be grasped by
considering two other cases, which are $\mu_{u}=\mu_{d}$
(and thus $\mu_{I}=0$), and $\mu_{s}=0$, which has been 
already considered in ref. \cite{Barducci:2004tt}.
In this section, we suggest a hypothesis for a kaon condensation driven mainly by light quark finite densities, with $\mu_s$ low or even zero. We recall that in the case of $\mu_s=0$ there is an equal number of strange quarks and anti-quarks, and in that case kaon condensation 
would concern a light particle (antiparticle), associated to the external field (the chemical potential),  and a strange anti-quark (quark) belonging to the sea.
In ref. \cite{Barducci:2004tt}, we had not inserted the pseudoscalar fields with strangeness content: the possibility of kaon condensation at $\mu_s=0$ would modify the phase diagram in the ($\mu_u,\mu_d$) plane of ref. \cite{Barducci:2004tt} with the insertion of regions with condensed kaons. 
However, except for this, no further modification
would be necessary, whereas the possibility of kaon condensation
at zero or low values of $\mu_{s}$ can be also studied for $\mu_{I}=0$,
which is therefore the last situation we examine.\\
From its definition in Eq. (\ref{eq:defmu}), we see that $\mu_{Y}$
measures the difference between the strange quark chemical
potential and $\mu_{q}$ (which at $\mu_{I}=0$ is the chemical
potential associated with one of the light flavors) similarly to
$\mu_{I}$ which measures the unbalance between the $u$ and $d$
quark chemical potentials. Therefore, since pion condensation
occurs, at $T=0$ and for instance at $\mu_{d}=0$, when $\mu_{u}\gtrsim
m_{\pi}$, similarly we could expect a kaon condensed phase at
$T=0$ and $\mu_{s}=0$ when $\mu_{q}\gtrsim m_{K}$. However, within
our approximations, this does not happen, the reason being that
$m_{K}$ is higher than the critical value of $\mu_{q}$ for the
melting of the dynamical part of $\langle {\bar
\Psi}_{f}\Psi_{f}\rangle$, $f=u,d$. In other words, when the
condensates of the light quarks start to decrease to values $\sim
m/\Lambda$, the effective kaon mass starts to increase and kaon
condensation always remains unfavored (the same thing does not happen at
$\mu_{q}=0$ because the melting of the dynamical part of $\langle
{\bar s}s\rangle$ occurs at higher values of $\mu_{s}$ and actually we find
kaon condensation as shown in the last section).
Nevertheless, since we do not have taken into account the effects
of di-quark condensates which could break chiral symmetry (as it happens for instance in the CFL phase), it
would be interesting to further analyze what happens with their
inclusion. We will analyze this case in further work.

For the time being, let us show, in Fig. \ref{fig:Diagmuqmus},
the $T=0$ phase diagram in the $(\mu_{q},\mu_{s})$ plane at
$\mu_{I}=0^{-}$, where $0^{-}$ stands for an infinitesimally small
negative value, which is necessary to select which are the kaons
that condense (in any case, within a neutron star a finite negative isospin chemical potential is present).
 The structure of this phase diagrams recalls that
of \cite{Barducci:2004tt} for pion condensation in the
$(\mu_{u},\mu_{d})$ plane, with a ``normal" phase in the middle,
characterized by high values of the scalar condensates
$\chi_{u},\chi_{d},\chi_{s}$. This phase is separated from the regions of chiral symmetry approximate restoration (relative to one or more flavors depending on the values of chemical potentials) which are characterized by the respective scalar field $\chi_{f}\sim
m_{f}/\Lambda$. Furthermore this region is separated through discontinuous transitions,
both from the ``normal" phase and from the regions of kaon
condensation. There are four different regions of this kind,
distinguishable by the type of the condensing kaon and by the values of the associated scalar condensates. Finally, the ``normal" phase turns
into a region of kaon condensation through the smooth rotation of
$\chi_{s}$ and one of the light flavors (according to the sign of
$\mu_{Y}$ one between $\chi_{u}$ or $\chi_{d}$ remains large) into
a kaon condensate, when $\mu_{Y}\gtrsim m_{K}/2$, with a
second order phase transition. We can see that the presence of a high value of $|\mu_q|$
 lowers the corresponding $|\mu_s|$ necessary to have kaon condensation: moreover, a value of $\mu_I\neq 0$ would favor a kaon condensation with lower values of $|\mu_q|$ and/or $|\mu_s|$.\\
Since we do not know exactly the critical value $\mu_q^C$ for chiral symmetry restoration, and since we are interested here in the case where $\mu_q^C>m_{K}$, instead of varying $\mu_q^C$ (for instance performing a different fit) we take the kaon mass as a free parameter, and we present, in Fig. \ref{fig:limsuk}, the same result as in Fig.
\ref{fig:Diagmuqmus}, for $m_{K}=300~\mbox{MeV}$, which is
the upper limit for having a kaon condensation at
$\mu_{s}=0$.\\

\section{Conclusions and further developments}\label{conc4}
In this Chapter we have presented a calculation of the QCD phase diagram obtained by using a three-flavor NJL model, without $U(1)_A$ breaking. We 
have considered the plane ($\mu_I,\mu_Y$) to study the competition of kaon and pion condensation at zero and finite temperature and to establish a comparison with a previous analysis performed in a chiral model \cite{Kogut:2001id}. The results of the microscopic model accurately reproduce
 those achieved in the chiral model only for low densities, since in the high strangeness density regime approximate chiral symmetry restoration (relative to the strange sector) occurs, disfavoring kaon with respect to pion condensation.
Moreover we have suggested the possibility of kaon condensation driven only by light quark densities: this situation could be of interest for the core of compact stars, where the densities could be approximatively associated with light flavors only.

To make this study more quantitative, we should include the effects we have up to now neglected.
In the first place, we have to consider di-quark condensation in the regime of high densities; those studies have been driven in recent years in the context of microscopic models. The last goal of this project would be taking into account meson condensation in superconducting regimes. The possibility of kaon condensation in the CFL phase (K0-CFL) \cite{Bedaque:2001je} has been recently proposed in the framework of a chiral model. It would be interesting to perform the same study but with a more fundamental model: the problem is that the structure of the one loop term of the effective potential results rather hard to deal with, even though conceptually simple.
 Actually, we have to compute the determinant of a 72x72 matrix, corresponding to 3 colors, 3 flavors, 4 Dirac, 2 Nambu-Gorkov indices.
 In the Nambu-Gorkov formalism one introduces a bi-spinor 
\begin{equation}\label{Nambugorkov}
\chi=\frac{1}{\sqrt{2}}\left(
\begin{array}{c}
\Psi\\
C\Psi*
\end{array}
\right)
\end{equation}
The result is that di-quarks can be included in the path-integral formalism by means of the calculation of a fermionic integral in Nambu-Gorkov space. The price to pay is the doubled dimension of the spinor itself.
\\
For di-quark fields, we should consider the following combinations:
\begin{equation}\label{diquarkkaon257}
s_{22},s_{55},s_{77}
\end{equation}
with
\begin{equation}\label{diquarkkaon}
s_{AA'}=\langle\Psi^TC\gamma_5\tau_A\lambda_{A'}\Psi\rangle
\end{equation}
where $\tau_A,\lambda_{A'}$ are Gell-Mann matrices acting in flavor, color spaces respectively.
In the pure CFL regime, these fields are all equal; here we consider that, due to chemical potential asymmetries or to the condensation of a flavor-mixing pseudoscalar field, they can be different.
In order to get the ground state, we have to minimize the effective potential with respect to at least seven variables (3 scalar condensates, 3 di-quark condensates, 1 meson condensate). 
Actually, pseudoscalar and di-quark condensates mix different flavors and colours, which turn out to be variables dependent on each other.
Obviously, the study gets cumbersome from the numerical side.\\
Once overcome this hurdle, it would be possible to study whether a kaon condensate could arise, at high densities, by considering only light quark densities. This would certainly be a remarkable results for the physics of stars.  
The competition of di-quark and meson condensates in a NJL model has been partially afforded in the papers
\cite{Buballa:2004sx,Forbes:2004ww,Warringa:2005jh} which followed our work.
 Eventually, the inclusion of the $\beta$-equilibrium as well as charge and color neutrality, should be important for reproducing theoretically the settings attained in the core of a compact star.
The inclusion of the 't Hooft determinant in the model would not dramatically modify the results we have obtained; moreover the restoration of $U(1)_A$ in the regime of high densities is expected.

\begin{figure}[htbp]
\begin{center}
\includegraphics[width=14cm]{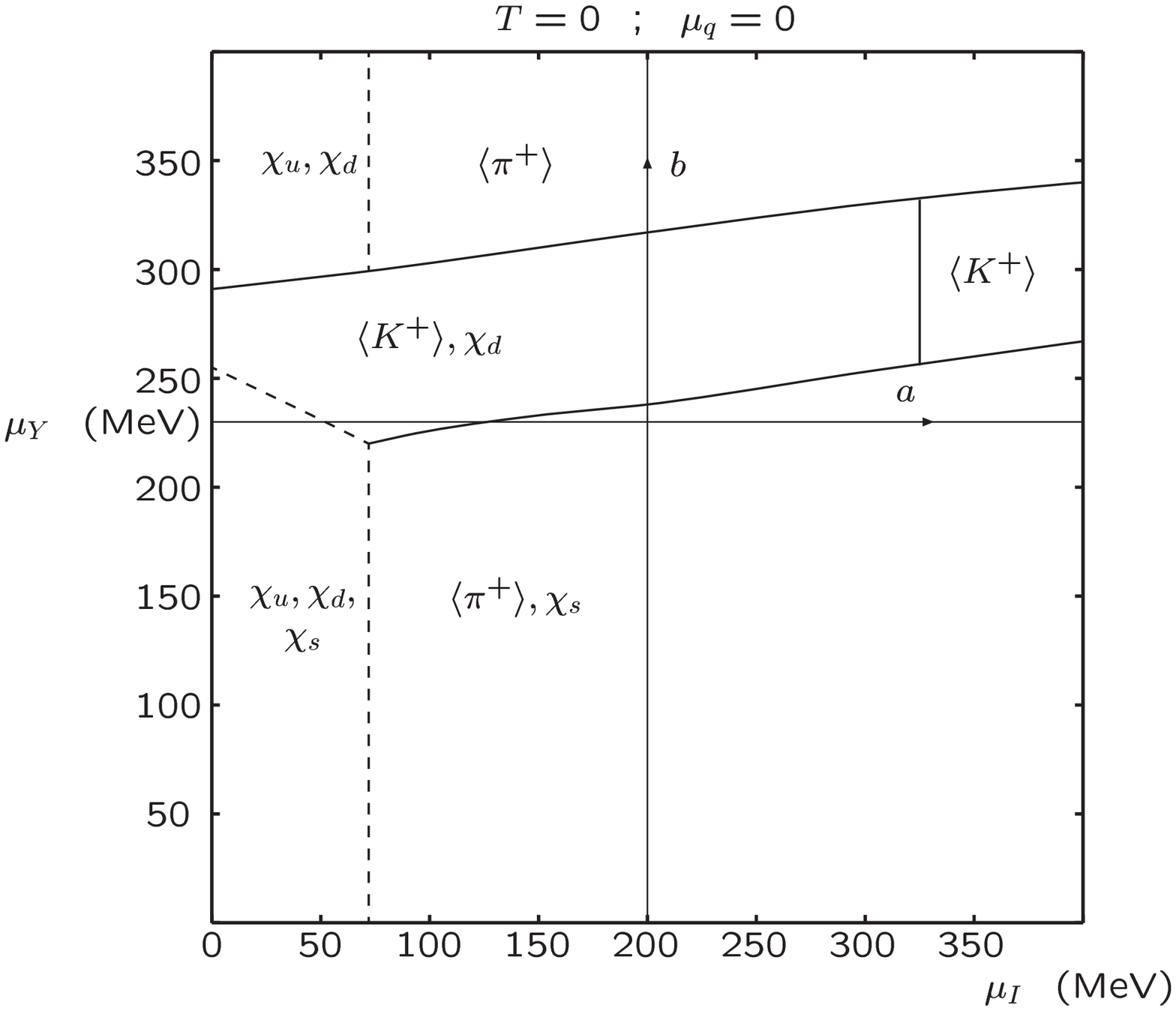}
\end{center}
\caption{\it Phase diagram for chiral symmetry restoration and meson condensation in the plane ($\mu_I,\mu_Y$) at $\mu_q=0$ and $T=0$. Different regions are specified by the non vanishing of a given condensate, whereas the others are vanishing ($\rho_{ud},\rho_{us}$) or order $\sim m_f/\Lambda$ ($\chi_u,\chi_d,\chi_s$). Dashed lines are for the continous vanishing of pseudoscalar fields, whereas solid lines are for discontinuous behaviours. The solid lines
a and b refer to specific paths at fixed values of $\mu_Y=230~\mbox{MeV}$ (line a) and $\mu_I=200~\mbox{MeV}$ (line b).} \label{fig:Diagmuimus}
\end{figure}

\begin{figure}[htbp]
\begin{center}
\includegraphics[width=14cm]{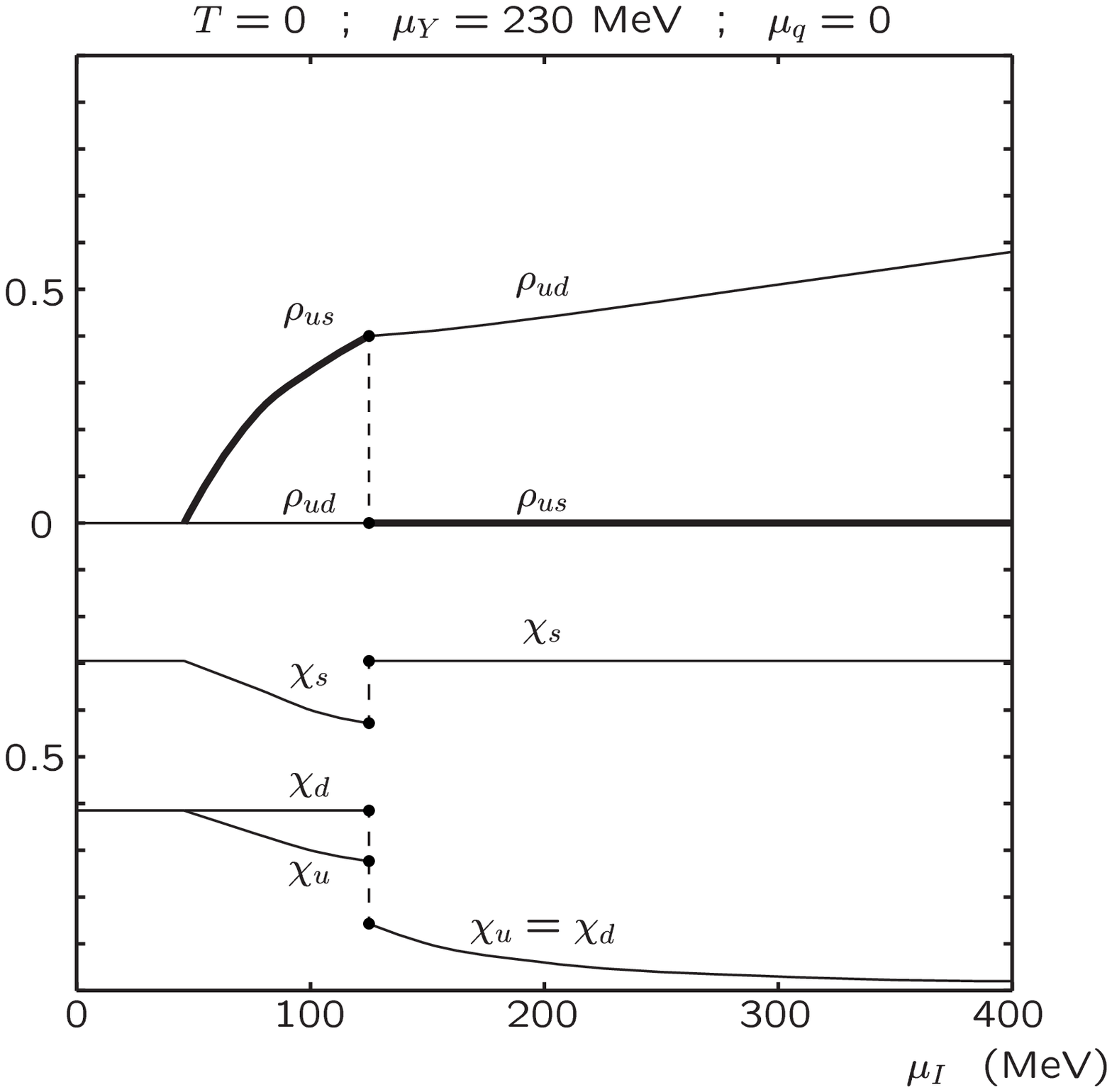}
\caption{\it Scalar and pseudoscalar condensates vs. $\mu_I$, for $T=0$, $\mu_q=0$ and $\mu_Y=230~{MeV}$. The path followed in the phase diagram of Fig. \ref{fig:Diagmuimus} is that of the solid line a.} \label{fig:CondT0mui230}
\end{center}
\end{figure}

\begin{figure}[htbp]
\begin{center}
\includegraphics[width=14cm]{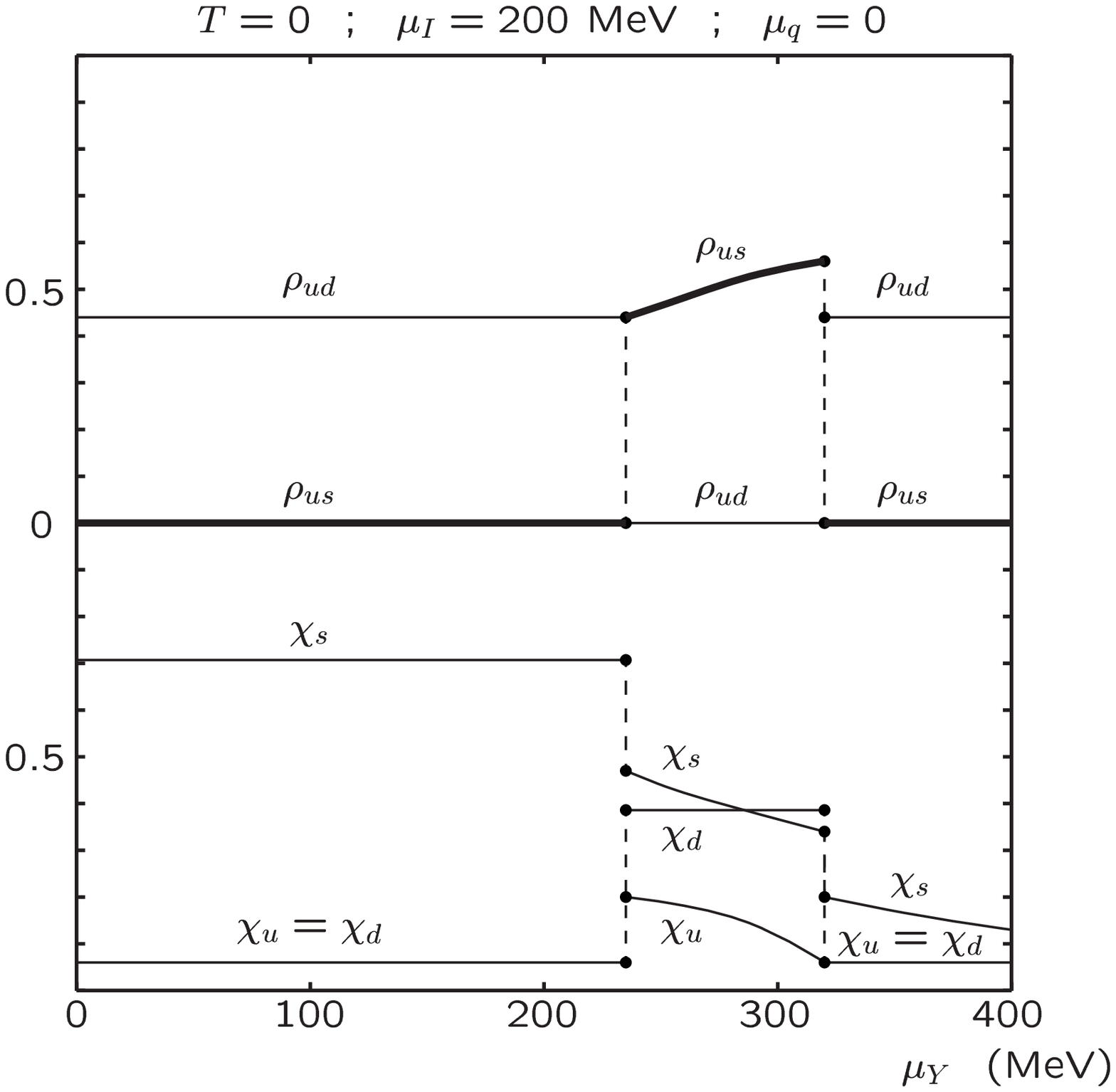}
\caption{\it Scalar and pseudoscalar condensates vs. $\mu_Y$, for $T=0$, $\mu_q=0$ and $\mu_I=200~{MeV}$. The path followed in the phase diagram of Fig. \ref{fig:Diagmuimus} is that of the solid line b.} \label{fig:condvert}
\end{center}
\end{figure}

\begin{figure}[htbp]
\begin{center}
\includegraphics[width=14cm]{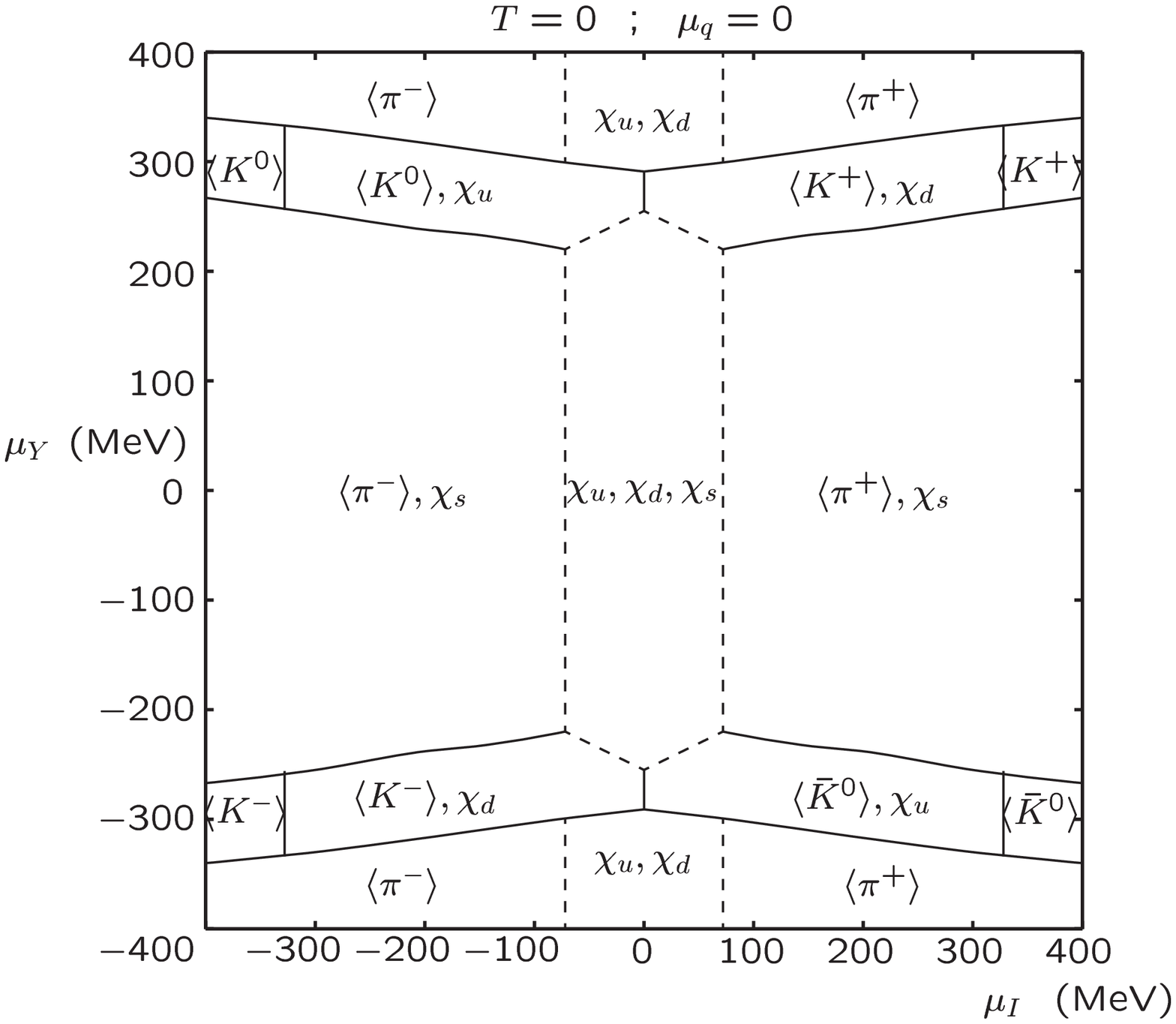}
\end{center}
\caption{\it Phase diagram for chiral symmetry restoration and meson condensation in the plane ($\mu_I,\mu_Y$) at $\mu_q=0$ and $T=0$. Different regions are specified by the non vanishing of a given condensate, whereas the others are vanishing ($\rho_{ud},\rho_{us},\rho_{ds}$) or order $\sim m_f/\Lambda$ ($\chi_u,\chi_d,\chi_s$). Dashed lines are for the continous vanishing of pseudoscalar fields, whereas solid lines are for discontinuous behaviours.} \label{fig:Diagallmuimus}
\end{figure}

\begin{figure}[htbp]
\begin{center}
\includegraphics[width=14cm]{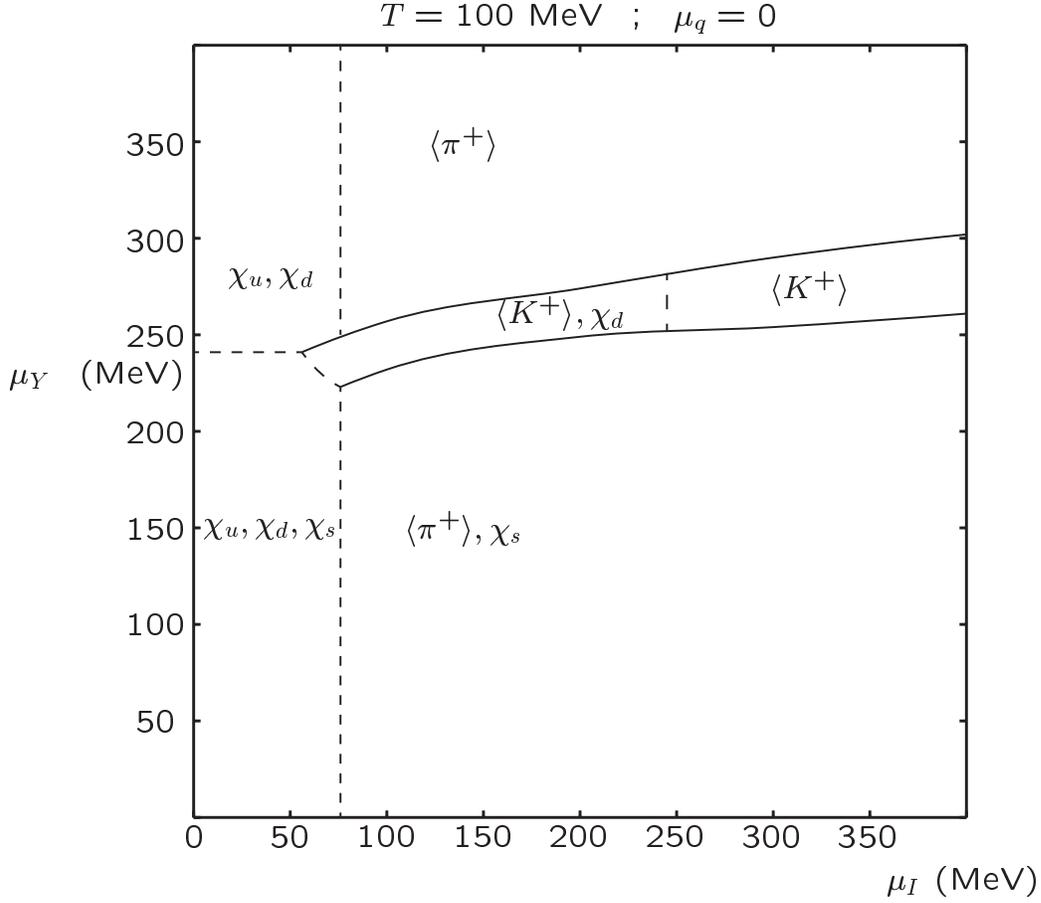}
\caption{\it Phase diagram for chiral symmetry restoration and meson condensation in the plane ($\mu_I,\mu_Y$) at $\mu_q=0$ and $T=100~MeV$, which is above the temperature of the tricritical point. Different regions are specified by the non vanishing of a given condensate, whereas the others are vanishing ($\rho_{ud},\rho_{us}$) or order $\sim m_f/\Lambda$ ($\chi_u,\chi_d,\chi_s$). Dashed lines are for the continous vanishing of pseudoscalar fields or for cross-over transitions for scalar fieds, whereas solid lines are for discontinuous behaviours.} \label{fig:DiagmuimusT100}
\end{center}

\end{figure}

\begin{figure}[htbp]
\begin{center}
\includegraphics[width=14cm]{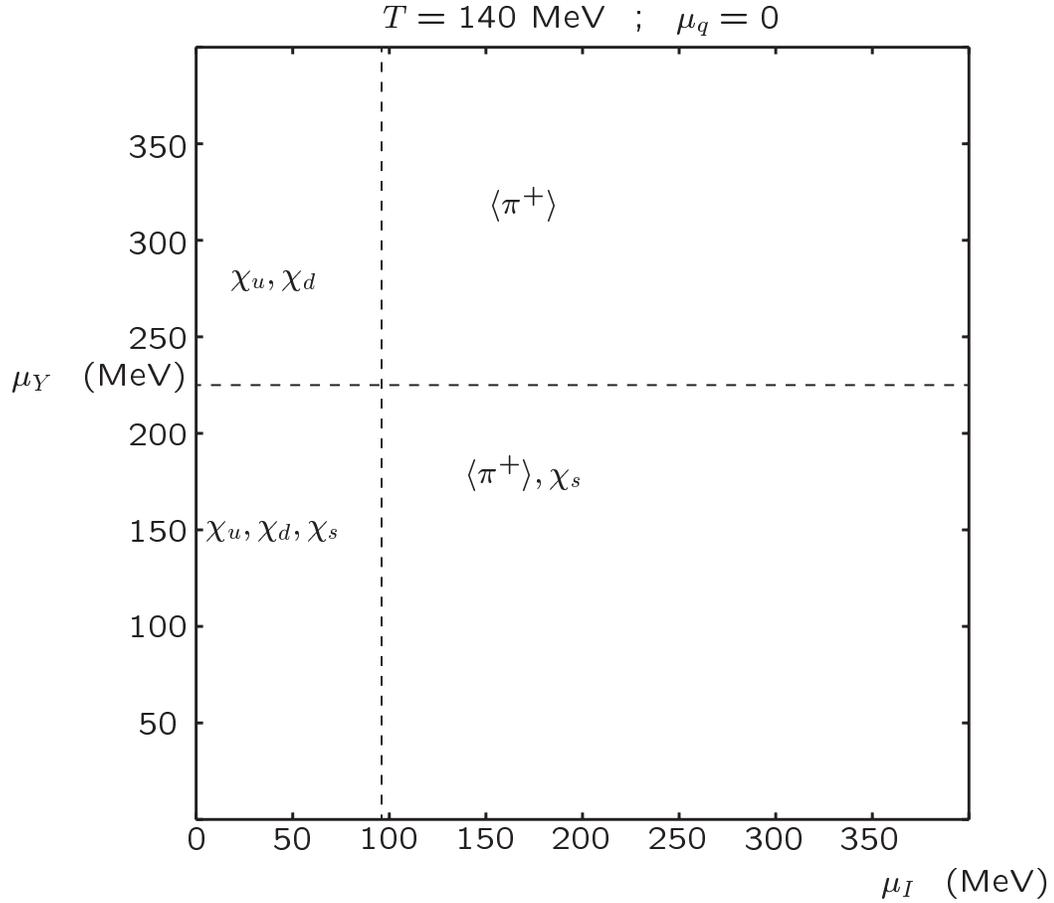}
\caption{\it Phase diagram for chiral symmetry restoration and pion condensation in the plane ($\mu_I,\mu_Y$) at $\mu_q=0$ and $T=140~MeV$, which is above the highest temperature $\sim~110~MeV$ to have kaon condensation. Different regions are specified by the non vanishing of a given condensate, whereas the others are vanishing ($\rho_{ud}$) or order $\sim m_f/\Lambda$ ($\chi_u,\chi_d,\chi_s$). Dashed lines are for the continous vanishing of pseudoscalar fields or for cross-over transitions for scalar fieds.} \label{fig:DiagmuimusT140}
\end{center}

\end{figure}

\begin{figure}[htbp]
\begin{center}
\includegraphics[width=14cm]{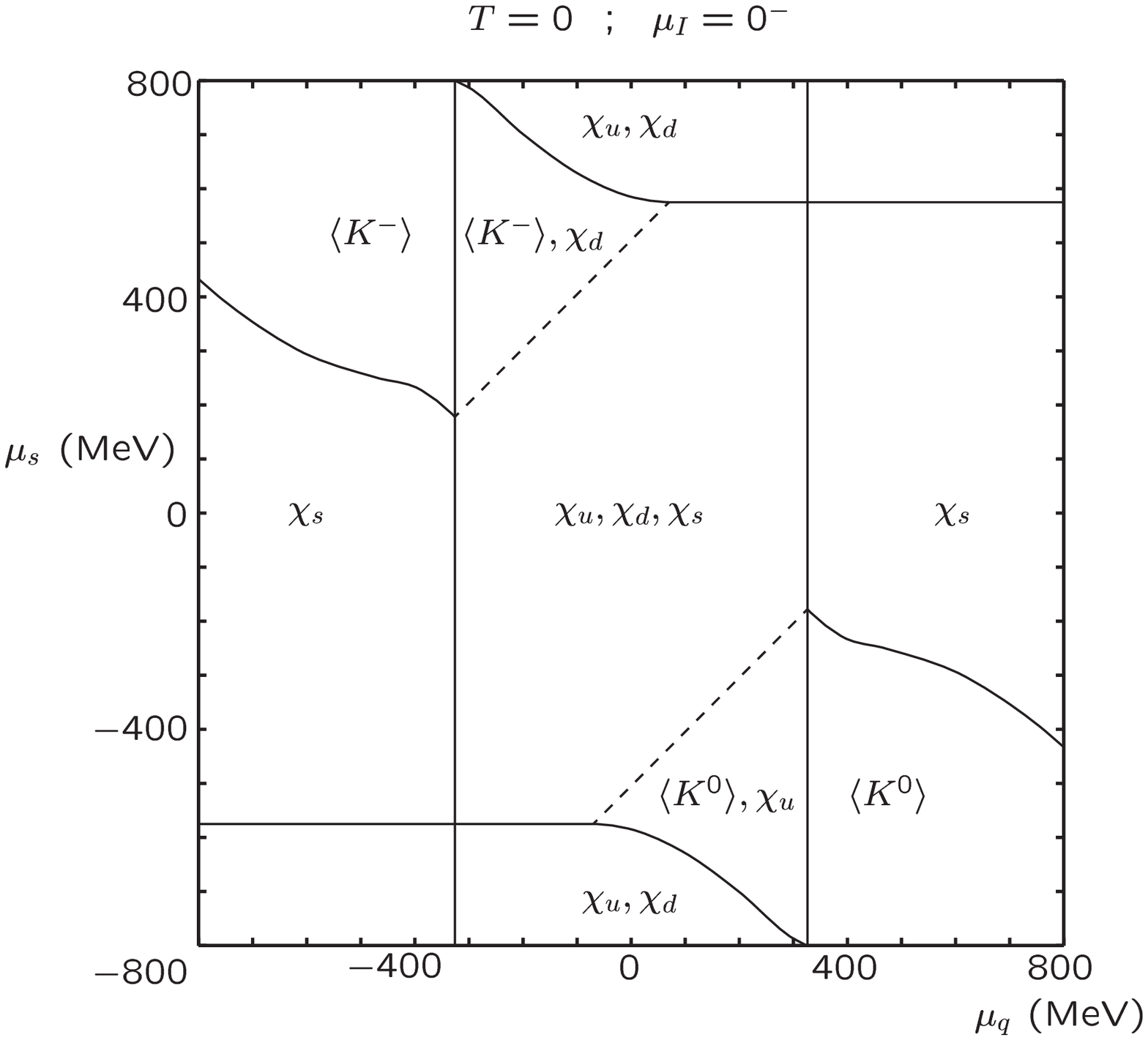}
\end{center}
\caption{\it Phase diagram for chiral symmetry restoration and kaon condensation in the plane ($\mu_q,\mu_s$) at $\mu_I=0^-$ and $T=0$. Different regions are specified by the non vanishing of a given condensate, whereas the others are vanishing ($\rho_{us},\rho_{ds}$) or order $\sim m_f/\Lambda$ ($\chi_u,\chi_d,\chi_s$). Dashed lines are for the continous vanishing of pseudoscalar fields, whereas solid lines are for discontinuous behaviours.}\label{fig:Diagmuqmus}
\end{figure}

\begin{figure}[htbp]
\begin{center}
\includegraphics[width=14cm]{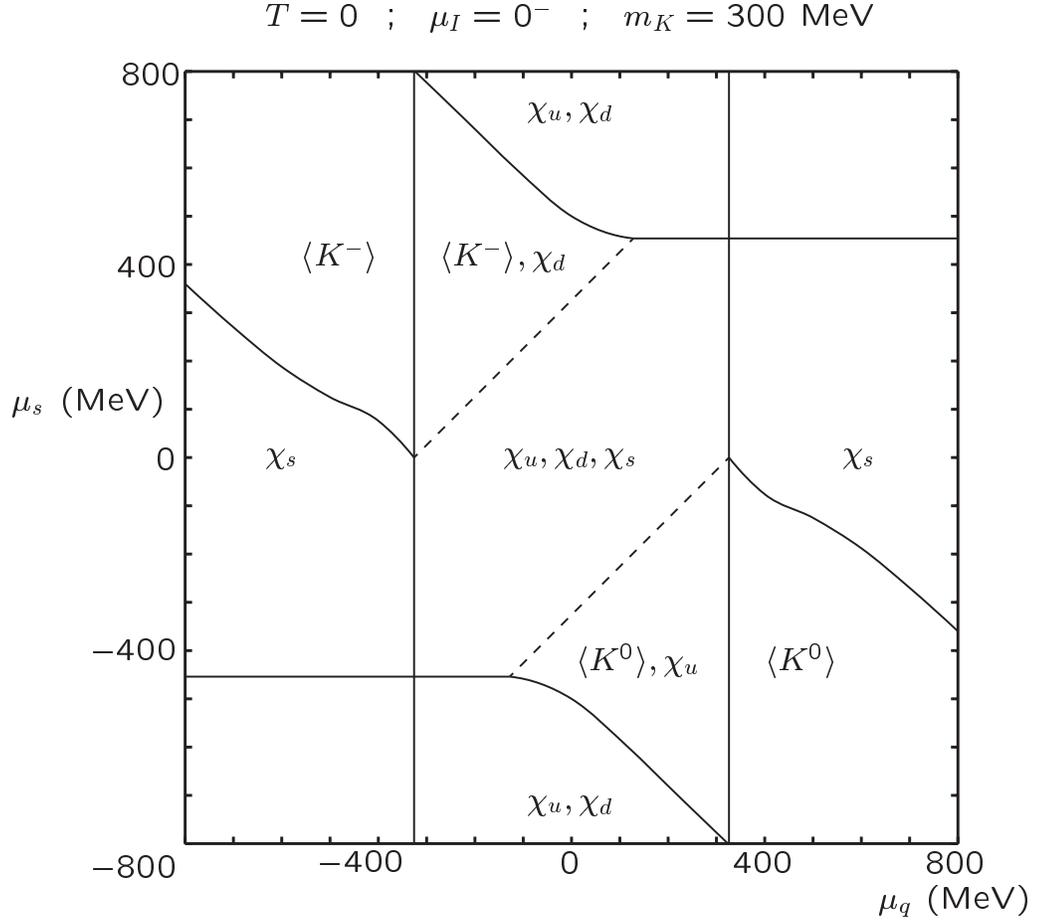}
\end{center}
\caption{\it Phase diagram for chiral symmetry restoration and kaon condensation in the plane ($\mu_q,\mu_s$) at $\mu_I=0^-$ and $T=0$, and for $m_K=300~MeV$. Different regions are specified by the non vanishing of a given condensate, whereas the others are vanishing ($\rho_{us},\rho_{ds}$) or order $\sim m_f/\Lambda$ ($\chi_u,\chi_d,\chi_s$). Dashed lines are for the continous vanishing of pseudoscalar fields, whereas solid lines are for discontinuous behaviours.}\label{fig:limsuk}
\end{figure}

\chapter{A study of the critical line in a 3-flavor NJL model with $U(1)_A$ breaking}

We employ a 3 flavor NJL model to stress some general remarks about the QCD critical line.
The dependence of the critical curve on $\mu_q=(\mu_u+\mu_d)/2$ and\\ $\mu_I=(\mu_u-\mu_d)/2$ is discussed. The quark masses are varied to confirm that, in agreement with universality arguments, the order of transition depends on the number of active flavors $N_f$. The slope of the critical curve vs. chemical potential is studied as a function of $N_f$. We compare our results with those recently obtained in lattice simulations to establish a comparison among different models.
We refer to the paper \cite{Barducci:2005ut}.

\section{Introduction}
\label{int4}

In recent years, the study of the QCD phase diagram by means of numerical lattice simulations has improved considerably. 
In particular, the presence of a critical ending point, first discovered within microscopical effective models \cite{Barducci:1989wi,Halasz:1998qr,Rajagopal:2000wf}, appears to be a solid feature \cite{Fodor:2001pe,Fodor:2004nz}, althought its exact location along the critical curve is still controversial. Moreover, the incoming realization of LHC enhances the interest of the scientific community toward this topic.\\
The main problem regarding the lattice study of the QCD phase diagram is related to the so called sign problem; the fermionic determinant is not positive definite at finite baryonic chemical potential, and therefore, to overcome this unwelcome feature, some suitable tricks are needed. Without going into detail, the methods commonly used are the study of QCD at imaginary chemical potential \cite{deForcrand:2002ci,D'Elia:2002gd,deForcrand:2003hx}, the reweighting procedure \cite{Fodor:2001pe} and the Taylor expansion in $\mu/T$ \cite{Allton:2002zi,Ejiri:2003dc}. 

The aim of this paper is to offer an overview of results concerning the behaviour of the critical line, obtained in the NJL model and directly comparable with recent lattice analyses. In its strong simplicity, the NJL model recovers the basic structure of the non perturbative dynamics ruling the problem.
Therefore, it can be trusted as a good toy model for the study of QCD phase diagram.

\section{The model}
\label{sec:physics4}

Let us consider the Lagrangian of the NJL model with three
flavors $u,d,s$, with current masses $m_u=m_d\equiv m$ and $m_s$
and chemical potentials $\mu_u,\mu_d,\mu_s$ respectively

\begin{eqnarray}\label{eq:njlagr4}
{\cal {L}}&=& {\cal {L}}_{0}+{\cal {L}}_{m}+{\cal {L}}_{\mu}+{\cal {L}}_{4}+{\cal {L}}_{6}\nonumber\\
&=&{\bar{\Psi}}i{\hat{\partial}}\Psi-{\bar{\Psi}}~\underline{\mathcal{M}}~\Psi~+~
\Psi^{\dagger}~\underline{\mathcal{A}}~\Psi~+~{G}\sum_{a=0}^{8}\left[\left(
{\bar{\Psi}}\lambda_{a}\Psi
\right)^{2}+\left({\bar{\Psi}}i\gamma_{5}\lambda_{a}\Psi\right)^{2}
\right]\\
&+&K\left[\mbox{det}{\bar{\Psi}(1+\gamma_5)\Psi+\mbox{det}\bar{\Psi}(1-\gamma_5)\Psi}\right]\nonumber 
\end{eqnarray}
where
\begin{equation}\label{eq.:fieldsuds} \Psi=\left(
\begin{array}{c} u\\ d\\s
\end{array} \right),~~~~~\underline{\mathcal{A}}=\left(
\begin{array}{c} \mu_u\\0\\0
\end{array} \begin{array}{c} 0\\ \mu_d\\0
\end{array}\begin{array}{c} 0 \\ 0\\ \mu_s
\end{array}\right),
~~~~\underline{\mathcal{M}}=\left(
\begin{array}{c} m\\0\\0
\end{array} \begin{array}{c} 0\\ m\\0
\end{array}\begin{array}{c} 0 \\ 0\\ m_s
\end{array}\right)
 \end{equation}

\bigskip\noindent $\underline{\mathcal{M}}$ is the current quark mass matrix which is
taken diagonal and $\underline{\mathcal{A}}$ is the matrix of the quark chemical
potentials. As usual $\lambda_0=\sqrt{\displaystyle{{2\over3}}}~
\mbox{\bf{I}}$ and
$\lambda_{a}~$, $~a=1,...,8~$ are the Gell-Mann matrices.

With respect to our previous studies, we consider here the 't Hooft term which breaks $U(1)_A$. It is the determinant term with coupling $K$ in eq. (\ref{eq:njlagr4}); for the three flavors case, it corresponds to a six fermion interaction. In the chiral case this term breaks the axial current conservation 
\begin{equation}\label{dethooft}
\partial_{\mu}J_5^{\mu}=-4N_f~K~\mbox{Im}(\mbox{det}(\bar{\Psi}(1-\gamma_5)\Psi))
\end{equation}
so that, by comparing with eq.(\ref{axialcharge}), we find for the topological charge density in this model \cite{Hatsuda:1994pi,Fukushima:2001hr}
\begin{equation}
Q(x)=-2K~\mbox{Im~det}(\bar{\Psi}(1-\gamma_5)\Psi))
\end{equation}
By working at the mean field level, the six
 fermion term can be recast into an effective four-fermion one. In such a way the Lagrangian (\ref{eq:njlagr4}) reduces to the usual NJL Lagrangian, apart from a redefinition of the four-fermion coupling constant G into a new set of effective ones, taking into account the flavor mixing arising from the 't Hooft term \cite{Hatsuda:1994pi,Klevansky:1992qe}. Each channel in the interaction (corresponding to different $\lambda_a$, and scalar/pseudoscalar sectors) has a different coupling, which depends on $G$, $K$ and on the average value of other fields.

In the following, we will work at the mean-field approximation.
Therefore, because we are now dealing with four fermion interactions only, we can calculate the effective potential by following
a similar procedure to that of Chapter four, taking care of the new set of couplings. The explicit calculation will not be shown.\\
If we limit ourselves to consider the three scalar condensates, and the pseudoscalar condensate in the light quark sector only, the one-loop effective potential we get is:
\begin{equation}\label{eq:poteff4}
 V=\frac{\Lambda^2}{8G}
(\chi_u^2+\chi_d^2+\chi_s^2)-\Lambda^3\frac{K}{16G^3}~\chi_u~\chi_d~\chi_s+\frac{\Lambda^2}{4G}(1-\frac{K\Lambda\chi_s}{8G^2})(\rho_{ud}^2)+V_{\mbox{log}}
\end{equation}
\vskip0.5cm
where

\begin{eqnarray} \label{eq:vlog4}
V_{\mbox{log}}=-{1\over\beta}\sum_{n=-\infty}
^{n=+\infty}\int{d^{3}p\over (2\pi)^{3}} ~\mbox{tr log} \left(
\begin{array}{ccc}
h_u & -\gamma_5 ~\Lambda~\rho_{ud}~(1-\frac{K\Lambda\chi_s}{8G^2})~&0\\
\gamma_5 ~\Lambda~\rho_{ud}~(1-\frac{K\Lambda\chi_s}{8G^2}) & h_d & 0\\
0& 0 & h_s
\end{array}
\right)
\nonumber
\end{eqnarray}
\begin{eqnarray}
\end{eqnarray}
\begin{eqnarray}
 \ \ \ \ \ \ \ \ \ \ \ ~h_f=(i\omega_n+\mu_f)\gamma_0~-~\vec{p}\cdot\vec{\gamma}~-~
M_f\nonumber
\end{eqnarray}
In eq. (\ref{eq:vlog4}) $\mbox{tr}$ means trace over Dirac, flavor and color indices
and $\omega_{n}=(2n+1)\pi/\beta$ are the Matsubara frequencies.
The dimensionless fields $\chi_{f}$ and $\rho_{ud}$ are
connected to the scalar and pseudoscalar condensates respectively by the following relations

\begin{eqnarray} \label{eq:fields4}
\chi_f &=& - ~4G~{\langle{\bar{\Psi}}_f\Psi_f\rangle\over \Lambda}\nonumber\\
\\
\rho_{ud} &=& -~2G
~{\langle{\bar{u}}\gamma_{5}d-{\bar{d}}\gamma_{5}u\rangle\over
\Lambda}\nonumber
\end{eqnarray}
and are variationally determined at the absolute minimum of the
effective potential. The constituent quark masses are
\begin{equation}\label{Masseconst}
M_i=m_i+\Lambda \chi_i-\Lambda^2\frac{K}{G^2}\frac{\chi_j~\chi_k}{8} ~~~(i\neq j \neq k)
\end{equation}
As an effect of the 't Hooft term, from eq.(\ref{Masseconst}) we can see that the constituent masses receive a flavor-mixing contribution from all of the three flavors. 

For the present application, we will consider the ``standard'' NJL model, without the form factor we had considered in the two previous Chapters. Since the model turns out to be non-renormalizable ($[G]=\mbox{energy}^{-2},[K]=\mbox{energy}^{-5}$), we introduce a hard cut-off $\Lambda$ on the three-momentum.

Through this Chapter, when we consider the physical case (with realistic values of meson masses and decay consant), we assume for the parameters the same values as in ref. \cite{Hatsuda:1994pi}
\begin{equation}\label{param1}
\Lambda=631.4~\mbox{MeV};~~~G~\Lambda^2=3.67;~~~K~\Lambda^5=-9.29;\\
\end{equation}
\begin{equation}\label{param2}
 ~~~~\hat{m}\equiv\frac{m_u+m_d}{2}=5.5~\mbox{MeV};~~m_s=135.7~\mbox{MeV}~~
\end{equation}

 To investigate regimes different than the real one, with a varying number of $N_f$ massless flavors, we will treat the quark masses as free parameters, by keeping coupling and cut-off scale fixed as in eq. (\ref{param1}). 

In the following, it will turn out to be more convenient to introduce the following linear combinations of chemical potentials:

\begin{equation}
\mu_q=(\mu_u+\mu_d)/2;~~~\mu_I=(\mu_u-\mu_d)/2;
\end{equation}

The quark chemical potential $\mu_q$ is just one third of the baryon chemical potential $\mu_q=\mu_B/3$.

\section{Behaviour of critical lines}

\subsection{Critical temperature dependence on baryon/isospin chemical potentials}
The aim of this section is to shed some light on the physics of QCD at finite baryon chemical potential ($\mu_q=\mu_B/3$), by comparing the physics at $\mu_q\neq 0$ and $\mu_I=0$ with that at $\mu_q=0$ and $\mu_I\neq 0$. In fact, the latter case can be studied on the lattice by means of standard importance sampling techniques. The connection of these two regimes could give a deeper understanding of the sign problem in the fermion determinant, and provide us with some procedure to check present simulations and possibly improve numerical algorithms.\\
We will consider here the dependence of the critical temperatures
\begin{equation}\label{crittemp}T_c(\mu_q)\equiv T_c(\mu_q; \mu_I=0)~~~;~~~T_c(\mu_I)\equiv T_c(\mu_I; \mu_q=0)\end{equation} 
for low chemical potentials, obtained by a mean-field analysis of the NJL model.
Obviously, mesons and baryons (and di-quarks) carry different spin and charges, and their properties depend differently on $\mu_q$ and $\mu_I$; for these reasons, one could expect the two curves $T_c(\mu_q)$ and $T_c(\mu_I)$, when starting by the same value at $\mu_q=\mu_I=0$, to be different, at least in the regime where bound states heavily influence the thermodynamics of the system. On the other hand, when the free-energy is mainly ruled by the constituent quarks, there is no reason to expect a dependence of the critical curve on the sign of chemical potentials (which fixes the sign of the total charges associated with the system). \\
Here, at the mean-field level, the effect of bound states is considered when we admit the formation of a pion condensate. Actually, in agreement with chiral models analyses, in the NJL model the pion effective mass dependence on chemical potentials can be analitically computed \cite{He:2005nk} (see Chapter IV). The charged pions chemical potential is exactly the double of the isospin chemical potential. For this reason, as $\mu_I$ is higher than some critical value ($m_{\pi}/2$ at $T=0$), a pion condensate starts to form; a similar effect happens when $\mu_q$ is higher than the critical value for di-quark condensation, which is expected to occur at values $\mu_q>400\div500~\mbox{MeV}$ (of course, before di-quark condensation, for $\mu_q\sim~m_N/3$ and low temperatures there should be the liquid-gas transition for the nucleons). Since in the following we will be interested mainly in the regime of relatively small chemical potentials (lower than $\sim 200~\mbox{MeV}$) we will neglect the latter possibility.\\
In fact, when the pion condensate is zero, the mean field effective potential is symmetric under $\mu_u\rightarrow-\mu_u, ~\mu_d\rightarrow-\mu_d$; this implies that $\mu_q\leftrightarrow\pm\mu_I$ is a symmetry of the problem. Therefore, for zero $\rho$, the two curves $T_c(\mu_q)$ and $T_c(\mu_I)$ have the same analytical dependence.\\
In Fig. \ref{fig:DiagTmu2+1}, \ref{fig:DiagTmui} the phase diagrams in ($\mu_q,T$), ($\mu_I,T$) spaces are shown; starting from the common value $T_0=201~\mbox{MeV}$, corresponding to $\mu_q=\mu_I=0$, the cross-over curves coincide up to the value of about $150~\mbox{MeV}$ for both chemical potentials. For higher $\mu_I$ (and temperatures lower than $\sim 200~\mbox{MeV}$) we are in the condensed pions phase; in agreement with \cite{He:2005nk} this regime will persist until $\mu_I\sim 860~\mbox{MeV}$, before the saturation regime takes place. In that regime the pion condensate vanishes.
The line $T_c(\mu_I)$ which defines the pion condensation phase should be in agreement with the calculation coming from a relativistic, interacting Bose gas. This line should be continuously connected with the critical curve of a BCS system, which should take place at higher chemical potential, according to perturbative QCD. 
The cross-over transition, according to the papers of Son and Stephanov \cite{Son:2000xc,Son:2000by}, should take place at a value $\mu_I\sim~400\mbox{MeV}$.
The saturation regime (i.e. the regime at high $\mu_I$ where the pion condensate vanishes), which is not expected from a perturbative calculation in the asymptotic high $\mu_I$ regime, is a feature of the present model, which is not reliable at high $\mu_I$. However, it is a feature common to the Random Matrix approach \cite{Klein:2003fy}.

On the other hand, if we follow the cross-over line $T_c(\mu_q)$, we will find, as expected, a critical ending point for $\mu_q=330~\mbox{MeV},T=42~\mbox{MeV}$, and a line of first order transitions for higher $\mu_q$.\\
By summarizing, at this level of calculation (mean field), as long as isospin chemical potential is lower than the critical value for pion condensation the curves $T_c(\mu_q)$ and $T_c(\mu_I)$ have the same analytical expression $T_c(\mu_q)=T_c(\mu_I)$. This could be interesting in the attempt to extend lattice results from $\mu_I\neq0$ and $\mu_q=0$ to $\mu_q\neq 0$ and $\mu_I=0$ (at least in the region of low chemical potentials).
Our conclusions appear to be in agreement with the authors of ref. \cite{Toublan:2004ks} (both from the lattice and from a hadron resonances gas model) .

\begin{center}
\begin{figure}[htbp]
\includegraphics[width=14cm]{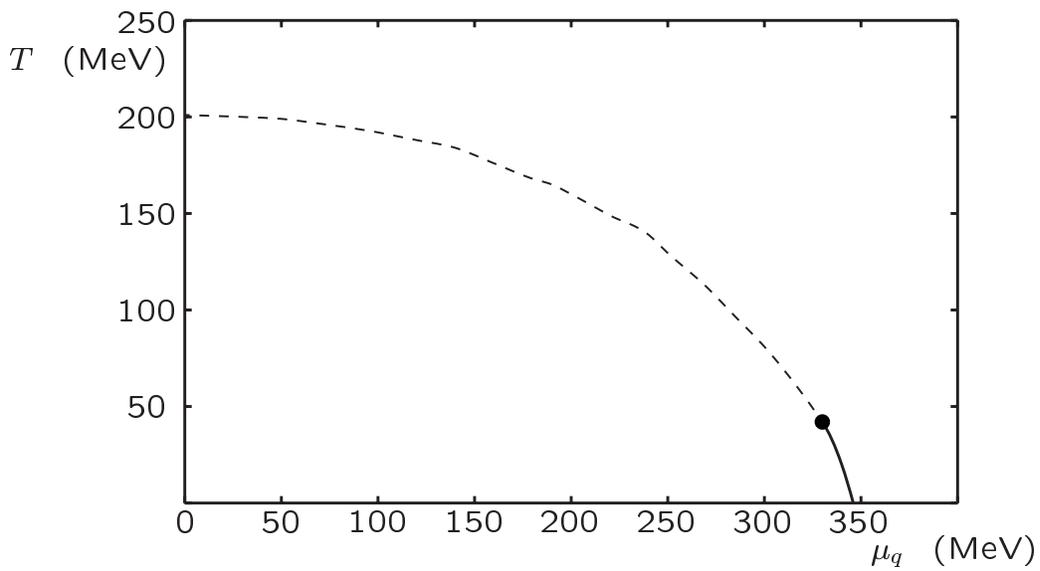}
\caption{\it Phase diagram in the plane ($\mu_q,T$), for the physical 2+1 case; $\mu_I$ and $\mu_s$ are set to zero. Dashed/solid lines indicate cross-over/first order transitions; consequently, the dot in the picture labels the critical ending point.} \label{fig:DiagTmu2+1}
\end{figure}
\end{center}

\begin{center}
\begin{figure}[htbp]
\includegraphics[width=14cm]{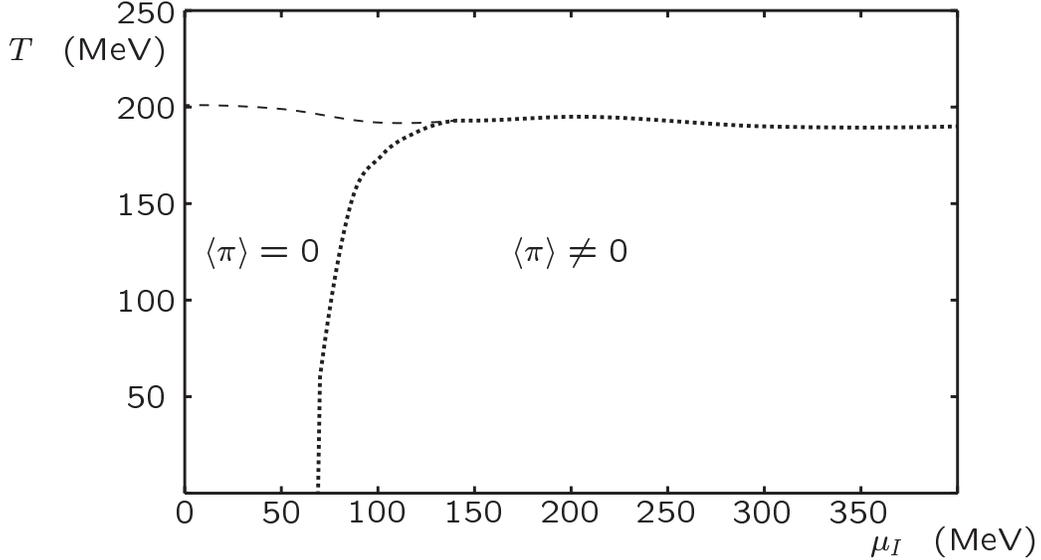}
\caption{\it Phase diagram in the plane $(\mu_I,T)$, for the physical 2+1 case; $\mu_q$ and $\mu_s$ are set to zero. The dashed line indicates a cross-over transition for the scalar condensates, whereas dotted line stands for genuine second order transition for the pion condensate. In the regions of a non-vanishing pion condensate, the discontinuous behaviour of the scalar condensate turns into a continuous one, therefore the critical ending point is not present in this case.} \label{fig:DiagTmui}
\end{figure}
\end{center}

\subsection{Order of the transition by varying $m_s$}

It is generally assumed from universality arguments \cite{Pisarski:1983ms} and lattice analyses \cite{Laermann:2003cv}, that the order of the phase transition by increasing the temperature at zero density can change as the 
quark mass values are varied. If in the realistic case (with physical quark masses) the zero density transition is expected to be a cross-over, lattice analyses seem to show a first order transition when the three light quark masses are small enough. In particular, by taking $m_u=m_d=0$, there should be a critical value for $m_s$ below which the transition turns into a discontinuous one: different lattice approaches find $m_s^C$ to be half of the physical value of the strange quark mass \cite{Brown:1990ev,Aoki:1998gi} or $m_s^C\sim 5\div10 ~ m_{u,d}~(\mbox{physical})$ \cite{Laermann:2003cv}.\\
To study this aspect in the NJL model, we start from the parameters fit of \cite{Hatsuda:1994pi}, and we take the quark masses as free parameters; namely, we take the four and six fermion couplings fixed so as to reproduce the phenomenology of the realistical physical situation. This way of proceeding could seem as rather arbitrary, but for any value of the masses considered here, we have verified that the output parameters have reasonable values (critical temperatures $130\div200~\mbox{MeV}$, light quark scalar condensates $(-250\div-240~\mbox{MeV})^3$, constituent light quark masses $250\div350~ \mbox{MeV}$). Of course our results will be strongly model dependent, both for the choice of a specific set of parameters and of a particular model itself. For instance, the NJL model provides an estimate of the position of the critical point at lower temperatures and higher chemical potentials with respect to those obtained in ladder-QCD and the ones from recent lattice simulations \cite{Fodor:2004nz}.

We can clearly understand the appearance of a first-order phase transition along the temperature axis if we consider a Ginzburg-Landau expansion of the free energy

\begin{equation}
F(\Phi)=a\Phi^2+b\Phi^4+c\Phi^3-m\Phi
\end{equation}
we are here considering three flavors with the same mass $m$, and $\Phi$ stands for the common value of the scalar fields. The relevant feature of the instanton-induced coupling of eq.(\ref{eq:poteff4}) is the appearance of a cubic term in the expansion. 
At $T=\mu=0$ we have $a<0$, $c<0$.
Following universality arguments \cite{Gavin:1993yk}, this can lead the transition to be first-order in the chiral case. This happens if the cubic term is still negative at the temperature where $a$ changes sign and becomes positive.
A finite mass term plays the role of ``smoothing'' the cubic term, so that for $m>m^C$ the transition turns into a cross-over.
This study will be the subject of a future work.

In Fig. \ref{fig:CondTms8} we show the behaviour of the light quark scalar condensates vs. temperature, in the $m_u=m_d=0$ limit, and for vanishing chemical potentials too. In the upper picture, $m_s$ is taken to be zero, and for a temperature of about $130~\mbox{MeV}$ there is a sharp first order transition. As we increase $m_s$ the discontinuity of the scalar condensates reduces (in the lower picture the case $m_s=8~\mbox{MeV}$ is shown), and when $m_s$ exceeds the critical value $m_s^C=10~\mbox{MeV}$, the zero temperature transition turns into a genuine second order transition. The value we get for $m_s^C$ is in any case smaller than the one from lattice predictions.

In Fig. \ref{fig:Diagmusmu} we plot the phase diagram in the ($\mu_q,m_s$) space, by taking $m_u=m_d$ fixed to zero, and $\mu_I=\mu_s=0$. The label $I/II$ indicates, for every couple $(\mu_q,m_s)$, whether, by increasing the temperature starting from zero, the transition is a first or a second order one. Actually, up to $m_s<m_s^C=10~\mbox{MeV}$, we have first order transitions for every value of $\mu_q$, and consequently there is no critical point; for $m_s$ slightly above the critical value, the critical point locates at a value of $\mu_q\sim200~\mbox{MeV}$. The critical value for $\mu_q$ grows together with $m_s$ until the strange quark decouples from the two light quarks and the critical point $\mu_q$ coordinate is independent on $m_s$, and lies at $\mu_q\sim~300~\mbox{MeV}$.\\
Finally we have studied whether, in agreement with lattice analyses, a first order transition persists when we consider a non zero but small $m_u=m_d$: this does not happen in the NJL model with our choice of parameters, for any value of $m_s$. A recent work based on the linear sigma model has found the critical value for $m_u=m_d=m_s$ to be $m_{crit}=40\pm20\ \mbox{MeV}$ \cite{Herpay:2005yr}.

\begin{center}
\begin{figure}[htbp]
\includegraphics[width=14cm]{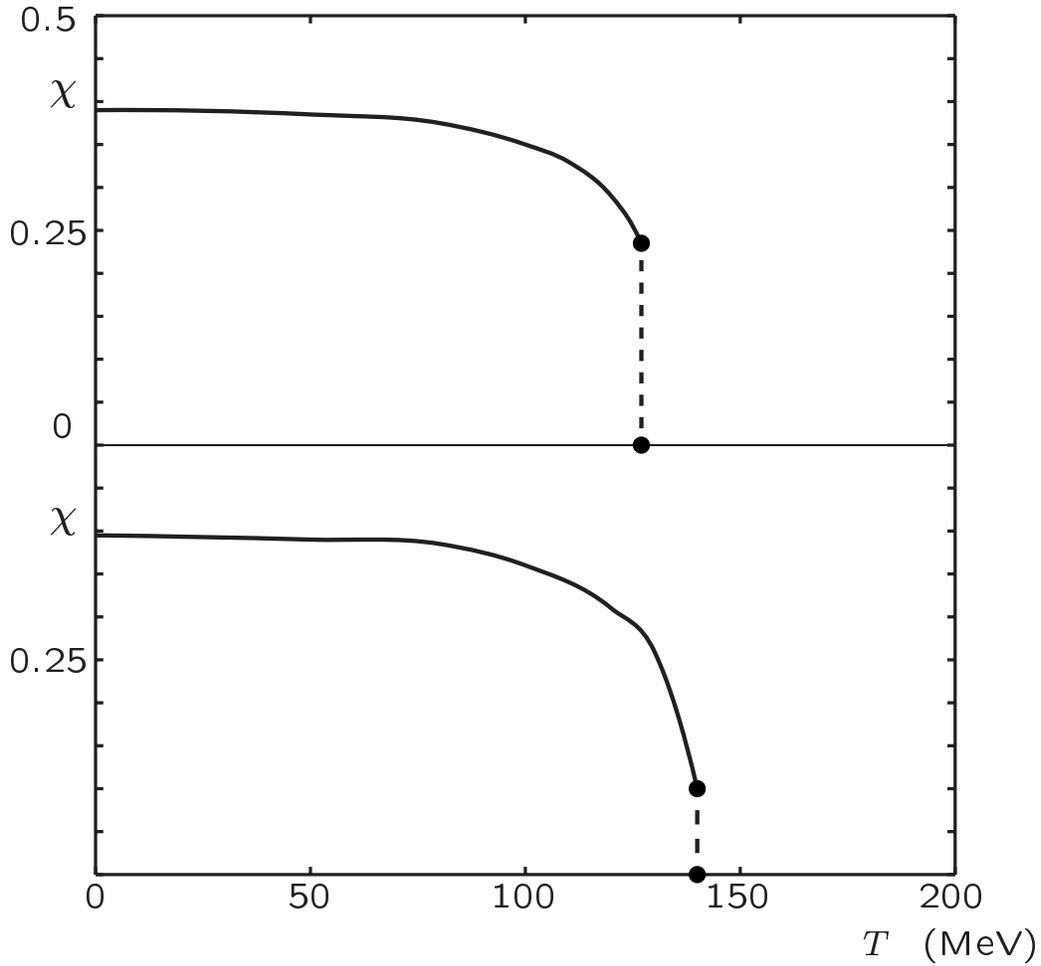}
\caption{\it Behaviour of the light quark scalar condensates as a function of temperature at zero chemical potentials and $m_u=m_d=0$, for $m_s=0$ (upper picture) and ${ m_s=8~\mbox{MeV}}$  (lower picture). In the upper picture, $\chi\equiv\chi_u=\chi_d=\chi_s$, in the lower $\chi\equiv\chi_u=\chi_d$. When the strange quark mass exceeds the critical value $m_s^C=10~\mbox{MeV}$, the discontinuous behaviour turns into a genuine phase transition.} \label{fig:CondTms8}
\end{figure}
\end{center}
\begin{center}
\begin{figure}[htbp]
\includegraphics[width=14cm]{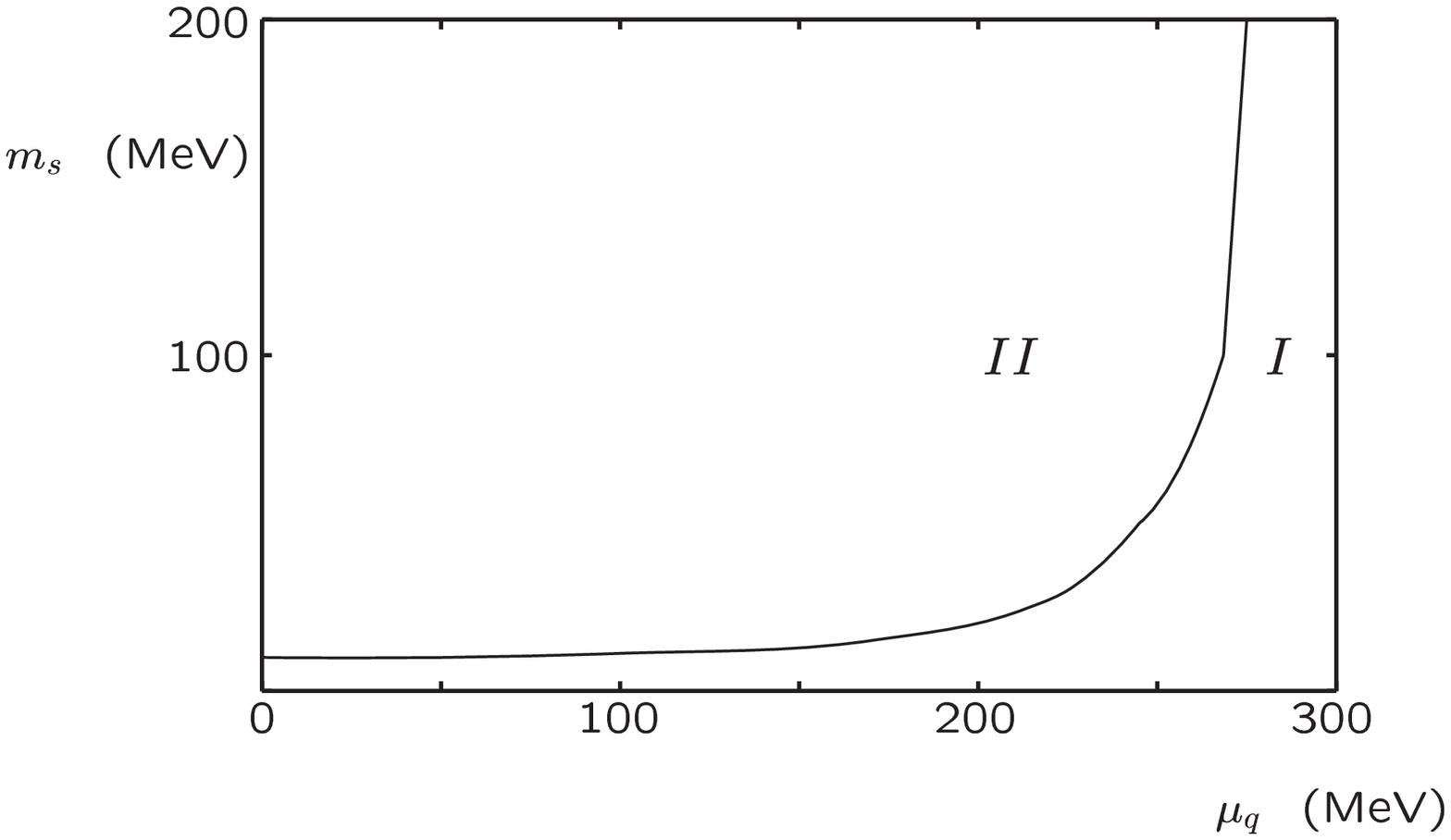}
\caption{\it Phase diagram in the ($\mu_q, m_s$) space, at $m_u=m_d=0$ and $\mu_I=\mu_s=0$. For every region in the diagram, the label $I$/$II$ means whether, by increasing temperature and starting from $T=0$, the transition is of first or second order. Actually, the line in the diagram separating the two different regions follows the critical point by varying $m_s$.} \label{fig:Diagmusmu}
\end{figure}
\end{center}

\subsection{Critical lines as a function of $N_f$}

Even though lattice analyses at finite density still present ambiguities in their different approaches, some general features about the critical line appear to be rather solid. 
In particular, if $T_0$ is the critical temperature for zero chemical potentials, the dependence of $(T/T_0)$ as a function of $(\mu/T_0)$ should be parabolic, (at least in the regime  $(\mu/T_0)<1$), i.e. of the form $(T/T_0)=1-\alpha (\mu/T_0)^2$.
Secondly, the $\alpha$ coefficient should depend on the number of flavors $N_f$, increasing with $N_f$; in fact, the curves relative to $N_f=2,2+1,3$ should be very close to each other and the one relative to $N_f=4$ should be steeper.\\
It is clear that a dependence of the $\alpha$ coefficient on $N_f$ must be related in any case to a coupling between the flavors; otherwise, the effective potential would become a sum of single flavors contributions, and the critical temperature would not depend on $N_f$. This fact can give us an idea about the strength of the coupling between the flavors at the phase transition, and of its possible reduction with temperature.
We will check the issue relative to the cases $2,2+1,3$ in the framework of the NJL model.\\
In this context, the aforementioned situations labeled by 2,3 are those with 2,3 massless flavors (in the case 2, $m_s$ is set to $5~\mbox{GeV}$ to decouple the strange quark); 2+1 is the physical case with realistic values of quark masses. In the following, $\mu$ will indicate a common value for the chemical potential equal for all the active flavors.

\begin{center}
\begin{figure}[htbp]
\includegraphics[width=14cm]{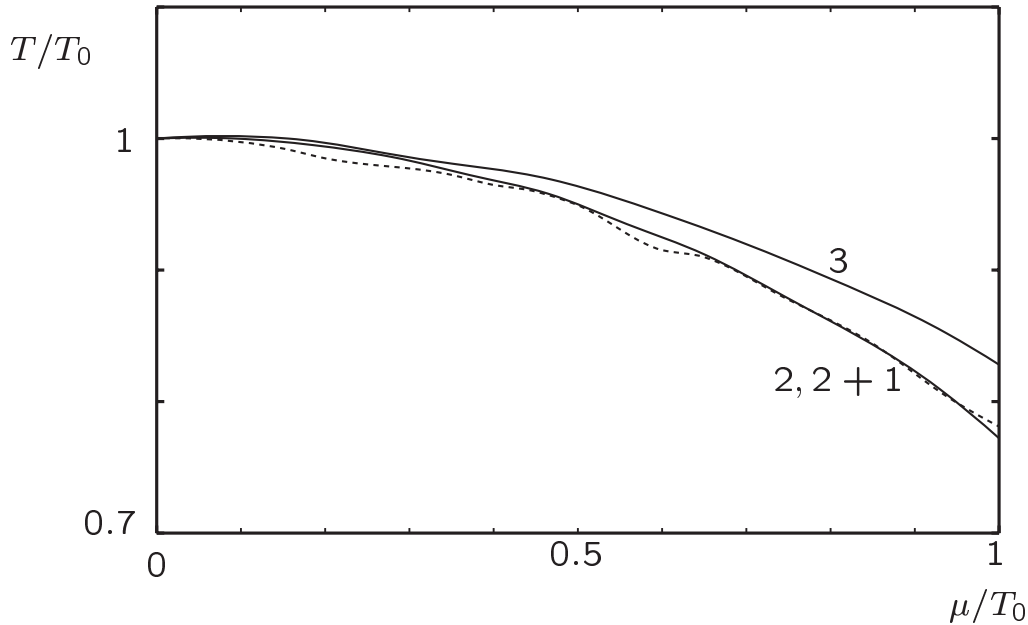}
\caption{\it Plot of the three cases $2,2+1,3$ in dimensionless units ($\mu/T_0,T/T_0$), for $\mu/T_0\leq1$. The curves relative to $2$ and $2+1$ (the dashed one) almost overlap. In disagreement with lattice simulations, the curve relative to $3$ stands slight above the others.} \label{fig:DiagTmuqadimflavors}
\end{figure}
\end{center}

In Fig. \ref{fig:DiagTmuqadimflavors} we show the phase diagrams relative to the cases $2+1,2,3$. Obviously, in the case $3$, we are in the situation where $m_s<m_s^C$ and we have only first order transitions. As we are varying the quark masses to consider different situations, $T_0$ has a large range of variation (from $130$ to $200~\mbox{MeV}$); it is impressive that, as we plot the phase diagrams in dimensionless units, these large differences almost cancel out. This is the most striking evidence that our approach, in its simplicity, has some validity.

When showing the three cases on the same diagram, we can clearly observe the surprising overlap of the results.
The agreement between $2$ and $2+1$ cases (apart from numerical instabilities related with $2+1$ case) is remarkable, and good between $2$ and $3$ cases; the slope of case $3$, with respect to the lattice previsions, is smaller than that of $2$, but the difference is slight. \\

The results we get for $\alpha$ are the following: \\
\begin{equation}
~~2~~~~~~~~~~~\alpha=0.1995\pm0.025~~~~~~~~~~~~~~~~~~~~~~~~~~~~~~~~~~~~~~~~~~\nonumber
\end{equation}
\begin{equation}
~~~~~~~2+1~~~~~~~~~\alpha=0.2496\pm0.0623~~~~~~~~~~~~~~~~~~~~~~~~~~~~~~~~~~~~~~~~~~
\end{equation}
\begin{equation}~~3~~~~~~~~~~~\alpha=0.1614\nonumber\pm0.011~~~~~~~~~~~~~~~~~~~~~~~~~~~~~~~~~~~~~~~~~~\end{equation}

The worse precision we obtain for the $2+1$ case depends on the larger error in the determination of the cross-over curve.\\

For completeness, we quote the results for $\alpha$ from lattice analyses: \\
\begin{eqnarray}&2&~~~~~~~\alpha=0.0507\pm0.0034;~\mu_u=\mu_d=\mu~ \cite{D'Elia:2002gd}\nonumber\\
&2&~~~~~~~\alpha=0.0504\pm0.0036;~\mu_u=\mu_d=\mu~ \cite{deForcrand:2003hx}\nonumber\\
&2&~~~~~~~\alpha=0.07\pm0.03;~\mu_u=\mu_d=\mu~ \cite{Ejiri:2003dc}\nonumber\\
&2+1&~~~~~~~\alpha=0.0288\pm0.0009;~\mu_u=\mu_d=\mu_s=\mu=\mu_B/3~ \cite{Fodor:2001pe}\\
&3&~~~~~~~\alpha=0.0610\pm0.0009;~\mu_u=\mu_d=\mu_s=\mu=\mu_B/3~ \cite{deForcrand:2003hx}\nonumber\\
&3&~~~~~~~\alpha=0.114\pm0.046;~\mu_s=0~ \cite{Ejiri:2003dc}\nonumber\\
&4&~~~~~~~\alpha=0.099;~ \mu_f=\mu=\mu_B/3~ \cite{D'Elia:2002gd}\nonumber
\end{eqnarray}

\vskip0.5cm
We find that our predictions for $\alpha$ are bigger than those obtained by lattice approaches, apart from \cite{Ejiri:2003dc} for $N_f=3$; in this case, the results are comparable. A recent study based on a hadron resonance gas model \cite{Toublan:2004ks} gives the result~~~~\\ $\alpha=0.17\pm0.01$ and this value is in a good agreement with our results. A study within a chiral quark model gives for $\alpha$ a value of about 0.1, extracted from Fig. 2 of \cite{Jakovac:2003ar}. A Statistical Hadronization model provides the result $\alpha=0.2299$ \cite{Becattini:2003wp}; in this case, the critical line must be associated with the freeze-out of hadrons rather than with chiral symmetry restoration. 

We have also studied the behaviour of the critical curve in ladder-QCD \cite{Barducci:1987gn} (in its version \cite{Barducci:2003un}); in this model there is no coupling between flavors, therefore it is independent on $N_f$.
We have found that the critical curve is flatter than that of NJL model, namely the coefficient $\alpha$ is much smaller, and hence closer to lattice predictions: $\alpha=0.0797\pm0.0056$.
We have attributed this feature to the lack of coupling between flavors in the model. On the other hand, since the value of the $\alpha$ coefficient that we find in the NJL model is slightly higher than the value found in ref. \cite{Toublan:2004ks} and sensibly higher than the values obtained from other lattice analyses, we can argue that by introducing an effective reduction of the $K$ coupling with temperature, as considered in ref. \cite{Hatsuda:1994pi,Costa:2005cz}, $\alpha$ would decrease at the same time. Therefore, by considering the following temperature dependence of the 't Hooft term
\begin{equation}K(T)=K_0~\mbox{exp}(-T/T_1)^2\end{equation}  
we have verified that reducing $T_1$ the critical  temperature at $\mu=0$ is also reduced; in this way, the curve gets flatter. By taking two different values for $T_1$, and considering for simplicity the case $2$, we find the value $\alpha=0.1186\pm0.0061$ in the case\\ $T_1=160~\mbox{MeV}$, and $\alpha=0.1084\pm0.0035$ in the case $T_1=100~\mbox{MeV}$.
In this way, the agreement with lattice is considerably improved; this can be considered an indirect proof of $U(1)_A$ effective restoration with temperature. 

For the sake of completeness, we have also studied the two flavor NJL model with the `t Hooft determinant. The model is not completely equivalent to the three flavor model in the limit of infinite $m_s$, since in the latter case a strange quark loop gives a contribution, proportional to the strange quark condensate, to the light quark constituent mass. In any case we do not expect a dramatic change in our results, with respect to previous expectations, for the behaviour of the critical curve.\\
We consider in this case the following expression for the interaction part of the Lagrangian:

\begin{equation}\label{NJL2flavors}
\mathcal{L}_{int}=G_1[(\bar{q}q)^2+(\bar{q}\vec{\tau}q)^2+(\bar{q}i\gamma_5q)^2+(\bar{q}i\gamma_5\vec{\tau}q)^2]+G_2[(\bar{q}q)^2-(\bar{q}\vec{\tau}q)^2-(\bar{q}i\gamma_5q)^2+(\bar{q}i\gamma_5\vec{\tau}q)^2]
\end{equation}
with
\begin{equation}
G_1=(1-\beta)G_0~~~;~~~G_2=\beta G_0
\end{equation}

For the two flavor case, the 't Hooft term writes as a four fermion interaction with coupling $G_2$. By looking at the $G_2$ term of eq.(\ref{NJL2flavors}), we see that combinations like 
\begin{equation}\label{U1breaking}
(\bar{q}q)^2-(\bar{q}i\gamma_5q)^2
\end{equation}
are not invariant under $U(1)_A$ transformations. 
On the other hand the $G_1$ term is $SU(2)_L\otimes SU(2)_R$ invariant.
The $\beta$ coefficient tells us how hard the flavor mixing is; it is maximal for $G_1=0$, namely for $\beta=1$.

For the choice of the parameters $G_1$ and $G_2$ we follow the approach proposed in ref. \cite{Hatsuda:1994pi,Frank:2003ve}.
In ref. \cite{Hatsuda:1994pi} the authors study the original two flavor Lagrangian, proposed by Nambu and Jona-Lasinio \cite{Nambu:1961fr}, with $G_1=G_2$ and therefore $\beta=0.5$. The value we find in this case for $\alpha$ is very similar with the result we obtained in the $SU(3)$ case in the limit $m_s\rightarrow\infty$: 
\begin{equation}\alpha=0.2107\pm0.0214\end{equation}

The authors of ref. \cite{Frank:2003ve} take instead $\beta$ as a free parameter. Here we furthermore consider the possible dependence of the $G_2$ coefficient on the temperature,\\ $G_2=G_2(T=0)~\mbox{exp} (-(T/T_1)^2)$.

If we take $G_2$ independent on the temperature (namely with $T_1=\infty$), the value we find for the $\alpha$ coefficient does not change by varying $\beta$: \begin{equation}\alpha=0.2142\pm0.0259\end{equation} This is again in agreement with previous analyses.
On the other hand, if we admit a restoration of the $U(1)_A$ symmetry with temperature, we find a dependence on the $\beta$ coefficient.\\ For $\beta=0.2$ we have \begin{equation}\alpha=0.1484\pm0.009 \ \ \mbox{for} \ \ T_1=160~\mbox{MeV}~~~\end{equation} and \begin{equation}\alpha=0.1196\pm0.0161 \ \ \mbox{for}\ \  T_1=100~\mbox{MeV}.\end{equation}
For $\beta=0.3$ we have \begin{equation}\alpha=0.1024\pm0.0042\ \ \mbox{for} \ \ T_1=160~\mbox{MeV}\end{equation} and \begin{equation}\alpha=0.08101\pm0.007 \ \ \mbox{for} \ \ T_1=100~\mbox{MeV}.\end{equation}

However, according to Shuryak \cite{Shuryak:1993ee} it is very unlikely that the restoration of $U(1)_A$ can occur before chiral symmetry restoration; therefore, the value $T_1=100\ \mbox{MeV}$ should not be taken too seriously. In any case, it appears clear that restoration of $U(1)_A$ symmetry can strongly influence the behaviour of the critical curve. In particular, the effect of the restoration of the axial symmetry with temperature could favour the splitting of the critical lines for the light quarks. It was pointed out in \cite{Frank:2003ve} that, for realistic values of isospin chemical potential, the strenght of the flavor-mixing interaction should prevent the separation of the critical lines.
On this subject, the inclusion of the flavor asymmetry due to the different masses should be taken into account too. We are currently performing an analysis which considers $\Delta m=(m_d-m_u)\sim m_u$ and a temperature-dependent coupling of the instanton induced interaction.

Moreover, a study of the NJL model with an imaginary chemical potential is under investigation. This study would allow a direct comparison with the same analyses performed on the lattice. \\
The mean-field NJL effective potential is symmetric for $\mu\rightarrow-\mu$, and due to its analytical properties depends on $\mu^2$ only, where here $\mu$ is a real chemical potential equal for all of the flavors.
The critical curve depends on $\mu^2$ too.
We can analytically prolungate $\mu\rightarrow~i\mu_{Im}$ to have that
\begin{equation}\label{Vimm}
V((i\mu_{Im})^2)=V(-\mu_{Im}^2)
\end{equation}
Therefore, in the regime of imaginary chemical potentials, we expect to find a critical curve which is the analytical prolongation to negative $\mu^2$ of our previous results.

\section{Conclusions}\label{Conc4}

In this application we have studied some general features of the QCD critical line in the framework of a NJL model. In section one, we have compared the physics at $\mu_q\neq 0$ with that at $\mu_I\neq 0$. In section two, we have varied quark masses to show that the order of finite temperature transition changes if we consider small enough masses. In section three, we have studied the dependence of the slope of the critical curve $T_c(\mu)$ on $N_f$. The fact that we have found a remarkable agreement between our analyses and lattice results may be taken as a proof that the NJL model can be trusted as a good toy model for describing the chiral transition. In addition, it is worth to stress that, since our analyses are performed at the mean-field level, we have basically computed the termodynamics corresponding to a gas of constituent quarks; on the other hand, Lattice-QCD analyses take into account bound states and fluctuations as well, therefore the successful comparison of the two approaches is even more impressive.

\chapter*{Conclusions}

In this work we have carried out a study of the phase diagram of QCD. We have started, in the Preface and in the first introductory Chapter, with a general survey on the QCD phase diagram. We have analyzed
the phenomenon of chiral symmetry breaking and shown its connection with the confinement problem. We have discussed the high temperature regime, where deconfinement should take place, and the low temperature and high densities regime, where color superconductivity is expected to occur. The presence of a critical point along the critical line has been highlighted.

 In the second Chapter we have introduced a subject which became rather popular in the last few years and which motivated our work: the phenomenon of Bose-Einstein condensation of mesons in the low temperature regime of the theory. 
According to several models, pion and kaon condensates are expected to form as the isospin and the hypercharge chemical potentials are high enough.
This phenomenon is correlated to the ``standard''  QCD critical behaviour defined by the study of chiral symmetry breaking and restoration, because the non-perturbative QCD vacuum influences meson properties too. In this Chapter a review on this subject has been given. We have also presented some general aspects of QCD at finite chemical potential which are of interest for the present discussion.

In order to carry out an original research project on this issue, we needed a model displaying the critical chiral behaviour and encompassing the possibility of describing meson condensates.
Actually, due to the non-perturbative regime of QCD, these issues cannot be faced directly within the fundamental theory.
A class of effective models reproduces a critical behaviour which is expected to be consistent with QCD.
Amongst them, two microscopic models, ladder-QCD and Nambu-Jona-Lasinio, have been our tool to investigate the problem.
In these models quarks are treated as fundamental degrees of freedom; chiral symmetry and meson condensation are studied on the same footing, as a dynamical effect, by varying the thermodynamic parameters. We have limited our analyses to the regime of moderate chemical potentials; superconducting effects have not been taken into account.

In the third Chapter we have considered pion condensation in the framework of the ladder-QCD model. This model is able to highlight the critical behaviour related to the chiral symmetry, and gave a prediction on the existence of a critical point along the $(T,\mu)$ critical line from the late '80s on. In the present application, we have found that the critical value of isospin chemical potential $\mu_I^C$ for having pion condensation (at $T=\mu_q=0$) is just half of the pion mass. This result is in agreement with many other analyses obtained within effective models and lattice simulations. We have also studied the effect of an isospin chemical potential on the splitting of the critical lines relative to the light quarks. This fact could be of great interest in heavy ion collision experiments, in the attempt of finding a definite signature for the transition to the deconfined phase of the theory.

In Chapters four and five, we have studied the subject of pion and kaon condensation by using the NJL model. The use of a model different from ladder-QCD was motivated by the technical difficulties encountered for extending the model at finite $(T,\mu)$, when considering meson condensates. The NJL model is simpler to deal with, and encompasses the relevant features for describing the problem as well. The results we have found confirm the possibility of having pion and kaon condensates in the regimes of high isospin and hypercharge densities, respectively.
In particular, in Chapter four we have performed an analysis of the two flavor model in the {\it whole} space of thermodynamic parameters $(T,\mu_q,\mu_I)$. In Chapter five we have extended the study to the third flavor, and studied the competition between pion and kaon condensates in relation with the behaviour of usual chiral condensates. We have proposed an original hypothesis for kaon condensation, by considering a chemical potential associated with light quarks only.
We remark that we have neglected in these works any superconducting effect; a natural extension of this analysis would require the combined presence of di-quark and meson condensates. 

In Chapter six, we have focused our attention to the critical line for chiral symmetry restoration, by neglecting meson condensates. 
We have studied the dependence of the critical temperature $T_c$ on quark and isospin chemical potentials. 
We have shown that the order of the zero chemical potential transition changes from second to first order if we consider low enough masses and a coupling between flavors induced by instantons ('t Hooft).
We have studied the dependence of the critical curve on a chemical potential, by varying the number of massless flavors $N_f$.
This study within a microscopic model, here the NJL model, is motivated by the growing interest that the lattice community has shown for this issue; actually, in recent years many possible algorithms have been proposed to overcome the difficulty (the sign problem) of the  fermion determinant at finite baryon chemical potential. Our results are directly comparable with Monte-Carlo simulations.
A study of a Ginzburg-Landau expansion of the free-energy, in order to highlight the cubic term induced by instantons, is under examination.
Another development regards the study of the critical line as a function of an imaginary chemical potential.
The effect of axial symmetry restoration
and of the explicit isospin breaking ($\Delta m=m_d-m_u$) on the behaviour of the light quarks critical lines is under study as well.

\addcontentsline{toc}{chapter}{Conclusions}
~
\newpage
~

\clearpage
\addcontentsline{toc}{chapter}{Bibliography}


\end{document}